# Emergence: from physics to biology, sociology, and computer science[a]

Ross H. McKenzie
School of Mathematics and Physics
University of Queensland
Brisbane, Australia
condensedconcepts.blogspot.com
r.mckenzie@uq.edu.au

**Abstract**

Many systems of interest to scientists involve a large number of interacting parts and the whole system can have properties that the individual parts do not. The system is qualitatively different to its parts. More is different. I take this novelty as the defining characteristic of an emergent property. Many other characteristics have been associated with emergence are reviewed, including universality, order, complexity, unpredictability, irreducibility, diversity, self-organisation, discontinuities, and singularities. However, it has not been established whether these characteristics are necessary or sufficient for novelty. A wide range of examples are given to show how emergent phenomena are ubiquitous across most sub-fields of physics and many areas of biology and social sciences. Emergence is central to many of the biggest scientific and societal challenges today. Emergence can be understood in terms of scales (energy, time, length, complexity) and the associated stratification of reality. At each stratum (level) there is a distinct ontology (properties, phenomena, processes, entities, and effective interactions) and epistemology (theories, concepts, models, and methods). This stratification of reality leads to semi-autonomous scientific disciplines and sub-disciplines. A common challenge is understanding the relationship between emergent properties observed at the macroscopic scale (the whole system) and what is known about the microscopic scale: the components and their interactions. A key and profound insight is to identify a relevant emergent mesoscopic scale (i.e., a scale intermediate between the macro- and micro- scales) at which new entities emerge and interact with one another weakly. In different words, modular structures may emerge at the mesoscale. Key theoretical methods are the development and study of effective theories and toy models. Effective theories describe phenomena at a particular scale and sometimes can be derived from more microscopic descriptions. Toy models involve minimal degrees of freedom, interactions, and parameters. Toy models are amenable to analytical and computational analysis and may reveal the minimal requirements for an emergent property to occur. The Ising model is an emblematic toy model that elucidates not just critical phenomena but also key characteristics of emergence. Many examples are given from condensed matter physics to illustrate the characteristics of emergence. A wide range of areas of physics are discussed, including chaotic dynamical systems, fluid dynamics, nuclear physics, and quantum gravity. The ubiquity of emergence in other fields is illustrated by neural networks, protein folding, and social segregation. An emergent perspective matters for scientific strategy, as it shapes questions, choice of research methodologies, priorities, and allocation of resources. Finally, the elusive goal of the design and control of emergent properties is considered.

**Table of Contents**

## 1. Introduction and Overview

This article clarifies what emergence is, its centrality to physics, its relevance to other fields of science, its role in scientific strategy and priorities, and the philosophical questions that it raises.

Emergence is related to the observation that the whole can be *qualitatively different* from the parts. In other words, a system composed of many interacting parts can have new properties that the individual parts do not have. This concept describes many fascinating phenomena at the heart of physics, chemistry, biology, psychology, economics, and sociology. A piece of solid gold is shiny, but a single gold atom is not. Water is wet, but a single water molecule is not. Wetness is a property of many water molecules that are in the liquid state. Similarly, the macroscopic properties that distinguish graphite and diamond (e.g., soft vs. hard, black vs. transparent) are emergent. Both diamond and graphite have identical constituents: carbon atoms. Yet they interact in different ways to collectively produce properties that the individual carbon atoms do not have.

On the one hand, scientific disciplines are very different from one another. They differ in the questions they seek to answer, the objects studied, the methods used, concepts developed, and the levels of certainty possible. On the other hand, when considered from an emergent perspective, there are common features across disciplines. Although the focus of this article is on physics I mention scientific challenges in other disciplines. Emergence is at the heart of some of the biggest questions in each discipline. I identify some of the scales and stratum in the discipline and how these are associated with sub-disciplines. The stratum can be viewed both in terms of ontology and epistemology. I characterise a system of interest and identify the parts and the interactions between those parts. I give examples of emergent properties, phenomena, and entities. Emergence is at the heart of many global issues: climate change, epidemics, poverty, mental illness, economic instability, political polarisation, surveillance, and misinformation on social media. This is because they concern large systems with many interacting parts. The whole often has novel properties that are not anticipated and are resistant to control. In this sense, they are examples of what social scientists call "collective action" or "wicked" problems.

### 1.1 Goals of this article

Clarify what emergence is and what its characteristics are. This is necessary because of the diversity and ambiguity of views about what emergence is.

Introduce physicists to the role of emergence in other fields including computer science, biology, and sociology. Stimulate greater cross-fertilisation between physics and other fields.

Illustrate how emergence is central to the biggest questions and challenges in the sciences and in society. Help physicists put their work in a larger context, appreciating similarities and differences with other fields.

Consider some of the practical implications of emergence for how science is done and for deciding scientific priorities.

Describe some of the philosophical issues associated with emergence.

The article aims to be pedagogical and hopefully inspirational. Referencing is not comprehensive, being selective, and is based on accessibility. More background, detail, and differing perspectives can be found elsewhere. Semi-popular books on emergence have been published by Holland,[1] Johnson,[2] Kauffman,[3] Laughlin,[4] and Morowitz.[5] Academic books have been written from the perspective of Complexity science by Jensen,[6] Economics by Krugman[7] and Schelling[8], Psychology and Sociology by Sawyer,[9] Philosophy by Batterman,[10] Physics by Bishop,[11] Computer Science by Fromm,[12] and a range of fields in a volume edited by Davies and Clayton.[13] There is also an article on Emergent Properties in *The Stanford Encyclopedia of Philosophy,*[14] and *The Routledge Handbook of Emergence*,[15] and an anthology of classic articles.[16]

### 1.2 Ten key ideas

Consider a system that is comprised of many interacting parts, involves multiple scales, or undergoes many iterations.

1. Many different definitions of emergence have been given. I take the **defining characteristic** of an **emergent property** of a system as **novelty,** i.e., the individual parts of the system do not have this property.

2. Many other characteristics have been associated with emergence, such as universality, order, complexity, unpredictability, irreducibility, diversity, self-organisation, discontinuities, and singularities. However, it has not been established whether these characteristics are necessary or sufficient for novelty.

3. Emergent phenomena are **ubiquitous** across scientific disciplines from physics to biology to sociology to computer science. Consequently, discipline boundaries become fuzzy with respect to the techniques used and the relevance of results obtained. Emergence is central to many of the biggest scientific and societal challenges today.

4. Reality is stratified. At each **stratum** (level) there is a distinct ontology (properties, phenomena, processes, entities, and effective interactions) and epistemology (theories, concepts, models, and methods). This stratification of reality leads to semi-**autonomous** scientific disciplines and sub-disciplines.

5. A common challenge is understanding the relationship between emergent properties observed at a macroscopic scale and what is known about a microscopic scale: the components and their interactions. A key and profound insight is to identify a relevant emergent **mesoscopic scale** (i.e., a scale intermediate between the macro- and micro- scales) at which new entities emerge and interact with one another weakly. In different words, a **modular structure** may emerge at the mesoscale.

6. Key theoretical methods are the development and study of **effective theories** and **toy models**. Effective theories describe phenomena at a particular scale. Sometimes they can be used to bridge the microscopic (or mesoscopic) and macroscopic scales. Effective theories are an accurate representation of reality in ways that toy models are not. Toy models involve minimal degrees of freedom, interactions, and parameters. Toy models are amenable to analytical and computational analysis and may reveal the minimal requirements for an emergent property to occur. The **Ising model** is an emblematic toy model that elucidates not just critical phenomena but also key characteristics of emergence.

7. **Condensed matter physics elucidates** many of the key features and challenges of emergence. Unlike brains and economies, condensed states of matter are simple enough to be amenable to detailed and definitive analysis but complex enough to exhibit rich and diverse emergent phenomena.

8. The perspective taken about emergence by individual scientists and communities matters for **scientific strategy**. It will shape questions, choice of research methodologies, priorities, and allocation of resources.

9. An emergent perspective that does not privilege the parts or the whole can address contentious issues and fashions in the humanities and social sciences, particularly around **structuralism.**

10. Emergence is central to important questions in **philosophy** including the unity of the sciences, the relationship between theory and reality, and the nature of consciousness.

### 1.3 An example: language and literature

Characteristics of emergence can be illustrated with an example given by Michael Polanyi: how literature emerges from language (Figure 1). At each level there are distinct components, rules of interaction, phenomena, and concepts. For example, grammar provides the rules about how words can interact to produce sentences.

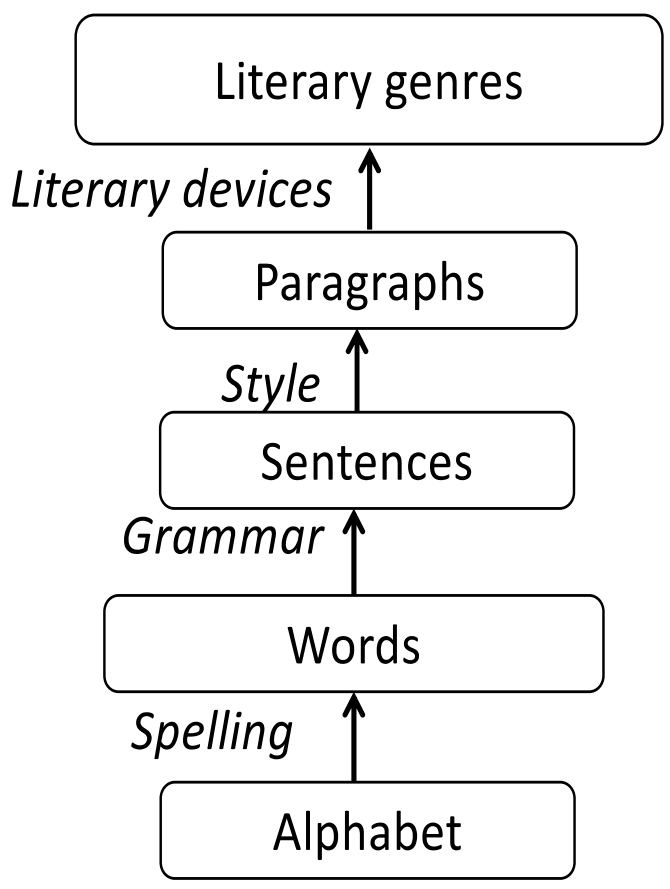

Figure 1. Emergence and the hierarchy of levels associated with language and literature. The boxes represent different components of a system. The rules for interactions between these components are given next to the arrows.

I now illustrate some characteristics of emergent properties with examples from language and literature, originally noted by Michael Polanyi.[17] Literary genres include the novel, drama, poetry, comedy, satire, romance, and tragedy. The concept of a romantic novel has no meaning below the top level. I once learnt the basics of French vocabulary and grammar and could read rudimentary sentences. However, this knowledge was not sufficient for me to conceive of different genres in French literature. A poem involves lines of verse in a particular relationship to each other. The concept of a literary genre may be independent of what language it is written in. It does not depend on details of the alphabet, spelling, and grammar that are used.

To illustrate this effective approach, I return to literature and language. Understanding a particular literary genre does not focus on the alphabet, spelling, or grammar. Rather, the genre may be best understood in terms of larger units such as paragraphs and literary devices. Again, great literary critics have insights about how to do this analysis.[b] This example of emergence is amenable to a more formal analysis through Chomsky's Universal Grammar.

Another example that is amenable to explaining emergence to non-scientists is that of geometry and was discussed by Luisi.[18] Consider how dots can be combined to produce lines, lines to produce planar shapes, and then solid shapes.

### 1.4 Scales and strata

Central to emergence is the idea of **scale**. Emergent properties only occur when scales become larger. Scales that are simply defined, and might be called extrinsic, include the number of parts in the system, length scale, time scale, or energy scale.

A more subtle scale, which might be called intrinsic, is a scale associated with an emergent property. This emergent scale is intermediate between that of the parts and that of the whole system, i.e., a mesoscale. Examples include the persistence length of a polymer molecule, the coherence length of a superconductor, the Kondo temperature for magnetic impurities in metals, and the mass of the W and Z bosons that mediate the weak nuclear force.
Emergent scales lead naturally to **strata** or hierarchies resulting from a separation of scales. I prefer to not use the term hierarchies as it might suggest that some levels (strata) are more important or fundamental than others. Figure 1 is an example for language and literature. There is stratum associated with different scientific disciplines, as discussed below. Strata also occur within individual disciplines, leading to sub-disciplines as will be discussed at many points in this paper.
At each stratum or level, there is a distinct ontology (what is real) and epistemology (how we know and describe reality). For ontology, there are distinct phenomena, entities, properties, processes, interactions, and relations. For epistemology, there are distinct concepts, theories, models, and scientific methods. The distinctness of levels means there is some autonomy within each level. What happens within the level can be described and understood without reference to other levels.[c]
An emergent property of some systems is that they are scale-free, with properties that depend on power laws over some range of scales, sometimes covering many orders of magnitude. In scale-free systems there is often fractal or self-similar structure where the system appears to have the same structure at each scale.

---

[b] This example is not a perfect analogy with physical systems in which emergent properties spontaneously emerge. The example of language and literature involves an external agent who creates the new entities within the constraints allowed by the rules.

[c] For example, with regard to biology, in 1945 Novikoff argued "the laws describing the unique properties of each level are qualitatively distinct, and their discovery requires methods of research and analysis appropriate to the particular level."[19]

Sometimes as the scales increase it is said that the complexity of the system increases. For example, humans are more complex than crystals which are more complex than atoms. However, defining complexity is challenging, as discussed below, and is ambiguous. For example, as one goes from the scale of quarks and gluons to a hydrogen atom, the system gets simpler in some sense.

### 1.5 The stratification of nature and scientific disciplines

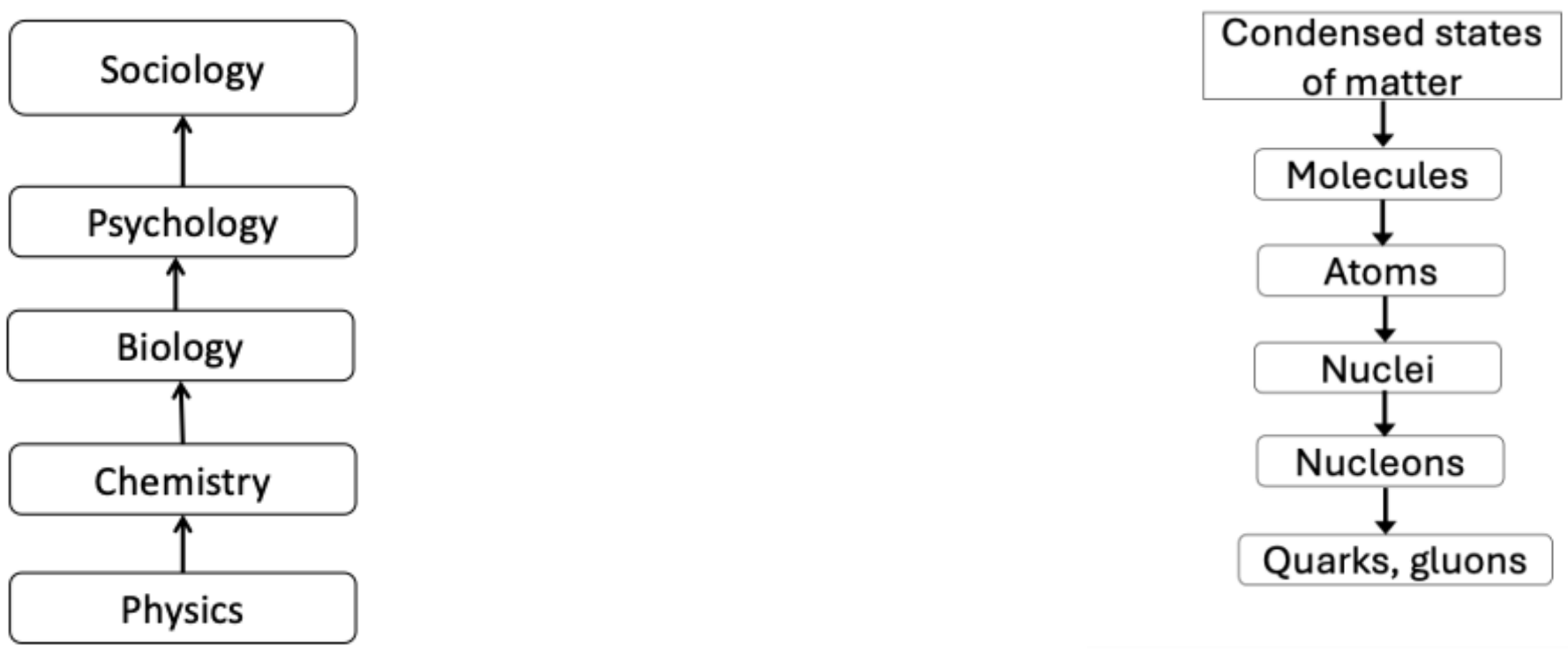

Figure 2a. The stratification of scientific disciplines. The vertical direction represents an increase in scale (number of atoms, length, or time).
Figure 2b. The stratification of objects of interest to physicists. There is a sub-discipline of physics associated with each stratum (level).

The process of considering complex systems in terms of their component parts and interactions leads naturally to a stratified view of reality. This illuminates the relationship between different scientific disciplines (Figure 2a).

At each level of the hierarchy, the objects of interest are composed of objects from the level below. For example, the molecules studied by chemists are composed of the atoms studied by physicists. At lower levels, there is a decrease in the degree of complexity of the systems (loosely defined as the number of variables required to give a complete description of the state of the system). In addition, the relevant time and length scales get smaller at lower strata.[d]

There is an ontology and an epistemology associated with each strata. The ontology includes objects, interactions between objects, and phenomena (states, processes, and properties). The epistemology includes laws, concepts, categories, organising principles, theories, and methods of investigation. In some sense, the ontology concerns things that happen in the system under study and the epistemology concerns things that happen inside the brains of the investigators. Sometimes the boundaries between strata are not well-defined. A stratum is well-defined if there is closure, a concept that is discussed further below.

[d] This increase in scale is not clear cut. For example, some sub-fields of physics, such as fluid dynamics, are concerned with similar time and length scales as biology. It is just that physics is also concerned with scales much smaller than all the other disciplines.

Within disciplines, there are also strata, associated with sub-disciplines. For example, in biology there are sub-disciplines associated with proteins, genes, cells, tissues, organs, organisms, and ecology. In physics, there are sub-disciplines associated with elementary particles, nuclei, atoms, molecules, solids, and fluids.

Viewing the disciplines as forming strata raises interesting questions. What is the exact relationship between the disciplines? Does the fact that physics is on the bottom mean it is the most fundamental discipline? If we fully understand things on one level, can we actually explain everything on the next higher level? For example, can all of biology really be explained solely in terms of chemistry?

*Up and/or down?*

Diagrams of strata often include arrows pointing between adjacent stratum. The direction of these arrows may be associated with an increase or decrease of scale, with causality, or with explanation.

In Figure 2a, the arrows point up to represent that entities and phenomena on one level emerge from the interactions between the entities on the next lowest level. On the other hand, in terms of the historical progression of science, the arrows tend to point down, as in Figure 2b. For example, in physics understandings of macroscopic phenomena such as thermodynamics and electromagnetism preceded understanding atoms which in turn preceded understanding sub-nuclear physics. Weinberg claims all the arrows of explanation point down.[20]

In Andrew Steane's picture of the explanatory relationship between physics, chemistry, and biology, he draws arrows pointing in both directions.[21] The up arrow is denoted "supports [allows and physically embodies the expression of]" and the down arrow is denoted "enarches [exhibits the structures and behaviours that make sense in their own terms and are possible within the framework of]."

There are subtle issues about the role of causality in connecting the different levels, particularly downwards. Consider the simple example of a gas inside a thermally insulated piston. Collisions between the molecules in the gas determine the pressure of the gas and exert a force on the piston wall. On the other hand, slowly moving the piston to reduce the volume of the gas with increase the average kinetic energy of the molecules in the gas.

Emergence suggests a scientific strategy for understanding complex systems, whether superconductors or viruses. Essential elements of that strategy include focusing on what experiments tell us, using a multi-faceted approach (a range of experimental, theoretical, and computational methods), and developing and understanding simple models that may capture the essential features of a phenomenon.

This strategy has implications for setting priorities, whether for an individual scientist (choice of topic and technique) or a funding agency (goals and budgets for different areas). It is best to invest in a portfolio of complementary approaches that look at different scales, from the microscopic to the macroscopic.

### 1.6 The scientific challenge of emergence

Over the last century the scientific strategy of *reductionism*, has been very successful in providing both a quantitative description and conceptual understanding of a wide range of natural phenomena. For example, biologists break organisms down to cells and then to membranes, proteins, and DNA. Physicists consider atoms as composed of nuclei and electrons, nuclei as composed of protons and neutrons, which in turn are composed of quarks. The advances made through scientific reductionism are many, such as discoveries of genetic information encoded in DNA and the molecular basis of genetics. The "zoo" of elementary particles produced by high energy collisions in particle accelerators can be understood in terms of just a few elementary particles, quarks and leptons. The structure and excitations of molecules and the dynamics of chemical reactions can be explained in terms of the laws of quantum theory.

But reductionism is only part of the story. I know the rules of chess but that does not make me a grand master. Knowing the constituents of a system of scientific interest and the laws that describe their interactions does not mean we can understand or predict the collective properties of the system. Indeed, this is central to the biggest challenges in the natural and social sciences today. Here are just a few examples.

Even if we know the exact geometrical arrangement of the atoms in a solid, it is extremely difficult to predict whether the material will be magnetic, a metal, an insulator, or a superconductor. This is why condensed matter physics is so full of surprising discoveries.

For more than two decades we have known the complete DNA sequence of the human genome. In principle, this contains all the information necessary to understand all diseases. However, understanding how different DNA sequences (i.e., different genes) lead to different biological properties remains a challenge. Knowing the sequence of amino acids that make up a specific protein molecule does not mean we can predict the biological function of that molecule. Knowing the properties of neurons in the brain and their interactions does not mean we can understand consciousness or what thoughts you are going to have. Knowing the psychological profiles of the individual members of a crowd does not mean we can predict whether it will become a violent mob.

## 2. Characteristics of emergent phenomena

There is no consensus about what emergence is, how to define it, or why it matters. In John Holland's book, *Emergence: from Chaos to Order*, he states that, "Despite its ubiquity and importance, emergence is an enigmatic, recondite topic, more wondered at than analysed… It is unlikely that a topic as complicated as emergence will submit meekly to a concise definition, and I have no such to offer."[1] Instead, Holland focuses on systems that can be described by simple rules or laws. The rules generate complexity: novel patterns that are sometimes hard to recognise and to anticipate.[e]

---

[e] Elsewhere, Holland[22] (page 4) stated that what distinguishes complexity from complicated is emergence, which is defined in terms of "the whole is more than the sum of the parts." This can occur in a system with "interactions where the aggregate exhibits properties not attained by summation", i.e., the interactions are non-linear. This yields levels of organisation and hierarchies, as emphasised by Herbert Simon.[23]

Below I seek to clarify what some of the important issues and questions are associated with defining emergence. I take a path that is intermediate between the precision of philosophers and the loose discussion of emergence by many scientists. In 2006, Degeut et al.[24] made an extensive literature survey of emergence definitions and presented five different representative definitions. My goals are clarity and brevity.

### 2.1 Novelty as the defining characteristic

Consider a system that is composed of *many* interacting parts or involves *many* scales. If the properties of the whole system are compared with the properties of the individual parts, *a property of the whole system is an emergent property if it is a property that the individual parts of the system do not have*. Emergent properties are *novel*. The system is qualitatively different from its parts. More is Different, as advocated by Anderson.[25]

Examples of properties of a physical system that are not emergent are volume, mass, charge, and number of atoms. These are additive properties and sometimes called resultant properties. The property of the whole system is simply the sum of the properties of the parts.[f]
Implicit in this definition is the concept of scale. Some sort of scale (for example, particle number, length, or energy) is used to define what the parts are and thus how many parts there are. In physics the scales may be microscopic and macroscopic, and perhaps an intermediate scale, the mesoscopic. For some physical systems, particularly early in research programs, what the mesoscale of interest is may not be at all obvious. For general systems, what the macroscale at which emergence occurs may not be obvious.

A subtle question is how many parts a system must have (i.e., how large it need be) for the system to have an emergent property. Coleman[26] followed Anderson's argument[27] , considering small collections of atoms of gold and niobium. A single atom of gold is not shiny but large metallic grains are. A small number of niobium atoms are not superconducting, only a bulk sample is. Kivelson and Kivelson[28] claimed that emergent properties can only be defined in the thermodynamic limit.

An alternative definition is that an emergent property is one that occurs in some states of the system but not in a state where the states of the parts are randomly assigned. This means that for physical systems in thermodynamic equilibrium, an emergent property is one that the system does not have at high temperatures. As a condensed matter physicist, my implicit perspective if often that of thinking of a system in thermal equilibrium with an environment.

The definition for emergence given above is formulated with physical systems in mind, particularly with many particles or degrees of freedom. In this article, I consider a broader range of systems (including social systems, artificial systems, and simple mathematical models) and so a broader sense of "many parts" and "interacting." Specifically, I include the case of many iterations, as occurs in dynamical systems and computer algorithms. Kadanoff emphasised how "the many times repeated application of quite simple laws" can lead to rich new structures.[29] Examples include cellular automata such as Conway's Game of Life. A deep neural network can learn to identify and classify objects, after being trained on a set of many objects. For Large Language Models (LLMs) the "size of the system" can be defined as

[f] Strictly speaking, mass is not additive if you allow for $E=mc^2$, since the binding energies between particles changes the total mass.

the run time, the size of the data set used for training, or the number of parameters used in the model.[30]

An alternative way of looking at emergence in terms of novelty is where there is a qualitative difference between local and global properties. This is helpful when considering dynamical systems, such as those that exhibit chaotic dynamics.
I don't claim that defining emergence in terms of novelty is necessarily better than definitions used by others. I won't catalogue or critique in detail alternative definitions here. This definition in terms of novelty is concrete and precise enough to help clarify other characteristics often associated with emergence. These characteristics are sometimes included in the definition of emergence by other authors. I have (mistakenly) done this in the past, in blog posts, talks, papers[31], and in my recent *Very Short Introduction* to condensed matter physics.[32]

There is more to emergence than novel properties, i.e., where a whole system has a property that the individual components of the system do not have. Here I focus on emergent properties, but in most cases "property" might be replaced with state, phenomenon, process, or entity. I now discuss twelve characteristics, besides novelty, that are sometimes associated with emergence. Some people include one or more of these characteristics in their definitions of emergence. However, I do not include them in my definition because some may not be necessary or sufficient for novel system properties.[g]

It may be helpful to think of the problem of defining the concept of emergence in terms of the philosophical idea of family resemblance, made popular by Ludwig Wittgenstein, and described in his posthumously published book *Philosophical Investigations* (1953), §65-71. He argued that something that could be thought to be defined by one essential feature may in fact be connected by a series of overlapping similarities, where no one feature is common to all of the things. He illustrated his argument with the example of what is a "game."

The first set of six characteristics discussed below might be classified as objective (i.e., observable properties of the system) and the second set as subjective (i.e., associated with how an investigator thinks about the system). In different words, the first set are mostly concerned with ontology (what is real) and the second set with epistemology (what we know). The first set of characteristics concern discontinuities, structure, modification of parts, universality, diversity, mesoscales, and structure. The second set concern self-organisation, unpredictability, irreducibility, downward causation, complexity, and closure. Some examples will be given to illustrate how it is possible to have novelty with or without some of these characteristics.

### 2.2 Objective characteristics: discontinuities, order, modification of parts, universality, diversity, modularity

1. *Discontinuities*

Quantitative changes in the system can become qualitative changes in the system. For example, in condensed matter physics spontaneous symmetry breaking only occurs in the thermodynamic limit (i.e., when the number of particles of the system becomes infinite). More is different. Thus, as a quantitative change in the system size occurs the order parameter becomes non-zero. In a system that undergoes a phase transition at a non-zero temperature, a small change in temperature can lead to the appearance of order and to a new state of matter. For a first-order phase transition there is discontinuity in properties such as the entropy and density. These discontinuities define a phase boundary in the pressure-temperature diagram. For continuous phase transitions the order parameter is a continuous function of temperature, becoming non-zero at the critical temperature. But the derivative with respect to temperature may be discontinuous and/or thermodynamic properties such as the specific heat and the susceptibility associated with the order parameter may approach infinite as the critical temperature is approached. New states of matter, including liquid crystals, antiferromagnetism, and superfluid $^3$He, have been discovered by observing discontinuities in thermodynamic quantities.

[g] In discussing emergence in economics, Lewis and Harper[33] pointed out the diversity of perspectives of what emergence is and sought to clarify the relationship between novelty, unpredictability, and irreducibility.

A discontinuity is not necessarily equivalent to a qualitative difference (novelty). This can be illustrated with the case of the liquid-gas transition. On the one hand, there is a discontinuity in the density and entropy of the system as the liquid-gas phase boundary is crossed in pressure-temperature diagram. On the other hand, there is no qualitative difference between a gas and a liquid. There is only a quantitative difference: the density of the gas is less than the liquid. Albeit sometimes the difference is orders of magnitude. The liquid and gas state can be adiabatically connected, i.e., one can smoothly deform one state into the other and all the properties of the system also change smoothly. Specifically, there is a path in the pressure-temperature phase diagram that can be followed to connect the liquid and gas states without any discontinuities in properties.

The ferromagnetic state also raises questions about the relationship between novelty and discontinuities. This was illustrated by a debate between Peierls[34,35] and Anderson about whether ferromagnetism exhibits spontaneous symmetry breaking. On the one hand, singularities in properties at the Curie temperature (critical temperature for ferromagnetism) only exist in the thermodynamic limit. Also, a small change in the temperature, from just above the Curie temperature to below, can produce a qualitative change, a non-zero magnetisation. Anderson argued that ferromagnetism did not involve spontaneous symmetry breaking, as in contrast to the antiferromagnetic state, a non-zero magnetisation (order parameter) occurs for finite systems at zero temperature and the magnetic order does not change the excitation spectrum, i.e., produce a Goldstone boson. Some condensed matter systems exhibit novelty without discontinuities as the temperature is lowered. Instead, there is a smooth crossover. Examples include spin ices, Kondo systems, Fermi liquids, and glasses. In social sciences, the term *tipping point* is often used to refer to a qualitative change produced by a quantitative change.[36]

## 2. *Order and Structure*

Emergent properties are often associated with the state of the system exhibiting patterns, order, or structure, terms that may be used interchangeably. This reflects that there is a particular relationship (correlation) between the parts which is different to the relationships in a state without the emergent property. In condensed matter physics, the order and structure is often associated with symmetry breaking, which Anderson considered to be the organisational principle for understanding the stratification of reality.[25]
Symmetry breaking is associated with a generalised rigidity. For example, applying a force to one surface of a solid results in all the atoms in the solid experiencing a force and moving together. The rigidity of the solid reflects a particular relationship between the parts of the system. Anderson stated[27] (p.49) that we "are so accustomed to this rigidity property that we don't accept its almost miraculous nature, that it is an "emergent property" not contained in the simple laws of physics, although it is a consequence of them." He argued that generalised rigidity was responsible for most of unique properties of ordered states.

*3. Modification of the parts*

Properties of the individual parts may be different in an emergent state. In a state of matter, where there is an order parameter associated with symmetry breaking, this modifies local properties. For example, in a ferromagnet, the probability of the magnetic moment of an atom pointing in a specific direction is different from in the paramagnetic state, where any direction is equally likely. In a superconductor, the local density of electronic states is different from in the metallic state. Thus, emergence is associated with novel properties at both the macroscopic and microscopic level. de Haan refer to the novel property at the lower level as the "conjugate" of the property at the higher level.[37]
In a crystal single-atom properties such as electronic energy levels change quantitatively compared to their values for isolated atoms. Properties of finite subsystems are also modified, reflecting a change in interactions between the parts. For example, in a molecular crystal the frequencies associated with intramolecular atomic vibrations are different to their values for isolated molecules. However, emergence is a sufficient but not a necessary condition for these modifications. In gas and liquid states, novelty is not present but there are still changes in the properties of the individual parts, such as the vibrational frequencies of molecules.

*4. Universality*

Many details do not matter. There are several dimensions to universality. They are related but not necessarily equivalent.

i. Many different systems exhibit the same phenomenon.
ii. Different phenomena are caused by essentially the same mechanism and so can be described by essentially the same theory.
iii. For a specific system some properties of the system are irrelevant to whether or not the system has the emergent property.
iv. The parameter dependence of some properties of the system has a universal functional form.

An example of i. is that superconductivity is present in metals with a diverse range of crystal structures and chemical compositions. An example of ii. is that supercurrents in superconductors and superfluids are both associated with the rigidity of the order parameter and the breaking of a global gauge symmetry. In different words, iii., is that the emergent property is independent of many of the details of the parts of the system. Robustness is an example of iii. If small changes are made to the composition of the system (for example replacing some of the atoms in the system with atoms of different chemical element) the novel property of the system is still present. In elementary superconductors, introducing non-magnetic impurity atoms has no effect on the superconductivity.

Universality is significant as it provides a basis for effective theories and toy models that can describe a wide range of phenomena at one stratum and requires little detailed information about lower strata. Universality also justifies the cross-disciplinary fertilisation of concepts and methods. For example, the concept of spontaneous symmetry breaking provides an organising principle for understanding phenomena in both condensed matter physics and elementary particle physics, even though these sub-disciplines concern phenomena at length, time, and energy scales that differ by many orders of magnitude.

Universality is both a blessing and a curse for theory. It is a blessing because solving a problem for one specific material can also solve it for whole families of materials. This universality is also of deep conceptual significance as understanding a general phenomenon is usually more powerful than just a specific example.

Universality can make it easier to develop successful theories because it means that many details need not be included in a theory to successfully describe an emergent phenomenon. Theoretical analysis becomes tractable. Effective theories and toy models can work even better than might be expected. Universality can make theories more powerful because they can describe a wider range of systems. For example, properties of elemental superconductors can be described by BCS theory and by Ginzburg-Landau theory, even though the materials are chemically and structurally diverse.

The curses of universality for theory are that it increases the problems of "multiple realisability", "under-determination of theory", "over-fitting of data" and "sloppy theories".[38–40] A theory can agree with the experiment even when the objects, interactions, and parameters used in the theory may be quite different from the actual ones. For example, the observed phase diagram of water can be reproduced, sometimes with impressive quantitative detail, by combining classical statistical mechanics with empirical force fields that assume water molecules are just rigid bodies composed of point electrical charges.

Suppose we start with a specific microscopic theory and calculate the macroscopic properties of the system, and they agree with experiment. It would then be tempting to think that we have the correct microscopic theory. However, universality suggests this may not be the case. For example, consider the case of a gas of weakly interacting atoms or molecules. We can treat the gas particles as classical or quantum. Statistical mechanics gives exactly the same equation of state and specific heat capacity for both microscopic descriptions. The only difference may be the Gibbs paradox [the calculated entropy is not an extensive quantity] which is sensitive to whether or not the particles are treated as identical or not. Unlike the zeroth, first, and second law of thermodynamics, the third law does require that the microscopic theory be quantum. Laughlin discussed[4] these issues in terms of "protectorates" that hide "ultimate causes". For example, the success of elasticity theory which assumes that matter is continuous hides the underlying discrete atomic structure of matter.

In some physical systems universality can be defined in a rigorous technical sense, making use of the concepts and techniques of the renormalisation group and scaling. These techniques provide a method to perform coarse graining, to derive effective theories and effective interactions, and to define universality classes of systems.

There are also questions of how universality is related to the robustness of strata, and the independence of effective theories from the coarse graining procedure.[39,41,42]

### 5. *Diversity with limitations*

Even when a system is composed of a small number of different components and interactions, the large number of possible stable states with qualitatively different properties that the system can have is amazing. Holland uses the term “perpetual novelty” to describe this[1]. Every snowflake is different. Water is found in 18 distinct solid states. All proteins are composed of linear chains of 20 different amino acids. Yet in the human body there are more than 100,000 different proteins and all perform specific biochemical functions. We encounter an incredible diversity of human personalities, cultures, and languages. A stunning case of diversity is life on earth. Billions of different plant and animal species are all an expression of different linear combinations of the four base pairs of DNA: A, G, T, and C.
This diversity is related to the idea that "simple models can describe complex behaviour". One example is Conway’s Game of Life. Another example is how simple Ising models with a few competing interactions can describe a devil's staircase of ground states or the multitude of different atomic orderings found in binary alloys. Condensed matter physics illustrates diversity with the many different states of matter that have been discovered. The underlying microscopics is “just” electrons and atomic nuclei interacting according to Coulomb’s law. A typical game of chess may potentially involve the order of $10^{50}$ move sequences. However, games can be understood in terms of “motifs”: recurring patterns (sequences of moves). There are about 50 common motifs.

The significance of this diversity might be downplayed by saying that it is just a result of combinatorics. In a system composed of many components each of which can take on a few states the number of possible states of the whole system grows exponentially with the number of components. But such a claim overlooks the issue of the stability of the diverse states that are observed. For example, for a chain of ten amino acids there are $10^{13}$ different possible linear sequences. But this does not mean that all these sequences will produce a functional protein, i.e., a molecule that will fold rapidly (on the timescale of milliseconds) into a stable tertiary structure and perform a useful biochemical function such as catalysis of a specific chemical reaction.[h] A broader question is, how does stability limit diversity?

### 6. *Modularity at the mesoscale*

Systems with emergent properties often exhibit structure at the mesoscale, i.e., intermediate between the micro- and the macro- scales. Furthermore, this structure reflects the emergence of entities that interact weakly with one another. In other words, the system can be viewed as a set of interacting modules, with each module composed of collections of micro-constituents.

The ancient Greeks showed how any mechanical machine could be constructed from six parts: lever, screw, inclined plane, wedge, wheel, and pulley. Modularity is also reflected in the motifs identified in chess, music, visual arts, literature, and genetics.

A key idea in condensed matter physics is that of quasiparticles. A system of strongly interacting particles may have excitations, seen in experiments such as inelastic neutron scattering and Angle Resolved PhotoElectron Spectroscopy (ARPES), that can be described as weakly interacting quasiparticles. These entities are composite particles, and have properties that are quantitatively different, and sometimes qualitatively different, from the microscopic particles. The existence of quasiparticles leads naturally to the technique of constructing an effective Hamiltonian [effective theory] for the system where effective interactions describe the interactions between the quasiparticles.

---

[h] Similarly given the large number of genes (each of which can be switched on or off), why are there not an incredibly large number of cell types? [Lewontin’s paradox].

In the Landau theory of phase transitions, Fisher[43] emphasised that the order parameter is at the mesoscopic scale and called this foliation.

The economist Herbert Simon argued that a characteristic of a complex system is that the system can be understood in terms of nearly decomposable units.[23,44] He identified hierarchies as an essential feature of complex systems, both natural and artificial. A key property of a level in the hierarchy is that it is nearly decomposable into smaller units, i.e., it can be viewed as a collection of weakly interacting units. The time required for the evolution of the whole system is significantly decreased due to the hierarchical character. The construction of an artificial complex system, such as a clock, is faster and more reliable if different units are first assembled separately and then the units are brought together into the whole. Simon argued that the reduction in time scales due to modularity is why biological evolution can occur on realistic time scales.

Laughlin et al. argued the mesoscale was key to understanding emergence in soft matter.[45] Rosas et al.[46] argued that emergence is associated with there being a scale at which the system is "strongly lumpable." Hartwell et al.[47] and Noble[48] highlighted how biological systems are modular, i.e., composed of simple interchangeable components.

Identifying the relevant scale and the corresponding modular units may be highly non-trivial and represent a significant breakthrough. Examples include atoms, DNA, and genes. The scale may not be spatial but relate to energy, momentum, number of particles, time, or connectivity.

As stated at the beginning of this section the six characteristics above might be associated with ontology (what is real) and objective properties of the system that an investigator observes and depend less on what an observer thinks about the system. The next six characteristics are arguably more subjective, being concerned with epistemology (how we determine what we believe is true). In making this dichotomy I do not want to gloss over the fuzziness of the distinction or two thousand years of philosophical debates about the relationship between reality and theory, or between ontology and epistemology.

### 2.3 Subjective characteristics: self-organisation, unpredictability, irreducibility, contextuality, complexity, closure

7. *Self-organisation*

Self-organisation is not a property of the system but a mechanism that a theorist says causes an emergent property to come into being. Self-organisation is also referred to as spontaneous order. In the social sciences self-organisation is sometimes referred to as an endogenous cause, in contrast to an exogenous cause, one that arises from outside the system. There is no external force or agent causing the order, in contrast to order that is imposed externally. For example, suppose that in a city there is no government policy about the price of a loaf of sliced wholemeal bread or on how many loaves that bakers should produce. It is observed that prices are almost always in the range of \$4 to \$5 per loaf, and that rarely bread shortages occur. This outcome is a result of the self-organisation of the free-market, and economists would say the price range and its stability has an endogenous cause. In contrast, if the government legislated the price range and the production levels that would be an exogenous cause.

Distinguishing exogeneous and endogenous causes is crucial to identifying emergent phenomena. It is not always obvious as defining the boundary of a system and its environment may be subjective at some level.
In physics, the periodicity of the arrangement of atoms in a crystal is a result of self-organisation and has an endogenous cause. In contrast, the periodicity of atoms in an optical lattice is determined by the laser physicist who creates the lattice and so has an exogenous cause.

Friedrich Hayek emphasised the role of spontaneous order in economics.[49,50] In biology, Stuart Kauffman equates emergence with spontaneous order and self-organisation.[51] Self-organisation shows how local interactions can produce global properties. In different words, short-range interactions can lead to long-range order. After decades of debate and study, the Ising model showed that this was possible.[52] Other examples of self-organisation include flocking of birds and teamwork in ant colonies. There is no director or leader but the system acts "as if" there is. As Adam Smith observed, economic systems act "as if" there is "an invisible hand" guiding them.[8]

Self-organisation does not necessarily involve order. For example, it may involve the production of apparent noise in a deterministic system, such as can occur in a achaotic dynamical system. Prior to the development of chaos theory, many scientists assumed that apparent randomness they observed in a system they were studying was external to the system rather than being intrinsic to it.

8. *Unpredictability*

Karl Popper said that "unpredictability in principle has always been considered as the salient point of emergence."[53] (p. 298) The biologist, Ernst Mayr defined emergence as "in a structured system, new properties emerge at higher levels of integration that could not have been predicted from a knowledge of the lower-level components."[54] (p.19). Recently, Ball also defined emergence in terms of unpredictability.[55],[56] (p.214). Sometimes unpredictability is expressed as "surprise" or "un-anticipation" of the observation of a phenomenon. Note that these are subjective criteria. One person may be surprised whereas another may not be. Broadly, in discussions of emergence, "prediction" is used in three different senses: logical prediction, historical prediction, and dynamical prediction.

*Logical prediction* (deduction) concerns when a scientist can *predict* (calculate) the emergent (novel) property of the whole system *solely* from a knowledge of all the properties of the parts of the system and their interactions. Logical predictability is one of the most contested characteristics of emergence. Sometimes "predict" is replaced with "difficult to predict", "extremely difficult to predict", "impossible to predict", "almost impossible to predict", or "possible in principle, but impossible in practice, to predict." These are not the same thing.

Philosophers distinguish between prediction that is "possible in principle, but difficult in practice" and "impossible in principle." Silberstein and McGeever equate these with *epistemological* emergence and *ontological* emergence, respectively.[57] Chalmers equates them with *weak* and *strong* emergence, respectively.[58]

After an emergent property has been discovered experimentally sometimes it can be understood in terms of the properties of the system parts. In a sense "pre-diction" then becomes "post-diction." An example is the BCS theory of superconductivity, which provided *a posteriori*, rather than *a priori*, understanding. In different words, development of the theory was guided by a knowledge of the phenomena that had already been observed and characterised experimentally. The phenomenon was a "Black swan", i.e., an unanticipated event with significant consequences, whose existence is rationalised after its occurrence.[59] Thus, a keyword in the statement above about logical prediction is "solely".

*Historical prediction*. Most new states of matter discovered by experimentalists were not predicted even though theorists knew the laws that the microscopic components of the system obeyed. Examples include superconductivity (elemental metals, cuprates, iron pnictides, organic charge transfer salts, …), superfluidity in liquid $^4$He, antiferromagnetism, quasicrystals, and the integer and fractional quantum Hall states.
There are a few exceptions where theorists did predict new states of matter. These include are Bose-Einstein Condensates (BECs) in dilute atomic gases and topological insulators, the Anderson insulator in disordered metals, the Haldane phase in even-integer quantum antiferromagnetic spin chains, and the hexatic phase in two dimensions. I note that prediction of BECs and topological insulators were significantly helped that theorists could predict them starting with Hamiltonians of non-interacting particles. Furthermore, all the successful predictions listed above involved working with effective Hamiltonians. None started with a microscopic Hamiltonian for a material with a specific chemical composition.

*Dynamical unpredictability* concerns what it means in chaotic dynamical systems, where it relates to sensitivity to initial conditions. On the one hand, it might be argued that this is not an example of emergence as most of the systems considered have only a few degrees of freedom. On the other hand, the chaotic behaviour emerges after many iterations and the state of the system can be viewed as a time series.
Finally, the unpredictability of emergent phenomena is connected to their resistance to control as will be discussed later.

### 9. *Irreducibility and singularities*

An emergent property cannot be reduced to properties of the parts, because if emergence is defined in terms of novelty, the parts do not have the property. Emergence is also associated with the problem of theory reduction. Formally, this is the process where a more general theory reduces in a particular mathematical limit to a less general theory. For example, in some sense quantum mechanics reduces to classical mechanics in the limit where Planck's constant goes to zero. Einstein's theory of special relativity reduces to Newtonian mechanics in the limit where the speeds of massive objects become much less than the speed of light. Theory reduction is a subtle and contentious philosophical problem. Primas claimed that "there exists not a single physically well-founded and non-trivial example" of theory reduction. Subtleties arise because the two different theories usually involve language and concepts that are "incommensurate" with one another. Subtleties can get lost in reductionist polemics such as "biology is just applied chemistry" and "chemistry is just applied physics."

Irreducibility is also related to the discontinuities and singularities associated with emergent phenomena. As emphasised independently by Primas[60] and Berry[61], singularities occur because the mathematics of theory reduction may involve singular asymptotic expansions. Primas illustrated this by considering a light wave incident on an object and producing a shadow. The shadow is an emergent property, well described by geometrical optics, but not by the more fundamental theory of Maxwell's electromagnetism. The two theories are related in the asymptotic limit that the wavelength of light in Maxwell's theory tends to zero. This example illustrates that theory reduction is compatible with the emergence of novelty. Primas also considered how the Born-Oppenheimer approximation, which is central to solid state theory and quantum chemistry, is associated with a singular asymptotic expansion (in the ratio of the mass of an electron to the mass of an atomic nuclei in the system).

Berry considered[61] several other examples of theory reduction, including going from general to special relativity, from statistical mechanics to thermodynamics, and from viscous (Navier-Stokes) fluid dynamics to inviscid (Euler) fluid dynamics. He has discussed in detail how the caustics that occur in ray optics are an emergent phenomenon and are associated with singular asymptotic expansions in the wave theory.[62] Batterman also considered the role of asymptotic reasoning in theory reduction.[10]

The philosopher of science Jeremy Butterfield showed rigorously that theory reduction occurred for four specific systems that exhibited emergence, defined by him as a novel and robust property.[42,63] Thus, novelty is not sufficient for irreducibility.

### *10. Contextuality and downward causation*

Any real system has a context. For example, a physical system has a boundary and an environment, both in time and space. In many cases the properties of the system are completely determined by the parts of the system and their interactions. Previous history and boundaries do not matter. However, in some cases the context may have a significant influence on the state of the system. Examples include Rayleigh-Bernard convection cells and turbulent flow whose existence and nature are determined by the interaction of the fluid with the container boundaries. A biological example concerns what factors determine the structure, properties, and function that a particular protein (linear chain of amino acids) has. It is now known that the only factor is not just the DNA sequence that encodes for the amino acid sequence, in contradiction to some versions of the Central Dogma of molecular biology.[56] Other factors may be the type of cell that contains the protein and the network of other proteins in which the particular protein is embedded. Context sometimes matters. Emergent properties are not determined solely by the microscopic components of the system and this has led to the idea of contextual emergence.[64]

Downward causation is sometimes equated with emergence, particularly in debates about the nature of consciousness. In the context of biology, Noble defines downward causation as when higher level processes can cause changes in lower level properties and processes [48]. For example, physiological effects can switch on and off individual genes or signalling processes in cells, as occurs with maternal effects and epigenetics.

### *11. Complexity*

Simple rules can lead to complex behaviour. This is nicely illustrated by cellular automata.[65] It is also seen in other systems with emergent properties. For example, the laws describing the properties of electrons and ions in a crystal or a large molecule are quite simple: Schrodinger's equation plus Coulomb's law. Yet from these simple rules, complex phenomena emerge: all of chemistry and condensed matter physics!

There is no agreed universal measure for the complexity of a system with many components. When someone says a particular system is "complex" they may mean there are many degrees of freedom and/or that it is hard to understand. "Complexity" is sometimes used as a buzzword, just like "emergence." Complexity means different things to different people. In *More is Different*, Anderson stated[25] that as one goes up the hierarchy of scientific disciplines the system scale and complexity increases. This makes sense when complexity is defined in terms of the number of degrees of freedom in the system (e.g., the size of the Hilbert space needed to describe the complete quantum state of the system or classically, the position and speed of all the constituent particles). On the other hand, from a coarse-grained perspective the system state and its dynamics may become simpler as one goes up the hierarchy. The equilibrium thermodynamic state of the liquid can be described completely in terms of the density, temperature, and the equation of state. At that level of description, the system is arguably a lot simpler than the quantum chromodynamics (QCD) description of a single proton. Thus, we need to be clearer about what we mean by complexity.

In 1999 the journal *Science* had a special issue that focussed on complex systems, with an introduction entitled, *Beyond Reductionism*.[68] Eight survey articles covered complexity in physics, chemistry, biology, earth science, and economics. Ladyman et al.[69] pointed out that each of the authors of these articles chose different properties to define what complexity is associated with. The different characteristics chosen included non-linearity, feedback, spontaneous order, robustness and lack of central control, emergence, hierarchical organisation, and numerosity. The problem is that these characteristics are not equivalent. If a specific definition for a complex system is chosen, the difficult problem then remains of determining whether each of the characteristics above is necessary, sufficient, both, or neither for the system to be complex as defined. This is similar with attempts to define emergence.

Ladyman et al. have systematically studied different definitions of a complex system. They compare different quantitative measures such as those based on information content (Shannon entropy, Kolmogorov complexity), deterministic complexity, and statistical complexity.[69,70] They distinguish different measures of complexity. There are three distinct targets of measures: methods used, data obtained, and the system itself. There are three types of measures: difficulty of description, difficulty of creation, or degree of organisation.

To complicate matters more, the effective theories that describe many emergent phenomena are simple in that the relevant equations are simple to write down and involve just a few variables and parameters. This contrasts with the underlying theories that may involve many degrees of freedom. For example, compare elasticity theory with the dynamical equations for all the atoms in a crystal. In other words, simplicity emerges from complexity.

### *12. Intra-stratum closure*

Rosas et al. recently considered emergence from a computer science perspective [46]. They defined emergence in terms of universality and discussed its relationship to informational closure, causal closure, and computational closure. Each of these are given a precise technical definition in their paper. Here I only give the sense of their definitions. In considering a general system they do not pre-define the micro- and macro- levels of a system but consider how they might be defined so that universality holds, i.e., properties at the macro-level are independent of the details of the micro-level.

*Informational closure* means that to *predict* the dynamics of the system at the macroscale an observer does not need any additional information about the details of the system at the microscale. Equilibrium thermodynamics and fluid dynamics are examples.

*Causal closure* means that the system can be *controlled* at the macroscale without any knowledge of lower-level information. For example, changing the software code that is running on a computer reliably controls the microstate of the hardware of the computer, regardless of what is happening with the trajectories of individual electrons in the computer.

*Computational closure* is defined in terms of "a conceptual device called the ε-machine [originally introduced by Shalizi and Crutchfield[71]]. This device can exist in some finite set of states and can predict its own future state on the basis of its current one... for an emergent system that is computationally closed, the machines at each level can be constructed by coarse-graining the components on just the level below: They are, `strongly lumpable.'" The latter property is similar to the characteristic of modularity at the mesoscale.

Rosas et al. showed that informational closure and causal closure are equivalent and that they are more restrictive than computational closure. It is not clear to me how these closures relate to novelty as a definition of emergence.

In the rest of this article, I discuss emergence in a range of scientific disciplines. For some specific scientific problems I identify some of the characteristics discussed above such as novelty, discontinuities, universality, and modularity at the mesoscale. The discussions are illustrative not exhaustive. First, I consider how these characteristics shape different methods of scientific investigation of systems with emergent properties.

## 3. Methods of investigation

The elements are choice of scale to focus on, differentiation and integration of parts, epistemology, effective theories, toy models, intellectual synthesis, discovering new systems, developing new experimental probes, and interdisciplinarity.

### 3.1 Choice of scales to focus on

A key is to decide at what scales a system of interest should be studied at, and then to use appropriate methods for each of these scales. There are often three scales of interest: micro, meso, and macro. What each of these scales are for a specific system may not be obvious, particularly in the early stages of an investigation. Even when they are clearly defined and agreed upon characterising a system at one of these scales requires multiple approaches and initiatives. The choice of scales shapes the choice and development of tools and methods, both experimental and theoretical, that can be used to study the system.

There can be ambiguity when it comes to choosing experimental probes to explore properties at a particular scale. On the one hand, nuclear physics is irrelevant to condensed matter physics, chemistry, and molecular biology. Nuclei can be viewed as merely point charges at the centre of atoms. Nevertheless, experimental methods based on physics at the nuclear scale, such as nuclear magnetic resonance (NMR), isotope labelling, and Mossbauer spectroscopy, have been fruitful in these fields. In some cases, these methods are relevant because of subtle effects such as when phenomena at the atomic scale slightly modify electric fields at the nuclear level.

### 3.2 Connecting stratum: bottom-up or top-down?

A major goal is to understand the relationship between different strata. Before describing two alternative approaches, top-down and bottom-up, I need to point out that in different fields these terms are used in the opposite sense. In this article, I use the same terminology that is traditionally used in condensed matter physics, chemistry,[72] in biology.[73] It is also consistent with the use of the term "downward causation" in philosophy. Top-down means going from long distance scales to short distance scales, i.e., going down in the diagrams shown in Figure 2. In contrast in quantum field theory of elementary particles and fields, or high-energy physics, "top-down" means the opposite, i.e., going from short to long distance length scales.[74] This is because practitioners in that field tend to draw diagrams with high energies at the top and low energies at the bottom.

*Bottom-up* approaches aim to answer the question: how do properties observed at the macroscale emerge from the microscopic properties of the system? History suggests that this question may often be best addressed by identifying the relevant mesoscale at which modularity is observed and connecting the micro- to the meso- and connecting the meso- to the macro.

*Top-down* approaches try to surmise something about the microscopic from the macroscopic. This has a long and fruitful history, albeit probably with many false starts that we may not hear about, unless we live through them or read history books. Kepler's snowflakes are an early example.[75,76] Before people were completely convinced about the existence of atoms, the study of crystal facets and of Brownian motion provided hints of the atomic structure of matter. Planck deduced the existence of the quantum from the thermodynamics of black-body radiation, i.e., macroscopic properties. Arguably, the first definitive determination of Avogadro's number was from Perrin's experiments on Brownian motion which involved macroscopic measurements. Comparing classical statistical mechanics to bulk thermodynamic properties gave hints of an underlying quantum structure to reality. The Sackur-Tetrode equation for the entropy of an ideal gas hinted at the quantisation of phase space.[77] The Gibbs paradox hinted that fundamental particles are indistinguishable. The third law of thermodynamics hints at quantum degeneracy. Pauling's proposal for the structure of ice was based on macroscopic measurements of its residual entropy. Pasteur deduced the chirality of molecules from observations of the facets in crystals of tartaric acid. Sometimes a "top-down" approach means one that focuses on the meso-scale and ignores microscopic details.

The top-down and bottom-up approaches should not be seen as exclusive or competitive, but rather complementary. Their relative priority or feasibility depends on the system of interest and the amount of information and techniques available to an investigator. Coleman has discussed the interplay of emergence and reductionism in condensed matter.[78] In biology, Mayr,[54] (p.20), advocated a "dual level of analysis" for organisms. In social science Schelling,[8] (p.13-14) discussed the interplay of the behaviour of individuals and the properties of social aggregates. In a classic study of complex organisations in business[79] understanding this interplay was termed differentiation and integration.

### 3.3 Identification of parts and interactions

To understand emergent properties of system a key step is identifying what the relevant parts are in the system and this is related to the choice of the microscopic scale. This is the agenda of methodological reductionism and has often been the first step in massive advances in science. In biology think of the discovery of cells, membranes, proteins, and DNA. But understanding emergent properties also requires a knowledge of the interactions between the parts.

There is a question as to whether place primacy on the parts or their interactions. In his Nobel Prize Lecture, John Hopfield noted[80] that his 1974 paper on kinetic proof reading in biosynthesis "was important in my approach to biological problems, for it led me to think about the function of the structure of reaction networks in biology, rather than the function of the structure of the molecules themselves. Six years later I was generalizing this view in thinking about networks of neurons rather than the properties of a single neuron." Jensen stated[6] that a distinguishing feature of complexity science is its emphasis on the network of interactions. What the parts are don't really matter. In contrast, consider the quark model for mesons and baryons, originally developed by Gell Mann and Zweig in the 1960's. Primacy was placed on the parts (quarks and their quantum numbers) not the interactions that held the quarks together.

### 3.4 Phase diagrams

Phase diagrams are ubiquitous in materials science. They show what states of matter are thermodynamically stable depending on the value of external parameters such as temperature, pressure, magnetic field, or chemical composition. However, they are only beginning to be appreciated in other fields. Recently, Bouchaud argued[81] that phase diagrams should be used more to understand agent-based models in the social sciences.

For theoretical models, whether in condensed matter, dynamical systems, or economics, phase diagrams can show how the state of the system predicted by the model has qualitatively different properties depending on the parameters in the model, such as the strength of interactions.

Phase diagrams illustrate discontinuities, how quantitative changes produce qualitative changes (tipping points), and diversity (simple models can describe rich behaviour). Phase diagrams show how robust and universal a state is, i.e., whether it only exists for fine-tuning of parameters. Theoretical phase diagrams can expand our scientific imagination, suggesting new regimes that might be explored by experiments. An example is how the phase diagram for QCD matter has suggested new experiments, such as at the Relativistic Heavy Ion Collider (RHIC). For dynamical systems, this will be illustrated with the phase diagram for the Lorenz model. It shows for what parameter ranges strange attractors exist.

Today, for theoretical models of strongly correlated electron systems it is common to map out phase diagrams as a function of the model parameters. However, this was not always the case. It was more common to just investigate a model for specific parameter values that were deemed to be relevant to specific materials. Perhaps, Anderson stimulated this new approach when, in 1961, he drew the phase diagram for the mean-field solution to his model for local moments in metals,[82] a paper that was partly the basis of his 1977 Nobel Prize.

At a minimum, a phase diagram should show the state with the emergent property and the disordered state. Diagrams that contain multiple phases may provide hints for developing a theory for a specific phase. For example, for the high-$T_c$ cuprate superconductors, the proximity of the Mott insulating, pseudogap, and non-Fermi liquid metal phases has aided and constrained theory development. Phase diagrams have also aided theory development for superconducting organic charge transfer salts.[83]

Phase diagrams constrain theories as they provide a minimum criterion of something a successful theory should explain, even if only qualitatively. Phase diagrams illustrate the potential and pitfalls of mean-field theories. On the positive side, they may get qualitative details correct, even for complex phase diagrams, and can show what emergent states are possible. Ginzburg-Landau and BCS theories are mean-field theories and work extremely well for many superconductors. On the other hand, in systems with large fluctuations, mean-field theory may fail spectacularly, and these systems are sometimes the most interesting and theoretically challenging systems.

### 3.5 Phenomenology, effective theories, toy models, and microscopic theories

With respect to the theory of emergent phenomena, people have different views about what theory is and the criteria used to evaluate the validity, value, or importance of different approaches to theory. Often the terms theory and model are used interchangeably.
In considering the challenges of inter-disciplinary research, Bouchaud distinguished three distinct models: phenomenological, fundamental, and metaphorical.[84]. Phenomenological models seek to parametrise observational data, mostly from the macro-scale. Metaphorical models are like the toy models I discuss below. Fundamental models include the effective theories I discuss and microscopic theory such as quantum theory.

There is a common, but not unique, historical sequence in the development of the theory of many specific emergent phenomena. The sequence is phenomenology leads to toy models, then effective theories, and finally the derivation (or at least justification) of the effective theory from microscopic theory. Although there are exceptions, this history might be helpful in facing the challenge of developing theory for any new system. It should also caution about grand schemes to rush towards reductionism (now partly driven by the seductive allure of increasing computational power).

The classification below is not hard and fast. The boundaries between the types of theory and modelling can be fuzzy.

*Phenomenology*

At the level of qualitative descriptions phenomenology may consist of classification of systems and their properties. At the quantitative level, phenomenology may comprise parametrisation of experimental data. These empirical relations provide a concrete challenge for theorists to explain. Phenomenology may suggest what data to gather, how to represent it, and interpret it. Phenomenology may provide a way to contrast and compare systems.

Examples of empirical relations include the power laws observed in diverse systems,[85] the dependence of the critical temperature of elemental superconductors on the square root of the atomic mass, the linear in temperature resistivity of the metallic state of optimally doped cuprate superconductors, and Guggenheim's scaling plot for liquid-gas critical points. Phenomenological models include Pippard's model for the non-local London equations in superconductors, the two-level systems theory of glasses, and the parton model for deep inelastic electron-proton scattering.

At best phenomenology may be a key clue that leads to a more microscopic understanding of the phenomena. At worst, empirical relations may be an artefact of curve fitting or data selection. For example, how many decades must a power law have for it to be significant?

*Effective theories*

An effective theory is valid at a particular range of scales. In physics, examples include classical mechanics, general relativity, Maxwell's theory of electromagnetism, and thermodynamics. The equations of some effective theories can be written down almost solely from consideration of symmetry and conservation laws. Examples include the Navier-Stokes equations for fluid dynamics and non-linear sigma models for magnetism. Some effective theories can be derived by the coarse graining of theories that are valid at a finer scale. For example, the equations of classical mechanics result from taking the limit of Planck's constant going to zero in the equations of quantum mechanics. The Ginzburg-Landau theory for superconductivity can be derived from the BCS theory. The parameters in effective theories may be determined from more microscopic theories or from fitting experimental data to the predictions of the effective theory.

Effective theories are useful and powerful because of the minimal assumptions and parameters used in their construction. For the theory to be useful it is not necessary to be able to derive the effective theory from a smaller scale theory, or even to have such a smaller scale theory. For example, thermodynamics is incredibly useful in engineering, independently of whether one has an atomic scale description of the materials involved. Even though there is no accepted quantum theory of gravity, general relativity can be used to describe phenomena in astrophysics and cosmology. Effective theories can also describe phenomena at the mesoscale and form a bridge between the micro- and macro- scales.

*Toy models*

In his 2016 Nobel Lecture on Topological Quantum Matter, Duncan Haldane said, "Looking back, … I am struck by how important the use of stripped down "toy models" has been in discovering new physics." [86]

Examples of toy models include the Ising, Hubbard, Kondo, Agent-Based Models, Schelling, Hopfield, and Sherrington-Kirkpatrick models. Most of these models will be discussed in this paper. I refer to them as "toy" models because they aim to be as simple as possible, while still capturing the essential details of a phenomena. At the scale of interest they are an approximation, neglecting certain degrees of freedom and interactions. In contrast, at the relevant scale, some effective theories might be considered exact because they are based on general principles such as conservation laws.

Toy models do not necessarily correspond to any real system. Consequently, specific toy models, particularly when they are originally proposed, have been criticized as simplistic, misleading, or of little value. Such criticism is more likely to be found among experimentalists who have spent a lifetime painstakingly studying all the details of a phenomenon and the components of the underlying system. For example, James Gleick recounts the resistance that Bernardo Huberman received when he presented a toy model for patterns of erratic eye movements associated with schizophrenia to a group of psychiatrists, biologists, and physicians.[87] John Hopfield had a similar experience with neuroscientists when he first proposed his neural network model for associative memory.

Historical experience suggests a strong justification for the proposal and study of toy models. They are concerned with qualitative understanding and not a quantitative description of experimental data. A toy model is usually introduced to answer basic questions about *what is possible*. In the context of biology, Servedio et al.[88] refer to such models as "proof of concept" models. What are the essential ingredients that are sufficient for an emergent phenomenon to occur? What details do matter? For example, the Ising model was introduced to see if it was possible for statistical mechanics to describe the sharp phase transition associated with ferromagnetism.

Batterman and Rice[89] discuss "minimal models", following terminology of Goldenfeld, and claim that their "explanatory structure has been almost universally misunderstood by philosophers." Their power is that they can explain universality. Holland argued that models are essential to understand emergence.[1] Scott Page's book *The Model Thinker* (and the associated online course, *Model Thinking*)[90] enumerated the value of simple models in the social sciences. An earlier argument for their value in biology was put by J.B.S. Haldane in his seminal article about "bean bag" genetics.[91] Simplicity makes toy models more tractable for mathematical analysis and/or computer simulation. The assumptions made in defining the model can be clearly stated. If the model is tractable then the pure logic associated with mathematical analysis leads to reliable conclusions. This contrasts with the qualitative arguments often used in the biological and social sciences to propose explanations. Such arguments can miss the counter-intuitive conclusions that are associated with emergent phenomena and the rigorous analysis of toy models.

Toy models can show what is possible, what are simple ingredients for a system sufficient to exhibit an emergent property, and how a quantitative change can lead to a qualitative change. In different words, what details do matter? In a toy model the ingredients to define are the degrees of freedom, the interaction parameters, and a "cost" function, such as a Hamiltonian, that determines stability, and possibly dynamics.

Toy models can provide guidance on what experimental data to gather and how to analyse it. Insight can be gained by considering multiple models as that approach can be used to rule out alternative hypotheses.[90,92]

Finally, there is value in the adage, "all models are wrong, but some are useful, " coined by George Box.[93] All models are wrong in the sense that they do not include all the details of the system, either of the components or their interactions. Nevertheless, they may still describe some phenomena and provide insights into what is sufficient for a phenomenon to occur. Historically, good toy models have been more useful than originally anticipated.

Due to universality, sometimes toy models work better than expected, and can even give a quantitative description of experimental data. An example is the three-dimensional Ising model, which was found to be consistent with data on the liquid-gas transition near the critical point. Although, a fluid is not a magnetic system and the atoms are not located on a periodic lattice, the analogy was bolstered by the mapping of the Ising model onto the lattice gas model. This success led to a shift in the attitude of physicists towards the Ising model. From 1920-1950, it was viewed as irrelevant to magnetism because it did not describe magnetic interactions quantum mechanically. This was replaced with the view that it was a model that could give insights into collective phenomena, including phase transitions.[94] From 1950-1965, the view diminished that the Ising model was irrelevant to describing critical phenomena because it oversimplified the microscopic interactions.[95]

An attractive feature of many toy models is that they are amenable to computer simulation because the degrees of freedom are discrete variables with only a small number of possible values. This allows the possibility of considering zillions of possible configurations of systems that cannot be humanly conceived or studied analytically. This has driven the development of Agent-Based Models in the social sciences.[96]

*Microscopic theories*

These start with the constituents at the microscopic level and their interactions. For example, in condensed matter physics and chemistry, constituents are electrons and nuclei, and they interact via the Coulomb interaction. The theory is the multi-particle Schrodinger equation with the Hamiltonian:

$$\begin{aligned}\hat{H} &= -\sum_{i=1}^{N_a} \frac{\hbar^2}{2M_i}\nabla^2_{\mathbf{R}_i} - \sum_{i=1}^{N_e} \frac{\hbar^2}{2m_e}\nabla^2_{\mathbf{r}_i} + \sum_{i=1}^{N_a}\sum_{j>i} \frac{Z_i Z_j e^2}{4\pi\epsilon_0 \left|\mathbf{R}_i - \mathbf{R}_j\right|} \\ &- \sum_{i=1}^{N_a}\sum_{j=1}^{N_e} \frac{Z_i e^2}{4\pi\epsilon_0 \left|\mathbf{R}_i - \mathbf{r}_j\right|} + \sum_{i=1}^{N_e}\sum_{j>i} \frac{e^2}{4\pi\epsilon_0 \left|\mathbf{r}_i - \mathbf{r}_j\right|}\end{aligned}$$

Laughlin and Pines mischievously dubbed this “The Theory of Everything” as *in principle* it describes almost all phenomena in chemistry and condensed matter physics.[67]

It might be claimed that due to the stratified nature of reality any microscopic theory should be considered an effective theory. However, what is macroscopic and microscopic is a subjective choice. If there is a mesoscopic scale of relevance then the associated effective theories may help bridge the mesoscopic and macroscopic, giving an understanding of phenomenology and macroscopic experiments. Microscopic theory provides a justification and possibly a parametrisation of effective theories. For example, in condensed matter physics strongly correlated electron models on a lattice can be parametrised by electronic structure calculations based on Density Functional Theory (DFT).[97]

### 3.6 Other issues

*Intellectual synthesis*

When a system has been studied by a range of methods (experiment, theory, computer simulation) and at a range of scales, a challenge is the synthesis of the results of these different studies. Value-laden judgements are made about the priority, importance, and validity of such attempts at synthesis. Often synthesis is relegated to a few sentences in the introductions and conclusions of papers. It needs to be more extensive and rigorous.

*New experimental probes and systems*

For known systems and emergent properties, there is the possibility of creating new methods and probes to investigate them at appropriate length and time scales. An example is the development of small angle neutron scattering for soft matter. The identification of new mesoscopic (emergent) scales may be intertwined with the development of new probes at the relevant scale. Sometimes development of the new probe leads to discovery of the scale. Other times, identification of the scale by theorists motivates the development of the new experimental probe.

New systems can be created and investigated in the hope of discovering new emergent properties, e.g., new states of matter. A more modest, but still important, goal is finding a new system that manifests a known emergent property but is more amenable to scientific study or technological application. Sometimes advances are made by identifying new classes of systems that exhibit an emergent phenomena or new members of a known class.

*Cultivating inter-disciplinary insights*

As emergent properties involve multiple scales they are often of interest to and amenable to study by more than one scientific discipline. Different disciplines may contribute complementary skills, methods, motivations, conceptual frameworks, and practical applications. For example, soft matter is of interest to physicists, chemists, biologists, and engineers. Protein folding which brings together physics, chemistry, molecular biology, and computer science. Neural networks brings together physics, neuroscience, psychology, and computer science.

Some of this interdisciplinarity is underpinned by association of emergence with universality. Two systems composed of very different components may exhibit collective phenomena with similar characteristics. This creates opportunities and challenges for inter-disciplinary collaboration. Cross-fertilisation is often not anticipated. For example, studies of spin glasses in condensed matter led to new methods and insights for protein folding, neuroscience, computer science, and evolutionary biology.

The extent that any of the above methods is developed, explored or sustained may be a matter of strategy and resource allocation, from that of the individual scientist to large funding agencies.

Organising principles and concepts provide a means to facilitate understanding, characterise phenomena, and organise knowledge. Examples include emergent scales, tipping points, phase diagrams, order parameters (collective variables), symmetry breaking, generalised rigidity, networks, universality, self-organised criticality, and rugged energy landscapes.

Before discussing how different characteristics of emergence and methods of investigation are manifest in specific scientific systems, I discuss the Ising model.

## 4. Ising models

The Ising model is emblematic of toy models that have been proposed and studied to understand and describe emergent phenomena. Although, originally proposed to describe ferromagnetic phase transitions, variants of it have found application in other areas of physics, in biology, economics, sociology, and neuroscience.

The general model is defined by a set of lattice points $\{i\}$, on each of which there is a "spin" $\sigma_i = \pm 1$ and a Hamiltonian

$$H = -\sum_{i,j} J_{ij}\sigma_i\sigma_j + h\sum_i \sigma_i$$

where h is the strength of an external magnetic field and $J_{ij}$ is the strength of the interaction between the spins on lattice sites i and j. The simplest models are where the lattice is regular, and the interaction is uniform and only non-zero for nearest-neighbour sites.

The Ising model illustrates many key features of emergent phenomena. Given the relative simplicity of the model, exhaustive studies over the past one hundred years have given definitive answers to questions that for more complex systems are often contentiously debated. I enumerate some of these insights: novelty, quantitative change can lead to qualitative change, spontaneous order, singularities, short-range interactions can produce long-range order, universality, mesoscopic scales, self-similarity, and that simple models can describe complex behaviour.

Most of these properties can be illustrated with the case of the Ising model on a square lattice with only nearest-neighbour interactions ($J_{ij} = J$). Above the critical temperature ($T_c$ = 2.25J), and in the absence of an external magnetic field (h=0) the system has no net magnetisation. For temperatures below $T_c$, a net magnetisation occurs, $\langle \sigma_i \rangle \neq 0$. For J > 0 (J < 0) this state has ferromagnetic (antiferromagnetic) order.

*Novelty*
The state of the system for temperatures below $T_c$ is *qualitatively different* from the state at higher temperatures or from the state of a set of non-interacting spins. For h=0 the Hamiltonian has the global spin-flip symmetry, $\sigma_i \rightarrow -\sigma_i$. However, the low-temperature state does not have this symmetry. Thus, the non-zero magnetisation is an emergent property, as defined by novelty. Thus, the low-temperature state is associated with *spontaneous symmetry breaking*, with long-range order, and with more than one possible equilibrium state, i.e., the magnetisation can be positive or negative.

*Quantitative change leads to qualitative change*
The qualitative change associated with formation of the magnetic state can occur with a small quantitative change in the value of the ratio T/J, i.e., either by decreasing T or increasing J, where the range of variation of T/J includes the critical value. Existence of the magnetic state is also associated with the quantitative change of increasing the number of spins from a large finite number to infinity. For a finite number of spins there is no spontaneous symmetry breaking.

*Singularities*
For a finite number of spins all the thermodynamic properties of the system are an analytic function of the temperature T and magnitude of the external field h. However, in the thermodynamic limit, thermodynamic functions become singular at T=$T_c$ and h=0. This is the critical point in the *phase diagram* of h versus T. Properties of the whole system, such as the specific heat capacity and the magnetic susceptibility, become infinite at the critical point. These singularities can be characterised by critical exponents, most of which have non-integer values. Consequently, the free energy of the system is not an analytic function of the variables T and h, over any domain containing their critical values, T=$T_c$ and h=0.

*Spontaneous order*
The magnetic state occurs spontaneously. The system self-organises, even in the presence of thermal fluctuations. There is no external field causing the magnetic state to form. There is long-range order, i.e., the value of spins that are infinitely far apart from one another are correlated. This also reflects a generalised rigidity (to be discussed later).

*Short-range interactions can produce long-range order.*
Although in the Hamiltonian there is no direct long-range interaction between spins, long-range order can occur for temperatures less than $T_c$. Prior to Onsager's exact solution of the two-dimensional model, published in 1944, many scientists were not convinced that this was possible.[52]

*Modification of single particle properties*
In the disordered state, the probability of a single spin being +1 or -1 is the same, in the absence of a magnetic field. In contrast, in the ordered state the probabilities are unequal.

*Universality*
The values of the critical exponents are independent of many details of the model, such as the value of J, the lattice constant and spatial anisotropy, and the presence of small interactions beyond nearest neighbour. Many details do not matter. Close to the critical point, the dependence of system properties on the temperature and magnetic field is given by universal functions. This is why the model can give a quantitative description of experimental data near the critical temperature, even though the model Hamiltonian is a crude description of the magnetic interactions in a real material. Furthermore, besides transitions in uniaxial magnets the Ising model can also describe transitions and critical behaviour in liquid-gas, binary alloys, and binary liquid mixtures.

*Structure at the mesoscale*
There are three important length scales associated with the model. Two are simple: the lattice constant (which defines the spatial separation of neighbouring spins), and the size of the whole lattice. These are the microscopic and macroscopic scale, respectively. The third scale is mesoscopic, emergent, and temperature dependent: the correlation length. This length is the distance over which spins are correlated with one another. This can also be visualised as the size of magnetisation domains seen in Monte Carlo simulations, as shown in the Figure below.

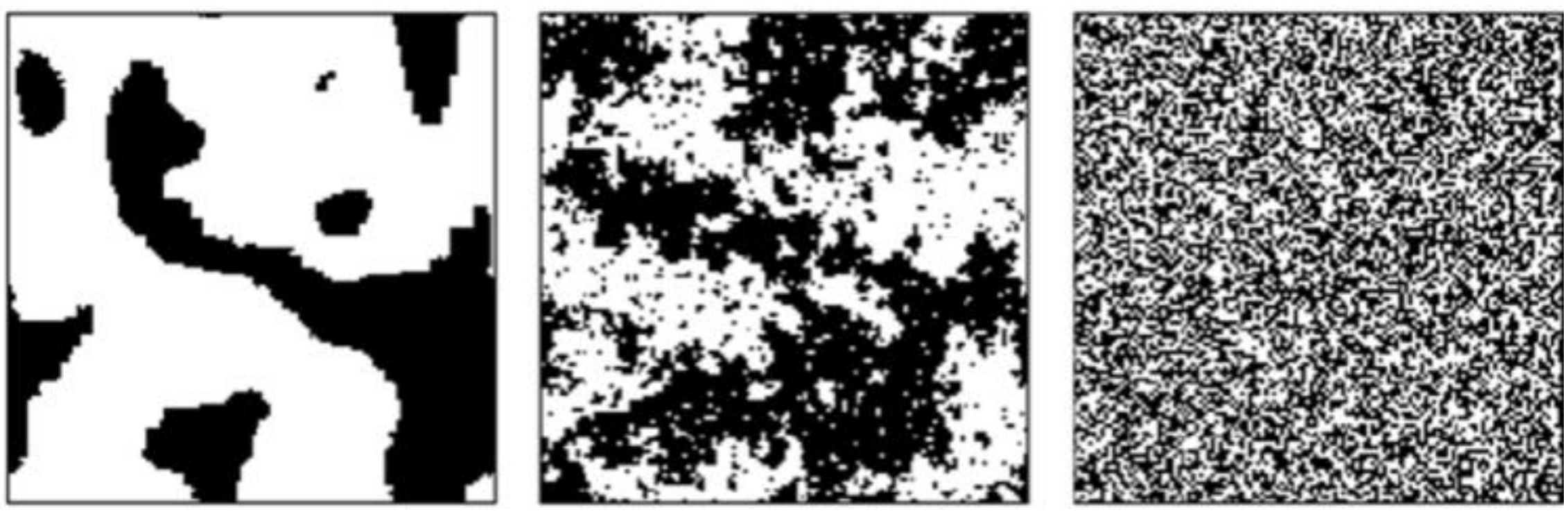

Figure 2. The left, centre, and right panels show a snapshot of a likely configuration of the system at a temperature less than, equal to, and greater than the critical temperature, Tc, respectively. The Figure is from Ref. [98]

Understanding the connection between the microscopic and macroscopic properties of the system requires studying the system at the intermediate scale of the correlation length. This scale also defines emergent entities [magnetic domains] that interact with one another weakly and via an effective interaction.

*Self-similarity*
At the critical temperature, the correlation length is infinite. Consequently, rescaling the size of the system, as in a renormalisation group transformation, the state of the systems does not change. The system is said to be scale-free or self-similar like a fractal pattern. This is the inspiration for the concept of self-organised criticality proposed by Per Bak to explain the prevalence of power laws in a wide range of physical, biological, and social systems.[99]

*Networks of interactions matter*
Properties of the system change when the network of interactions changes. This can happen when the topology or dimensionality of the lattice changes or when interactions beyond nearest neighbours are added with magnitudes comparable to the nearest-neighbour interactions. This can change the relationships between the parts and the whole. Sometimes details of the parts do matter, even producing qualitative changes. For example, changing from a two-dimensional rectangular lattice to a linear chain the ordered state disappears for any non-zero temperature. For antiferromagnetic nearest-neighbour interactions changing from a square lattice to a triangular lattice removes the ordering at finite temperature and leads to an infinite number of ground states at zero temperature. Thus, some microscopic details, such as the network of interactions, sometimes do matter. Knowing the components in a system is not sufficient; knowledge of how the components interact is also required.

The main point of this example is that to understand a large complex system we need to keep both the parts and the whole in mind. It is not either/or but both/and. Furthermore, there may be an intermediate scale, at which new entities emerge. Later I briefly discuss how these observations are relevant to debates about structuralism versus functionalism in biology, psychology, social sciences, and the humanities. I argue that they are trying to defend intellectual positions (and fashions) that contradict what the Ising model shows. One cannot preference either the whole, the parts, or intermediate structures.

*Potential and pitfalls of mean-field theory*
Given its simplicity, sometimes mean-field works surprisingly well. Other times it fails spectacularly. This can be illustrated by the Ising model on a hypercubic lattice of dimension d with nearest-neighbour interactions. Mean-field theory predicts a phase transition at a non-zero temperature for all dimensions. For d=1 this is qualitatively wrong. For d=2 and 3, this is qualitatively correct but quantitatively incorrect. The critical exponents are different from those given by mean-field theory. For d =4 and larger mean-field theory gives the correct critical exponents.

*Diversity: Simple models can describe complex behaviour*
Consider an Ising model with competing interactions, i.e. the neighbouring spins of a particular spin compete with one another and with an external magnetic field to determine the sign of the spin. An example is an Ising model on a hexagonal close packed (hcp) lattice with nearest neighbour antiferromagnetic interactions and an external magnetic field. The nearest-neighbour interactions are frustrated in the sense that if the energy of two neighbouring spins is minimised by being antiparallel a third adjacent spin must be parallel to one of these spins, frustrating the antiferromagnetic interaction. The hcp lattice can be viewed as layers of hexagonal (triangular) lattices where each layer is displaced relative to adjacent layers.

This model has been studied by materials scientists as it can describe the many possible phases of binary alloys, $A_xB_{1-x}$, where A and B are different chemical elements (for example, silver and gold) and the Ising spins on site i has value +1 or -1, corresponding to the presence of atom A or B on that site. The magnetic field corresponds to the difference in the chemical potentials of A and B and so is related to their relative concentration.[100] A study of this model[101] on the hexagonal close packed lattice at zero temperature found rich phase diagrams including 32 stable ground states with stoichiometries, including A, AB, $A_2B$, $A_3B$, $A_3B_2$, and $A_4B_3$. Of these structures, six are stabilized by purely nearest-neighbour interactions, eight by addition of next-nearest neighbour interactions. With multiplet interactions, 18 more distinct structures become stable. Even for a single stoichiometry, there can be multiple possible distinct orderings (and crystal structures).

A second example is the Anisotropic Next-Nearest Neighbour Ising (ANNNI) model on a cubic lattice. It supports a plethora of ordered states, even though the model has only two parameters. A graph of the wavevector of the ground state versus the anisotropy parameter or temperature has a fractal structure, and is known as the Devil's staircase.[102,103]
These two Ising models illustrate how relatively simple models, containing competing interactions (described by just a few parameters) can describe rich behaviour, particularly a diversity of ground states. They also illustrate how sometimes networks of interactions do matter, leading to qualitatively different behaviour.

*Dynamical formulations*
The original formulation of the Ising model has no dynamics and the focus was on properties in thermodynamic equilibrium. However, it can also be formulated with simple discrete dynamics and stochastic processes. These dynamical models can describe non-equilibrium phenomena such as metastability, hysteresis, domain nucleation, and coarsening. This formulation has been fruitful in applying the model to problems in finance and economics, as discussed later. One version, known as Glauber dynamics, is defined by

$$\sigma_i(t) = sign\left(\sum_j J_{ij}\sigma_j(t-1) + h + \eta_i(t)\right)$$

where $\eta_i(t)$ is randomly distributed. This equation belongs to the class of stochastic dynamical models of interacting particles which have been much studied mathematically and extensively by Liggett.[104]

*Cross-disciplinarity*
Beyond magnetism, Ising-like models are toy models in other fields, some of which will be discussed below. Examples include $Z_2$ lattice gauge theory, the Edwards-Anderson model for spin glasses, the Hopfield neural network model for associative memory, Hinton's Boltzmann machine for artificial intelligence, the random field Ising model in economics, and the Eaton-Munoz model for protein folding.

## 5. Stratification of physics

*Scales, strata, and subdisciplines*

There is a stratum of sub-disciplines of physics, illustrated in Table 1. For each stratum, there are a range of length, time, and energy scales that are relevant. There are distinct entities that are composed of the entities from lower strata. These composite entities interact with one another via effective interactions that arise due to the interactions present at lower strata and can be described by an effective theory. Each sub-discipline of physics is semi-autonomous. Collective phenomena associated with a single stratum can be studied, described, and understood without reference to lower strata.

Table entries are not meant to be exhaustive but to illustrate how emergence is central to understanding sub-fields of physics and how they are related and not related to other sub-fields. Sub-disciplines can also be defined sociologically. For each sub-discipline there are departments, courses, textbooks, professional organisations, conferences, journals, and funding programs.

| Sub-field of physics | Scale, length (metres) | Scale, energy (eV) | Entities | Effective interactions | Effective Theory | Collective Phenomena |
|---|---|---|---|---|---|---|
| Cosmology | $10^{21}$ - $10^{26}$ | | Cosmic microwave background, galaxies, dark matter, dark energy | Curvature of space-time | Equation of state, General relativity | Uniformity and isotropy |
| Stellar | $10^{8}$ - $10^{12}$ | | Protons, neutrons, helium, electrons, neutrinos | Buoyancy, Radiation pressure, Fermi degeneracy pressure | Equation of state | Neutron stars, white dwarfs, supernovae, |
| Earth science | $10^{3}$ - $10^{8}$ | | Clouds, oceans, currents, tectonic plates | | Climate models | Earthquakes, tornadoes, global warming |
| Fluid dynamics | $10^{-3}$ - $10^{3}$ | | Vortices, Rayleigh-Benard cells, Boundary layers, Thermal plumes | Buoyancy, convection, Kolmogorov | Navier-Stokes equation, mixing length model | Turbulence, pattern formation |
| Condensed Matter | $10^{-10}$ - $10^{-2}$ | $10^{-4}$ - 10 | Electrons, nuclei, Quasiparticles | Screened Coulomb, pseudopotentials, electron-phonon, BCS attraction | Ginzburg-Landau, Fermi liquid | States of matter, Topological order, Spontaneous Symmetry Breaking |
| Molecular | $10^{-10}$ - $10^{-8}$ | $10^{-3}$ - 10 | Molecules | van der Waals, H-bonds, ionic, and covalent | Lennard-Jones, Potential energy surface | Chemical bonding and reactions |
| Atomic | $10^{-11}$ - $10^{-9}$ | 1-100 | Atoms | | Hartree-Fock | Periodic table |
| Nuclear | $10^{-15}$ | $10^{3}$ - $10^{8}$ | Neutrons, protons, nuclei | Yukawa, Nuclear mean-field | Shell model, Interacting Boson, liquid drop | Magic numbers, fission, fusion, beta-decay |
| Elementary particles and fields | $10^{-15}$ - $10^{-15}$ | $10^{6}$ - $10^{12}$ | Quarks, leptons, photons, gluons, W bosons, Higgs boson | Gluon exchange | Standard model | Mesons, Hadrons, Spontaneous Symmetry Breaking, |
| | | | | | | Quark confinement, Asymptotic freedom |
| Early universe | $10^{-8}$ - $10^{12}$ | $10^{24}$ -$10^{28}$ | Topological defects, Inflatons | | GUTs | Anti-matter annihilation, phase transitions, inflation |
| Quantum gravity | $10^{-35}$ | $10^{28}$ | Gravitons | | | Hawking radiation, Black hole evaporation |

Table 1. Sub-disciplines of physics and the associated scales, effective interactions and theories, and phenomena. Time scales generally decrease from the top rows to the bottom rows of the Table.

Sometimes similar physics occurs at vastly different scales. For example, Baym discusses similarities of the physics associated with cold atomic gases and quark-gluon plasmas.[105] These similarities occur in spite of the fact that the relevant energy scales in the two systems differ by more than 20 orders of magnitude.

## 6. Classical physics

### 6.1 Theory of elasticity

This is an effective theory that describes smooth small distortions of solids over length scales much larger than atomic scales. It can be used to illustrate several aspects of emergence, particularly relating to the connection between macro- and micro- scales.

*Novelty.* Elasticity (rigidity) is an emergent property of a solid. The atoms that make up the solid do not have this property. Shear rigidity is a property that liquids do not have. Elasticity gives rise to the propagation of sound through a crystal.

*Intra-stratum closure*. The theory gives a complete description of phenomena on long length scales. There is no need for information from shorter length scales or microscopic details. Historically, the theory was completed before microscopic details such as crystal structures or quantum theory were known.

.

*Universality*. The theory is independent of the chemical composition of the solid or the origin of physical forces between the constituent atoms. The same elasticity theory is the continuum limit of both classical and quantum theories for the interactions and dynamics of the constituent atoms. Laughlin and Pines emphasised that this universality of the theory hides the underlying microscopic physics, calling it a protectorate.[67]

Energy functional has a simple form, being quadratic in gradients of the local deviations of the solid from uniformity. For a cubic crystal there are three independent elastic constants, $C_{11}$, $C_{12}$, and $C_{44}$. The latter determines the shear modulus.

Macro- properties can provide hints of the micro- properties. Marder gave two examples.[106] [The first is on p.290 and the second on p. 305 in his first edition].

First, relationships between the elastic constants constrain the nature of microscopic forces. Cauchy and Saint Venant showed that if all the atoms in a crystal interact through pair-wise central forces then $C_{44}=C_{12}$. However, in a wide range of elemental crystals, one finds that $C_{12}$ is one to three times larger than $C_{44}$. This discrepancy caused significant debate in the 19th century but was resolved in 1914 by Born who showed that angular forces between atoms could explain the violation of this identity. From a quantum chemical perspective, these angular forces arise because it costs energy to bend chemical bonds.

Second, even before the discovery of x-ray diffraction by crystals, there were strong reasons to believe in the microscopic periodic structure of crystals. The first paper on the dynamics of a crystal lattice was by Born and von Karman in 1912. This preceded the famous x-ray diffraction experiment of von Laue that established the underlying crystal lattice. In 1965, Born had the following reflection.

> "The first paper by Karman and myself was published before Laue's discovery. We regarded the existence of lattices as evident not only because we knew the group theory of lattices as given by Schoenflies and Fedorov which explained the geometrical features of crystals, but also because a short time before Erwin Madelung in Göttingen had derived the first dynamical inference from lattice theory, a relation between the infra-red vibration frequency of a crystal and its elastic properties.... Von Laue's paper on X-ray diffraction which gave direct evidence of the lattice structure appeared between our first and second paper. Now it is remarkable that in our second paper there is also no reference to von Laue. I can explain this only by assuming that the concept of the lattice seemed to us so well established that we regarded von Laue's work as a welcome confirmation but not as a new and exciting discovery which it really was."

Distinctly different types of sound, distinguished by their speed and their polarisation, can travel through a particular crystal are macroscopic manifestations of the specific translational and rotational symmetries that the crystal has at the microscopic level. The highest symmetry crystal is a cubic crystal. It has three distinct types of sound. In a crystal with no rotational or mirror symmetry there are many more different sound types. Hence, measurement of all the different types of sound in a crystal can provide information about the symmetry of the crystal.

The theory only describes small deviations from uniformity and so cannot describe objects that occur at the mesoscale, such as disinclinations and dislocations, that are topological defects. A hint of this failure was provided in the 1920s when it was found that the observed hardness of real materials was orders of magnitude smaller than predicted by the theory.

### 6.2 Thermodynamics

*Novelty*. Temperature and entropy are emergent properties. Classically, they are defined by the zeroth and second law of thermodynamics, respectively. The individual particles that make up a system in thermodynamic equilibrium do not have these properties. Kadanoff gave an example of how there is a qualitative difference in macro- and micro- perspectives. He pointed out how deterministic behaviour can emerge at the macroscale from stochastic behaviour at the microscale.[107] The many individual molecules in a dilute gas can be viewed as undergoing stochastic motion. However, collectively they are described by an equation of state such as the ideal gas law.
Temperature is a classical observable in quantum statistical mechanics. Primas gave a technical argument, involving C* algebras, that temperature is emergent: it belongs to an algebra of contextual observables but not to the algebra of intrinsic observables.[60] Following this perspective, Bishop argues that temperature and the chemical potential are (contextually) emergent.[11] Temperature is a macroscopic entity that is defined by the zeroth law of thermodynamics.

*Intra-stratum closure.* The laws of thermodynamics, the equations of thermodynamics (such as TdS = dU + pdV), and state functions such as S(U,V), provide a complete description of processes involving equilibrium states. A knowledge of microscopic details such as the atomic constituents or forces of interaction is not necessary for the description.

*Irreducibility*. A common view is that thermodynamics can be derived from statistical mechanics. The philosopher of science, Ernst Nagel claimed this was an example of theory reduction[108] and this is often cited by other philosophers. However, Nagel's argument is contentious. It is based solely on the case of classical ideal gas for which temperature is directly related to the average kinetic energy of the molecules in the gas. For a quantum system, such a simple relationship does not exist. For example, two different systems can have the same temperature but the average kinetic energy of their constitutent particles can be different.

David Deutsch claimed that the second law of thermodynamics is an "emergent law": it cannot be derived from microscopic laws, like the principle of testability.[109] Lieb and Yngvason stated that the derivation from statistical mechanics of the law of entropy increase "is a goal that has so far eluded the deepest thinkers."[110] In contrast, Weinberg claimed that Maxwell, Boltzmann, and Gibbs "showed that the principles of thermodynamics could in fact be deduced mathematically, by an analysis of the probabilities of different configurations… Nevertheless, even though thermodynamics has been explained in terms of particles and forces, it continues to deal with emergent concepts like temperature and entropy that lose all meaning on the level of individual particles."[111] (pages 40-41) I agree that *thermodynamic properties* (e.g., equations of state, and the temperature dependence of heat capacity, and phase transitions) can be deduced from statistical mechanics. However, *thermodynamic principles*, such as the second law, are not thermodynamic properties. These thermodynamic principles are required to justify the equations of statistical mechanics, such as the partition function, that are used to calculate thermodynamic properties. Thus, claiming that temperature is an emergent entity and irreducible may be related to the question as to whether the zeroth law can be derived from microscopic laws. Sklar has discussed the philosophical issues associated with the reduction of thermodynamics to statistical mechanics.[112] Baldovin et al. recently reviewed the physics issues associated with the foundations of statistical mechanics.[113]

Earlier, I mentioned how macroscopic thermodynamics provided hints of underlying microscopic physics through the Third law, Gibbs paradox, and the Sackur-Tetrode equation.

### 6.3 Brownian motion

*Novelty.* The random behaviour of Brownian motion emerges from underlying deterministic dynamics at the microscale.[107] The diffusion equation breaks time-reversal symmetry whereas the underlying microscopic dynamics has time-reversal invariance.

Sethna discussed how Brownian motion illustrates emergence.[114] Consider a large number of steps of the same size in random directions. The associated random walks have a scale invariant (fractal or self-similar) *structure* in time and space. i.e., if the time is rescaled by a factor N and the length is rescaled by a factor $\sqrt{N}$, then the walk looks the same. Secondly, the distance from the original position is given by a probability distribution that is a solution to the diffusion equation. Both properties are *universal*, i.e., they are independent of most details of the random walk and so apply to a wide range of systems in physics, chemistry, and biology.

Kadanoff noted the universality of random walks in terms of the independence from the underlying theoretical model. He stated[107]

> "The random walk model can be derived from more fundamental models of molecular scattering. The scattering events could be realized in at least three different ways: using classical mechanics (fully deterministic), using quantum mechanics (partially deterministic), or prespecifying the probabilities of scattering. Thus the probabilistic single-particle model, the random walk, can be equally well obtained from a many-particle model that is entirely deterministic, partially so, or not deterministic at all. Real gases will all show the same behavior independent of the detailed laws governing the scattering."

In Schrodinger's influential book, *What is life?,*[115] he used Brownian motion as an example of the principle of "order from disorder", that physical laws on a large scale are due to chaos on a small scale.

There are several different *effective theories* used to describe Brownian motion. These include the diffusion equation, Fokker-Planck equation, and Langevin equation.

### 6.4 Fluid dynamics

Weather involves many scales of distance, time, and energy. Describing weather means making decisions about what range of scales to focus on. Key physics involves thermal convection which reflects an interplay of gravity, thermal expansion, viscosity and thermal conduction. This can lead to Rayleigh-Bénard convection and convection cells.

*Novelty.* Turbulent flow is qualitatively different from normal uniform flow. In the transition to turbulent heat flow meso-scale structures emerge, such as plumes, flywheels, jets, and boundary layers.

*Discontinuities.* There are qualitative changes in fluid flow when a dimensionless parameter such as Rayleigh number or Reynolds number passes through some critical value.

Kadanoff describes three levels of description for convective turbulence, including laws unique to each level.[116] Multiple scales are associated with multiple entities:
-the molecules that make up the fluid

-small volumes of fluid that are in local thermodynamic equilibrium with a well-defined temperature, density, and velocity
-individual convection cells (rolls)
-collections of cells.
At each scale, the corresponding entities can be viewed as emerging from the interacting entities at the next smallest scale. Hence, at each scale the entities are collective degrees of freedom.

*Unpredictability.* In principle, a complete description, including the transition to turbulence, is given by the equations of fluid dynamics, including the Navier-Stokes equation. Despite the apparent simplicity of these equations, making definitive predictions from them remains elusive.

*A toy model* is the lattice gas on a hexagonal lattice. It can describe the complex structures seen for incompressible fluid flow behind a cylinder for a Reynolds number of about 300.[117]

*Universality.* Similar properties occur for fluids with diverse chemical composition and origins of the molecular forces responsible for fluid properties such as viscosity or thermal conductivity. Emergent properties often depend on dimensionless parameters that are combinations of several physical parameters, such as the Reynolds number or the dimensionless heat flow Nu. There are universal scaling laws, such as relating dimensionless heat flow to temperature difference or the probability distribution for temperature fluctuations in a cell.[116] For turbulence, Kolmogorov proposed a scaling law for the probability distribution of the scales at which energy is transferred from larger eddies to smaller ones without significant viscous dissipation.[118]

*Singularities.* Consider the incompressible Euler equations in the limit that the viscosity tends to zero. Onsager argued that there is non-zero energy dissipation in this limit and that the velocity field does not remain differentiable.[119–121] This singularity is central to Kolmogorov's theory of turbulence. Another class of singularities occurs when a mass of fluid forms a thin neck and that neck can breaks so that the fluid breaks into two pieces. There are then singularities in the mathematical solutions to the fluid dynamic equations.[122] This is also relevant to fluid flow on a surface that is partially wet and dry.[123]

### 6.5 Pattern formation

Patterns in space and/or time form in fluid dynamics (Rayleigh-Bernard convection and Taylor-Couette flow), laser physics, materials science (dendrites in formation of solids from liquid melts), biology (morphogenesis), and chemistry (Belousov-Zhabotinsky reactions). Most of these systems are driven out of equilibrium by external constraints such as temperature gradients.

*Novelty.* The parts of the system can be viewed as the molecular constituents or small uniform parts of the system. In either case, the whole system has a property (a pattern) that the parts do not have.

*Discontinuity.* The system makes a transition from a uniform state to a non-uniform state when some parameter becomes larger than a critical value.

*Universality.* Similar patterns such as convection rolls in fluids can be observed in diverse systems regardless of the microscopic details of the fluid. Often there is a single parameter, such as the Reynolds number, which involves a combination of fluid properties, that determines the type of patterns that form. Cross and Hohenberg[124] highlighted how the models and mechanisms of pattern formation across physics, chemistry, and biology have similarities. Turing's model for pattern formation in biology associated it with concentration gradients of reacting and diffusing molecules. However, Gierer and Meinhardt[125] showed that all that was required was a network with competition between short-range positive feedback and long-range negative feedback. This could occur in circuit of cellular signals. This illustrates the problem of protectorates.

*Self-organisation.* The formation of a particular pattern occurs spontaneously, resulting from the interaction of the many components of the system.

*Effective theories.* A crystal growing from a liquid melt can form shapes such as dendrites. This process involves instabilities of the shape of the crystal-liquid interface. The interface dynamics are completely described by a few partial differential equations that can be derived from macroscopic laws of thermodynamics and heat conduction.[126]

*Diversity.* Diverse patterns are observed, particularly in biological systems. In toy models, such as the Turing model, with just a few parameters, a diverse range of patterns, both in time and space, can be produced by varying the parameters. Many repeated iterations can lead to a diversity of structures. This may result from a sensitive dependence on initial conditions and history. For example, each snowflake is different because as it falls it passes through a slightly different environment, with small variations in temperature and humidity, compared to others.[127]

*Toy models.* Turing proposed a model for morphogenesis in 1952 that involved two coupled reaction-diffusion equations. Homogeneous concentrations of the two chemicals become unstable when the difference between the two diffusion constants becomes sufficiently large. A two-dimensional version of the model can produce diverse patterns, many resembling those found in animals.[128] However, after more than seventy years of extensive study, many developmental biologists remain sceptical of the relevance of the model, partly because it is not clear whether it has a microscopic basis. Kicheva et al.,[129] argued that "pattern formation is an emergent behaviour that results from the coordination of events occurring across molecular, cellular, and tissue scales." Other toy models include Diffusion Limited Aggregation, due to Witten and Sander,[130,131] and Barnsley's iterated function system for fractals that produces a pattern like a fern.[132]

### 6.6 Caustics in optics

Michael Berry highlighted how caustics illustrate emergence.[61,133,134] He stated that "A caustic is a collective phenomenon, a property of a family of rays that is not present in any individual ray. Probably the most familiar example is the rainbow."

Caustics are envelopes of families of rays on which the intensity diverges. They occur in media where the refractive index is inhomogeneous. For example, the cells seen in bodies of sunlit water occur due to an interplay of the uneven air-water interface and the difference in the refractive index between air and water. For rainbows, key parameters are the refractive index of the water droplets and the size of the droplets. The caustic is not the "rainbow", i.e., the spectrum of colours, but rather the large light intensity associated with the bow.

Caustics illustrate several characteristics of emergent properties: novelty, singularities, hierarchies, new scales, effective theories, and universality.

*Novelty.* The whole system (a family of light rays) has a property (infinity intensity) that individual light rays do not.

*Discontinuities.* A caustic defines a spatial boundary across which there are discontinuities in properties.

*Irreducibility and singular limits.* Caustics only occur in the theory of geometrical optics which corresponds to the limit where the wavelength of light goes to zero in a wave theory of light. Caustics (singularities) are not present in the wave theory.

*Hierarchies.* These occur in two different ways. First, light can be treated at the level of rays, scalar waves, and vector waves. At each level, there are qualitatively different singularities: caustics, phase singularities (vortices, wavefront dislocations, nodal lines), and polarisation singularities. Second, treating caustics at the level of wave theory, as pioneered by George Bidell Airy, reveals a hierarchy of non-analyticities, and an interference pattern, reflected in the supernumerary part of a rainbow.

*New (emergent) scales.* An example is the universal angle of 42 degrees subtended by the rainbow, that was first calculated by Rene Descartes. Airy's wave theory showed that the spacing of the interference fringes shrinks as $\lambda^{2/3}$, where $\lambda$ is the wavelength.

*Effective theories.* At each level of the hierarchy, one can define and investigate effective theories. For ray theory, the effective theory is defined by the spatially dependent refractive index and the ray action.

*Universality.* Caustics exist for any kind of waves: light, sound, and matter. They exhibit "structural stability". They fall into equivalence (universality) classes that are defined by the elementary catastrophes enumerated by Thom and Arnold.[133] Any two members of a class can be smoothly deformed into one another. K is the number of parameters needed to define the class and the associated polynomial.

## 6.7 Chaos theory

The origins of the field lie with two toy models: Lorenz's model and the logistic iterative map. In 1975, Robert May published an influential review entitled, "Simple mathematical models with very complicated dynamics."[135] I now review the two models and how they illustrate how emergent properties occur in systems in which there are many iterations.

*Lorenz model*
This model was studied by the meteorologist Edward Lorenz in 1963, in a seminal paper, "Deterministic Nonperiodic Flow."[136] Under the restrictive conditions of considering the dynamics of a single convection roll the model can be derived from the full hydrodynamic equations describing Rayleigh-Bernard convection. Lorenz's work and its impact is beautifully described in James Gleick's book *Chaos: The Making of a New Science*.[87]

The model consists of just three coupled ordinary differential equations:

$$\dot{x} = -\sigma x + \sigma y \ , \qquad \dot{y} = rx - y - xz \ , \qquad \dot{z} = -bz + xy$$

The variables x(t), y(t), and z(t) describe, respectively, the amplitude of the velocity mode, the temperature mode, and the mode measuring the heat flux Nu, the Nusselt number. x and y characterize the roll pattern. The model has three dimensionless parameters: r, σ, and b. r is the ratio of the temperature difference between the hot and cold plate, to its critical value for the onset of convection. It can also be viewed as the ratio of the Rayleigh number to its critical value. σ is the Prandtl number, the ratio of the kinematic viscosity to the thermal diffusivity. σ is about 0.7 in air and 7 in water. Lorenz used σ = 10. b is of order unity and conventionally taken to have the value 8/3. It arises from the nonlinear coupling of the fluid velocity and temperature gradient in the Boussinesq approximation to the full hydrodynamic equations.

The model is a toy model because for values of r larger than $r_c$ (defined below) the approximation of three modes for the full hydrodynamic equations describing thermal convection is not physically realistic. Nevertheless, mathematically the model shows its most fascinating properties in that parameter regime.

*Novelty*

The model has several distinct types of long-time dynamics: stable fixed points (no convection), limit cycles (convective rolls), and most strikingly a chaotic strange attractor (represented below). The chaos is reflected in the sensitive dependence on initial conditions.

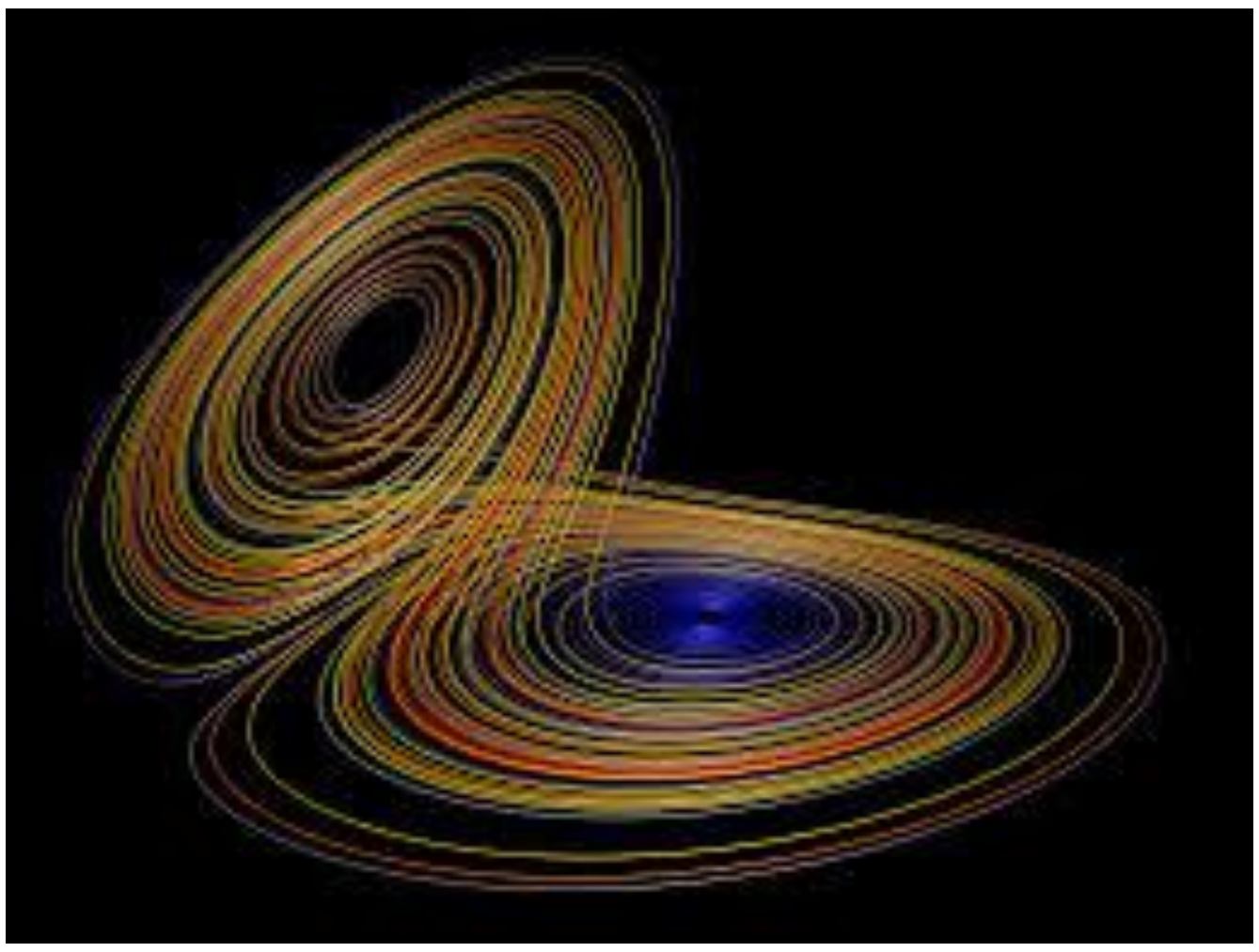

Figure. Image of the strange attractor associated with the Lorenz model.

Briefly, a strange attractor is a curve of infinite length that never crosses itself and is contained in a finite volume. This means it has a fractal structure and a non-trivial Hausdorff dimension, calculated by Viswanath[137] to be 2.0627160.

*Discontinuities*

Quantitative changes lead to qualitative changes. For r < 1, no convection occurs. For r > 1, convective rolls develop, but these become unstable for

$$r > r_c = \sigma \frac{\sigma + b + 3}{\sigma - b - 1}$$

and a strange attractor develops.

*Phase diagram*

Lorenz only considered one set of parameter values [r =28, σ=10, and b=8/3]. This was rather fortunate, because then the strange attractor was waiting to be discovered. The phase diagram shown below maps[138] out the qualitatively different behaviours that occur as a function of σ (vertical axis) and r (horizontal axis).

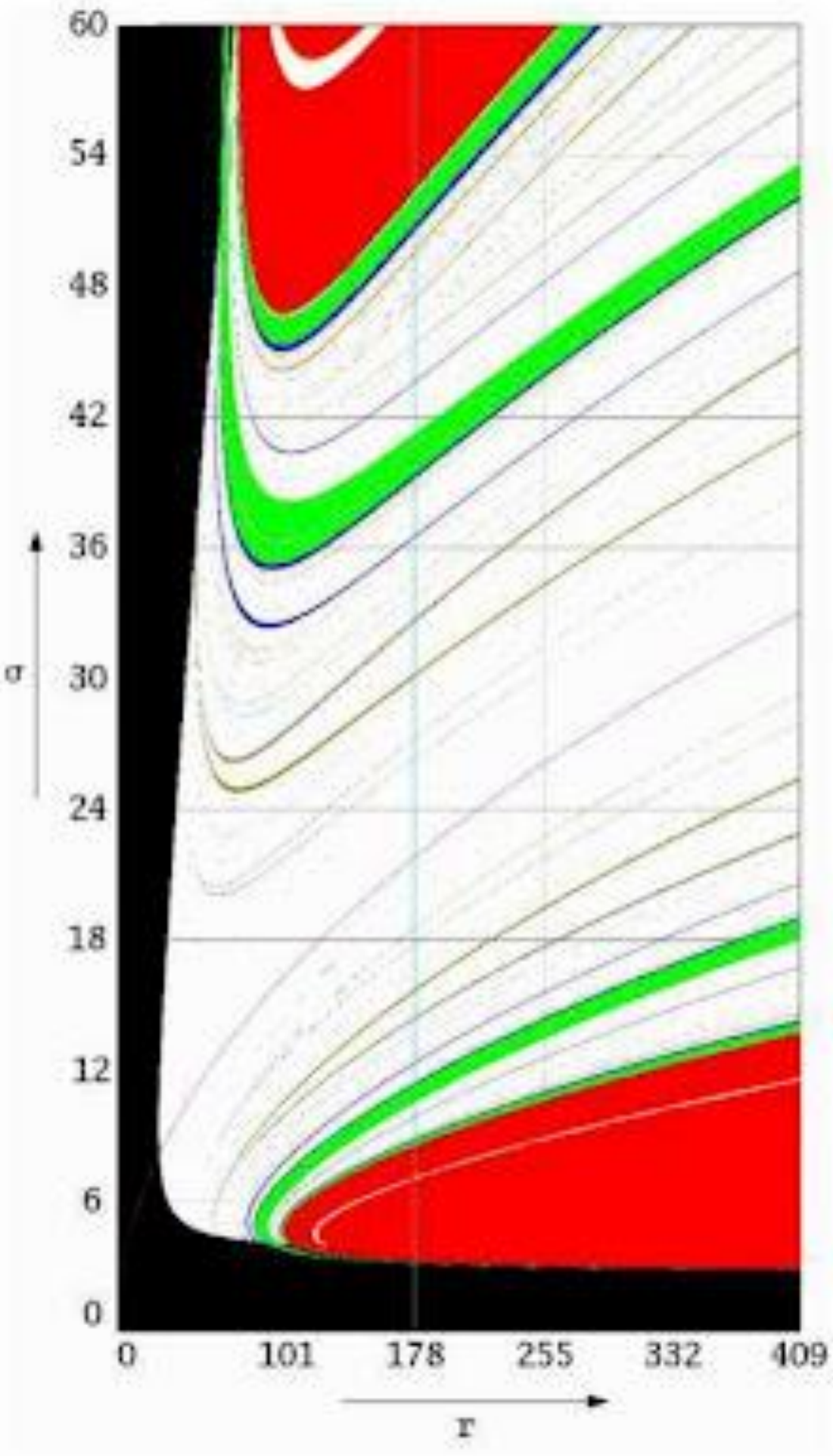

Figure. Phase diagram for the Lorenz model. Different phases are the fixed points P± associated with convective rolls (black), orbits of period 2 (red), period 4 (green), period 8 (blue), and chaotic attractors (white).

*Universality*

The details of the molecular composition of the fluid and the intermolecular interactions are irrelevant beyond how they determine the three parameters in the model. Hence, qualitatively similar behaviour can occur in systems with a wide range of chemical compositions and physical properties. The strange attractor also occurs for a wide range of parameters in the model.

*Unpredictability*
Although the system of three ordinary differential equations is simple, discovery of the strange attractor and the chaotic dynamics was unanticipated. Furthermore, the dynamics in the chaotic regime are unpredictable, given the sensitivity to initial conditions.

*Top-down causation*
The properties and behaviour of the system are not just determined by the properties of the molecules and their interactions. The external boundary conditions, the applied temperature gradient and the spatial separation of the hot and cold plates, are just as important in determining the dynamics of the system, including motion as much smaller length scales.

*Logistic iterative map*

The iterative map was studied by Feigenbaum who discovered universal properties that he explained using renormalisation group techniques developed to describe critical phenomena.[139] The state of the system is defined by infinite discrete "time series" $\{x_0, x_1, \dots . x_m, \dots\}$. The "interaction" between the $x_m$ is defined by the iterative map

$$x_{m+1} = \mu \, x_m(1 - x_m)$$

where the parameter $\mu$ can be viewed as the strength of the interactions.

*Novelty.*
Ordered states are periodic orbits to which the system converges regardless of the initial state $x_0$. These states are qualitatively different from one another, just like different crystals with different symmetry are different from one another. The chaotic orbits are qualitatively different from the periodic orbits.

*Discontinuities, tipping points, and phase diagram.*
Quantitative change in μ leads to qualitative change in behaviour. As μ increases "period doubling" occurs. At $\mu = \mu_n$, there is a transition from and orbit with period n to one with period 2n. For $\mu > \mu_\infty$ the dynamics is chaotic, i.e., there is no periodicity and there is sensitivity to the initial value, $x_0$. For larger μ, periodicity reappears for some finite range of μ, before chaotic behaviour recurs. The associated bifurcation diagram also exhibits self-similarity.

*Universality.*
Feigenbaum numerically observed the scaling relationship

$$\mu_n - \mu_\infty \sim \delta^{-n}$$

where $\delta$= 4.6692... is the first Feigenbaum constant. He conjectured this was true for any iterative map, $x_{n+1} = f(x_n)$ where f(x) is of similar shape to the logistic map, and gave a justification inspired by the renormalisation group. This conjecture was rigorously proven by Collet, Eckmann, and Lanford.[140] They also showed how logistic map could describe transition to turbulence. Libchaber performed experiments showing the period doubling transition to turbulence in Rayleigh-Bernard convection cells, observing a value of $\delta = 4.4 \pm 0.1$.[141]

### 6.8 Sychronisation of oscillators

When a large number of harmonic oscillators with different frequencies interact with one another they can undergo a transition to a collective state in which they all oscillate with the same frequency. This phenomenon has been seen in a diverse range of systems including arrays of Josephson junctions, neural networks, fireflies, and London's Millenium bridge.[142,143]

Kuramoto proposed a toy model[144] that can describe the essential aspect of the phenomena. It describes a phase transition that occurs at a critical value of the coupling strength. For weak coupling the oscillators are incoherent, oscillating at different frequencies. Above the critical coupling they oscillate in phase at the same frequency.

### 6.9 Arrow of time

Time has a direction. Microscopic equations of motion in classical and quantum mechanics have time-reversible symmetry. But this symmetry is broken for many macroscopic phenomena. This observation is encoded in the second law of thermodynamics. We experience the flow of time and distinguish past, present, and future. The arrow of time is manifest in phenomena that occur at scales covering many orders of magnitude. Below, six different arrows of time are listed in order of increasing time scales. The relationship between some of them is unclear. These are discussed by Leggett in chapter 5 of *The Problems of Physics*[145] and more recently reviewed by Ellis and Drossel.[146]

*Elementary particle physics.* CP violation is observed in certain phenomena associated with the weak nuclear interaction, such as the decay of neutral kaons observed in 1964. The CPT symmetry theorem shows that any local quantum field theory that is invariant under the "proper" Lorentz transformations must also be invariant under combined CPT transformations. This means that CP violation means that time-reversal symmetry is broken. In 1989, the direct violation of T symmetry was observed.

*Electromagnetism.* When an electric charge is accelerated an electromagnetic wave propagates out from the charge towards infinity. Energy is transferred from the charge to its environment. We do not observe a wave that propagates from infinity into the accelerating charge, i.e., energy being transferred from the environment to the charge. Yet this possibility is allowed by the equations of motion for electromagnetism. There is an absence of the "advanced" solution to the equations of motion.

*Thermodynamics.* Irreversibility occurs in isolated systems. They tend towards equilibrium and uniformity. Heat never travels from a cold body to a hotter one. Fluids spontaneously mix. There is a time ordering of the thermodynamic states of isolated macroscopic systems. The thermodynamic entropy encodes this ordering.

*Psychological experience.* We remember the past and we think that we can affect the future. We don't think we can affect the past or know the future.

*Biological evolution.* Over time species adapt to their environment and become more complex and more diverse.

*Cosmology.* There was a beginning to the universe. The universe is expanding not contracting. Density perturbations grow independent of cosmic time. Hawking et al.[147] used this to define the direction of time.

The problem of how statistical mechanics connects time-reversible microscopic dynamics with macroscopic irreversibility is subtle and contentious. Lebowitz claimed that this problem was solved by Boltzmann, provided that the distinction between typical and average behaviour are accepted, along with the Past Hypothesis.[148] This was advocated by Albert[149] and states that the universe was initially in a state of extremely low entropy. Wallace discussed the need to accept the idea of probabilities in law of physics and that the competing interpretations of probability as frequency or ignorance matter.

The Past Hypothesis is fascinating because it connects the arrow of time seen in the laboratory and everyday life (time scales of microseconds to years) to cosmology, covering timescales of the lifetime of the universe ($10^{10}$ years) and the "initial" state of the universe, perhaps at the end of the inflationary epoch ($10^{-33}$ seconds). This also raises questions about how to formulate the Second Law and the concept of entropy in the presence of gravity and on cosmological length and time scales.[150,151]

## 7. Quantum-classical boundary

Classical physics emerges from quantum physics in the limit that Planck's constant tends to zero. How this happens is a source of much debate and study, both from a scientific and philosophical point of view. Berry points out that this limit is a singular asymptotic expansion.[61]

*Quantum measurement problem*

One of the biggest challenges in the foundations of physics is the quantum measurement problem. It is associated with a few key (distinct but related) questions.

i. How does a measurement convert a coherent state undergoing unitary dynamics to a "classical" mixed state for which we can talk about probabilities of outcomes?

ii. Why is the outcome of an individual measurement always definite for the "pointer states" of the measuring apparatus?

iii. Can one derive the Born rule, which gives the probability of a particular outcome?

*Emergence of the classical world from the quantum world via decoherence*

A quantum system always interacts to some extent with its environment. This interaction leads to decoherence, whereby quantum interference effects are washed out.[152] Consequently, superposition states of the system decay into mixed states described by a diagonal density matrix. A major research goal of the past three decades has been understanding decoherence and the extent to which it does provide answers to the quantum measurement problem. One achievement is that decoherence theory seems to give a mechanism and time scale for the "collapse of the wavefunction" within the framework of unitary dynamics. However, this is not the case because decoherence is not the same as a projection (which is what a single quantum measurement is). Decoherence does not produce definite outcomes but rather statistical mixtures. Decoherence only resolves the issue if one identifies ensembles of measured states with ensembles of the decohered density matrix (the statistical interpretation of quantum mechanics). Thus, it seems decoherence only answers the first question above, but not the last two. On the other hand, Zurek has pushed the decoherence picture further and given a “derivation” of the Born rule within its framework, with an additional assumption that he dubs quantum Darwinism.[153],[154] Decoherence does *not* solve the quantum measurement problem: measurements always produce definite outcomes.

One approach to solving the quantum measurement problem is to view quantum theory as only an approximate theory. It could be an effective theory for some underlying theory valid at time and length scales much smaller than those for which quantum theory has been precisely tested by experiments. This is sometimes known as the Emergent Quantum Mechanics (EmQM) hypothesis.

*Emergence of quantum theory from a “classical” statistical theory*

Einstein did not accept the statistical nature of quantum theory and considered it should be derivable from a more “realistic” theory. In particular, he suggested “a complete physical description, the statistical quantum theory would …. take an approximately analogous position to the statistical mechanics within the framework of classical mechanics.”[155]

Einstein's challenge was taken up in a concrete and impressive fashion by Stephen Adler in a book, “Quantum Theory as an Emergent Phenomenon: The Statistical Mechanics of Matrix Models as the Precursor of Quantum Field Theory”, [156] published in 2004. A helpful summary is given in a review by Pearle.[157]

The starting point is "classical" dynamical variables $q_r$ and $p_r$ which are NxN matrices, where N is even. Half of these variables are bosonic, and the others are fermionic. They all obey Hamilton's equations of motion for an unspecified Hamiltonian H. Three quantities are conserved: H, the fermion number N, and (importantly) the traceless anti-self-adjoint matrix,

$$\tilde{C} \equiv \sum_B [q_r, p_r] - \sum_F \{q_r, p_r\}$$

where the first term is the sum for all the bosonic variables of their commutator, and the second is the sum over anti-commutators for the fermionic variables.

Quantum theory is obtained by tracing over all the classical variables with respect to a canonical ensemble with three (matrix) Lagrange multipliers [analogues of temperature and chemical potential in conventional statistical mechanics] corresponding to the conserved quantities H, N, and C. The expectation values of the diagonal elements of C are assumed to all have the same value, hbar!

An analogy of the equipartition theorem in classical statistical mechanics (which looks like a Ward identity in quantum field theory) leads to dynamical equations (trace dynamics) for effective fields. To make these equations look like regular quantum field theory, an assumption is made about a hierarchy of length, energy, and "temperature" [Lagrange multiplier] scales, which cause the Trace dynamics to be dominated by C rather than H, the trace Hamiltonian. Adler suggests these scales may be Planck scales. Then, the usual quantum dynamical equations and the Dirac correspondence of Poisson brackets and commutators emerge. Most of the actual details of the trace Hamiltonian H do not matter; another case of universality, a common characteristic of emergent phenomena.

The "classical" field C fluctuates about its average value. These fluctuations can be identified with corrections to locality in quantum field theory and with the noise terms which appear in the modified Schrodinger equation of "physical collapse" models of quantum theory.[158],[159]

More recently, theorists including t'Hooft and Preskill have investigated how quantum mechanics can emerge from other deterministic systems. Underlying deterministic systems considered include Hamilton-Randers systems defined in co-tangent spaces of large dimensional configuration spaces,[160] neural networks,[161] cellular automata,[162] fast moving classical variables,[163] and the boundary of a local classical model with a length that is exponentially large in the number of qbuits in the quantum system.[164] The fact that quantum theory can emerge from such a diverse range of underlying theories again illustrates universality. In most of these versions of EmQM the length scale at which the underlying theory becomes relevant is conjectured to be of the order of the Planck length.

The question of quantum physics emerging from an underlying classical theory is not just a question in the foundations of physics or in philosophy. Slagle points out[165] that Emergent Quantum Mechanics may mean that the computational power of quantum computers is severely limited. He has proposed a specific experimental protocol to test for EmQM. A large number d of entangling gates (the circuit depth d) is applied to n qbits in the computational basis, followed by the inverse gates. This is followed by measurement in the computational basis. The fidelity should decay exponentially with d, whereas for EmQM will decay much faster above some critical d, for sufficiently large n.

Slagle and Preskill justified their work on EmQM as follows.[164]

> "A characteristic signature of emergence at low energy or long time and distance scales is that the resulting physics is typically well described by remarkably simple equations, which are often linear (e.g., the harmonic oscillator) or only consist of lowest order terms in an effective Lagrangian."

In principle, it is possible that quantum mechanics is also only an approximate description of reality. Indeed, Schrödinger's equation is a simple linear differential equation, suggesting that it might arise as the leading approximation to a more complete model."

Consistent with experience in other areas of physics it is natural to think of QM as emerging in the limit of low energy or long distance. However, it could also emerge in the limit of reduced quantum complexity. In other words, QM may not hold for Hilbert spaces of high complexity. Experimental tests of QM, such as Bell inequality violations, have been restricted to relatively simple and small Hilbert spaces, and problems of low computational complexity.[164] However, it is not clear how to measure quantum complexity. It may not be just a question of the size of the Hilbert space, but perhaps its geometry. For just two entangled qubits the geometry is complex.[166]

Independent of experimental evidence, EmQM provides an alternative interpretation to quantum theory that avoids the thorny issues such as the many-worlds interpretation.

## 8. Quantum condensed matter physics

A central question of condensed physics is how do the physical properties of a state of matter *emerge* from the properties of the atoms of which the material is composed and their interactions? There are three dimensions to the question. One is at the macroscale to characterise the physical properties of a specific state of matter. A second dimension is to characterise the microscopic constituents of materials exhibiting this state. The third dimension is to connect the macroscopic and microscopic properties. Usually, this connection is done via the mesoscale. Other important questions include whether all possible states of matter can be classified; and if so, how.

Emergence describes why condensed matter physics works as a unified discipline.[32] As a result of universality there are concepts and theories that can describe diverse phenomena in materials that are chemically and structurally diverse. For example, although superfluidity occurs in liquids and superconductivity in solids there are unifying concepts that can describe both.

### 8.1 States of matter

Condensed matter physics is about the study of different states of matter. There are a multitude of states: liquid, crystal, liquid crystal, glass, superconductor, superfluid, ferromagnetic, antiferromagnetic, quantum Hall, … Each state has qualitatively different properties. Even within this list, there are a multitude of different states. For example, there are different liquid crystal states: nematic, chlorestic, smectic… This incredible *diversity* is characteristic of emergence.

A state of matter has emergent properties as the constituent parts of the system (atoms or molecules) do not have these properties. Novel properties may include spontaneous symmetry breaking, long-range order, generalised rigidity, topological defects, and distinct low-energy excitations (quasiparticles).

Diamond and graphite are distinct solid states of carbon. They have *qualitatively different* physical properties, at both the microscopic and the macroscopic scale. Graphite is common, black, soft, and conducts electricity moderately well. In contrast, diamond is rare, transparent, hard, and conducts electricity very poorly. Their crystal structures are different and the network of interactions between carbon atoms are different.

*Novelty.* In many cases, symmetry can be used to define the *qualitative difference* between different states and the nature of the ordering in the state. An order parameter can be used to quantify this qualitative difference. It only has a non-zero value in the ordered state. New entities are associated with states of matter, such as topological defects.

*Self-organisation* is reflected in the fact that for states in thermodynamic equilibrium as the system is cooled below a transition temperature it self-organises into a state with the long-range order associated with the broken symmetry. This why it is said that the symmetry is *spontaneously* broken.

*Discontinuities.* As external parameters such as temperature, pressure, or magnetic field are varied discontinuities or singularities occur in thermodynamic properties such as the specific heat capacity or density, as the system undergoes a phase transition, i.e., transforms into the ordered state. These discontinuities can be used to map out a *phase diagram* which shows under what conditions (external parameters) the different states are thermodynamically stable.

*Universality i*s evident in that materials that have different chemical composition or crystal structure can be found in the same state of matter. An example is superconductivity. More examples are given later.

Solids are rigid and it costs energy to distort the shape of a finite solid with external stresses. This contrasts with liquids whose shape changes to that of its container. In general, there is a rigidity associated with an ordered state of matter. Disturbing the relative spatial arrangement of the constituents of the system has an energy cost. This defines an *emergent length scale*, which is often *mesoscopic*. In superconductors this scale is the coherence length.

Landau's theory of phase transitions provides an *effective theory* for all states of matter associated with a broken symmetry. It is a valid at the length scale associated with the generalised rigidity. Based on the symmetry of the order parameter, the free energy functional associated with the theory can be written down using group theoretical analysis.

*Modularity at the mesoscale.* In some states of matter, properties are determined by topological defects such as vortices in superfluids and dislocations in solids. The spatial size of these defects is typically of the order of the length scale associated with the generalised rigidity and much larger than the microscopic scale.

Metallic materials that are used in industry and studied by metallurgists are not perfectly ordered crystals. Many of their properties, such as plasticity and the rate of crystal growth from a melt, are determined by the presence and behaviour of topological defects such as dislocations and disinclinations. This is an example of how macroscopic behaviour is largely determined by and best understood in terms of the mesoscale and not the atomic scale.

*Unpredictability.* As discussed previously, very few states of matter have been predicted prior to their experimental discovery. Their existence in principle might be predicted based on Landau's theory. However, predicting whether such a state will be exhibited by a material with a specific chemical composition is much harder.

## Topological order

Until a few decades ago it was believed that Landau's theory provided a complete classification of all states of matter. However, it turns out there are states of matter that do not exhibit a broken symmetry, but rather topological order. It is present in several different states of matter, some of which will be discussed further below. Examples include the quantum Hall effect, Haldane quantum spin chains, topological insulators, and the Ising $Z_2$ lattice gauge theory. A signature of topological order in quantum many-body systems is that the degeneracy of the ground state depends on the topology of space, such as the genus of the space, e.g., for a two-dimensional surface whether it has the topology of sphere, donut, or pretzel. The genus is a topological invariant, meaning that it is unchanged by continuous distortions of the surface. The topological character of the state means that there are qualitative differences between states and provides a natural explanation of universality as many of the details do not matter.

An example of an *effective theory* for a state with topological order is the topological field theory of Wen and Zee that includes a Chern-Simons topological term to describe the edge currents in a fractional quantum Hall state.[167]

*Critical phenomena*

The study of systems close to a critical point in their phase diagram has provided important insights into emergence. The renormalisation group (RG) provides a concrete and rigorous tool to connect scales, construct effective theories, and define universality classes. The RG can be used to do coarse graining and define collective degrees of freedom, such as associated with order parameters. The RG is needed to treat the effect of thermal fluctuations in the local order that are neglected by Landau's theory which is a mean-field theory. Fluctuations can lead to qualitatively different properties.

Consider the case of a fluid near the liquid-gas critical point. There are a range of different scales at which the system can be studied: from the macroscopic scale of the whole fluid down to the microscopic scale of atoms and molecules. Of particular relevance is the mesoscopic scale of blobs of gas inside liquid. Near the critical point these blobs increase in size leading to critical opalescence where incident light is scattered sufficiently for the fluid to take on a milky appearance.

With the range of scales there is a hierarchy of effective theories: quantum chemistry, van der Waals, hard sphere fluid, Ising lattice gas, Landau theory, and the Wilson (renormalisation group) theory. The effective theory at each strata can be derived, or at least justified, from the theory at the strata below.

*Toy models*

There is a long list of toy models that have provided important insights into strongly correlated electron systems. Examples include the Anderson model for a magnetic impurity atom in a metal, the Hubbard model for the Mott insulator-metal transition, the Kondo model for a spin in a metal, the AKLT model for an antiferromagnetic spin-1 chain, the Luttinger model for a one-dimensional interacting fermion gas, and the Haldane model for a topological insulator. The emphasis in the design of these models is simplicity and tractability for analysis, not faithfulness to the microscopic details of real materials. The key attribute of most of these models is that they exhibit a specific emergent phenomenon and that this can be turned on and off by varying a parameter in the model or temperature. Study of several specific toy models led to predictions of new states of matter.

## 8.2 Quasiparticles

In many-body systems the modularity that occurs at the mesoscale is exhibited by the existence of quasiparticles. They are observable entities and can be the basis of effective theories.

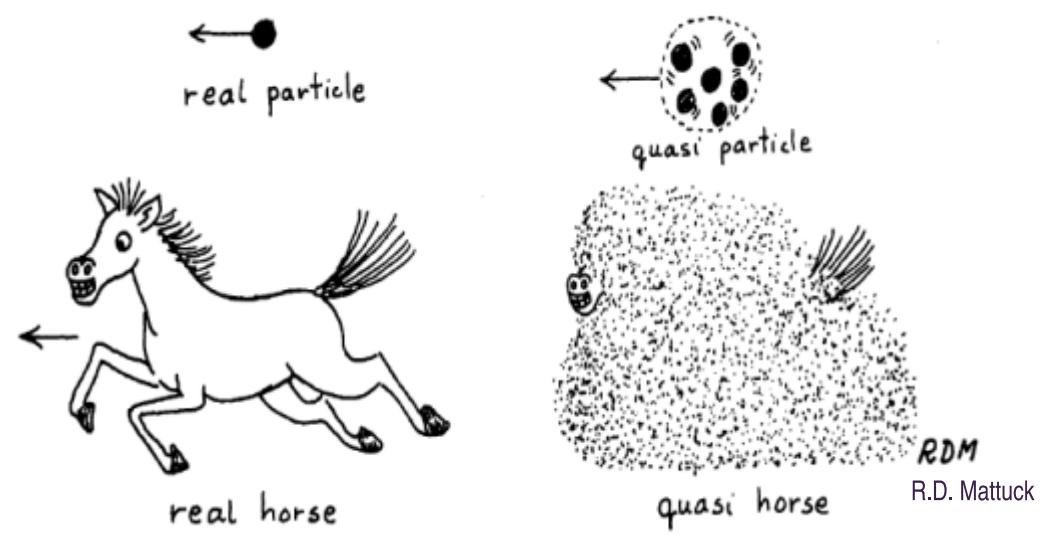

Figure 3. Cartoon representing the concept of a quasiparticle in a many-particle system.[168]

The concept of *quasiparticles* is illustrated by an analogue in Figure 3. When a horse gallops through the desert it stirs up a dust cloud that travels with it. The motion of the horse cannot be separated from the accompanying dust cloud. They act as one entity. Consider a system consisting of many interacting particles. When one particle moves it carries with it a "cloud" of other particles. This composite entity is referred to as a quasiparticle. Often it is easier to understand the system in terms of the quasiparticles rather than the individual particles. This is because the quasiparticles interact with one another via an *effective interaction* that may be much weaker than the interaction between the original constituent particles.

Like the constituent particles in the system quasiparticles have properties such as charge, mass, and spin. However, these properties for a single quasiparticle may be different from those of the individual particles, thus exhibiting the *novelty* characteristic of emergence. An example is holes in semiconductors; a system of many electrons in a crystal acts collectively to produce a hole (the absence of a single electron), a quasiparticle with the opposite charge to that of a single electron. More striking examples, such as in the fractional quantum Hall effect, are discussed below.

Formulating an effective theory in terms of quasiparticles requires profound physical insight. Landau was one of the first theoretical physicists to take this approach, introducing the idea of quasiparticles in his theories of superfluidity in $^4$He and of liquid $^3$He.

*Liquid $^3$He* is composed of $^3$He atoms that are fermions. In the liquid state they interact strongly with one another. Landau's Fermi liquid theory describes the system at temperatures below about 1 K (the degeneracy temperature) in terms of weakly interacting quasiparticles. The quasiparticles are fermions, but their effective mass is many times the mass of an isolated $^3$He atom. Experiments show that as the pressure increases from 1 bar to the melting pressure of 30 bar, the effective mass ratio increases from about 3 to 6. The quasiparticles are not particles: they have a finite lifetime and so their energy is not precisely defined, but the uncertainty in their energy is much less than their energy. For quasiparticles with the Fermi energy, the uncertainty tends to zero as the temperature approaches zero.

The effective Hamiltonian describing Landau's Fermi liquid only describes low energy excitations relative to the ground state. The energy region of validity is at most the renormalised Fermi energy, which is related to the degeneracy or "coherence" temperature. The Fermi liquid in $^3$He also exhibits a new collective excitation zero sound, a density wave predicted by Landau to exist at frequencies much larger that the quasiparticle relaxation rate.

Landau's Fermi liquid theory provides a justification for the band theory of simple metals, such as elemental metals. It justifies the Sommerfeld and Bloch models, which as a first approximation ignore any interaction between the electrons in a metal. They are surprisingly successful as the bare interaction energy due to Coulomb repulsion is comparable to the Fermi energy which is a measure of the kinetic energy of the electrons. Fermi liquid theory can be derived by a renormalisation group approach. The quasiparticles and effective interactions emerge when high energy degrees of freedom are "integrated out" in a functional integral approach.[169]

Landau's Fermi liquid theory provides an example of *adiabatic continuity*, which Anderson identified as one of the organising principles of condensed matter physics.[27] Consider a Hamiltonian in which the interaction between the fermions is proportional to a variable parameter λ. Then as λ is increased from zero to large values there is no qualitative change in the character of the excitation spectrum. The quantum numbers of charge and spin of the excitations do not change. In different words, as λ is varied there is no phase transition in the ground state of the system. This is not always the case. Below, I discuss cases where the quasiparticles (excitations) can have different quantum numbers to the constituent particles.

*Universality.* At temperatures below the degeneracy temperature $T_F$ properties of the Fermi liquid are universal functions of $T/T_F$. Fermi liquid theory can describe liquid $^3$He, electron liquids in elemental metals, and degenerate Fermi gases of ultracold atoms.

Quasiparticles are only well-defined at some energy scale, usually a scale much smaller than typical energies in the system. In Fermi liquids this is known as the coherence temperature. This can be seen by certain experimental signatures. For example, a Drude peak in the optical conductivity is only seen for temperatures less than the coherence temperature.[170]

*Superfluid $^4$He*. Landau proposed that the low-lying excitations were quasiparticles, which were density fluctuations, with an excitation spectrum that was linear for small wavevectors (phonons) and had a local minimum as a function of wavevector. He dubbed the excitations associated with the minimum as rotons. Later Feynman interpreted the roton as a vortex ring. According to Wen,[171] there is an *effective interaction* between rotons with the same distance dependence as a dipole-dipole interaction in electrostatics.
In a system with a broken symmetry associated with continuous symmetry, there are low-energy excitations with a linear dispersion relation. These are Goldstone bosons and their number and character reflect the underlying order.[i] Examples include phonons in crystals and magnons in antiferromagnets.

Quasi-particles weakly interact with one another and can be clearly seen in experiments such as inelastic neutron scattering, as shown in the Figure below.[173] On the theory side, the existence of quasi-particles is evident by the presence of well-defined peaks in spectral densities and the location of the maximum as a function of frequency and wavevector defines a dispersion relation, just as would be seen for a non-interacting particle.

---

[i] The number of Nambu-Goldstone modes (NGMs) can be lower than that of broken symmetry generators, and the dispersion of NGMs is not necessarily linear. Examples include the phonons in crystals of skyrmions, Wigner crystals in a magnetic field, and spinor BECs.[172]

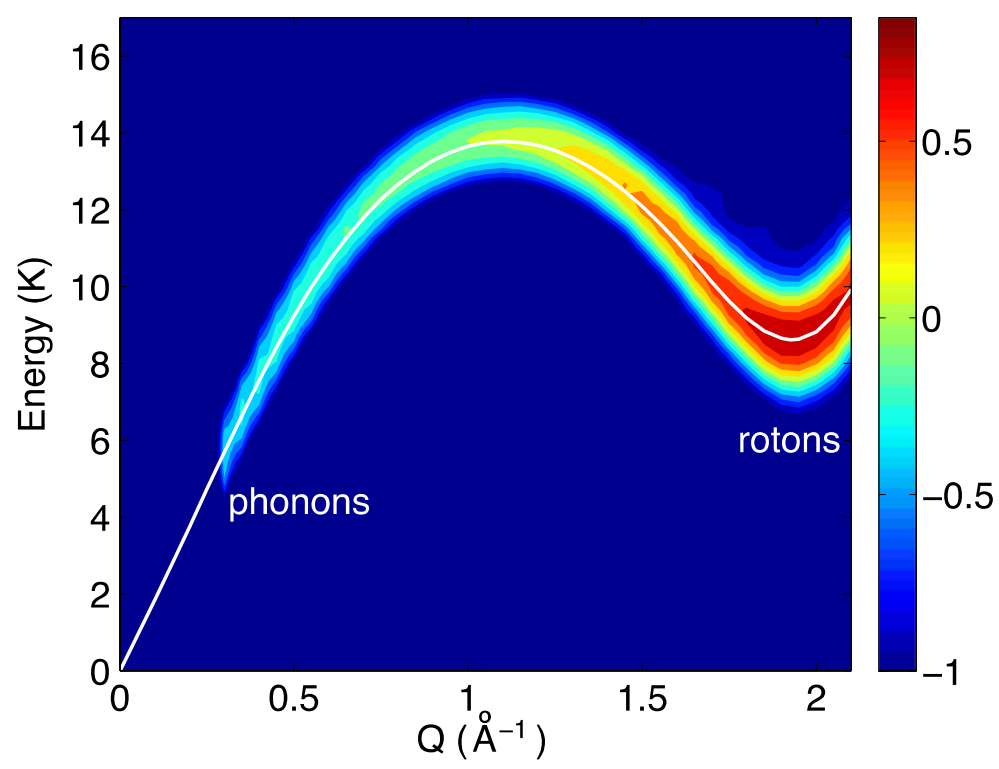

FIG. 1 (color online). The neutron scattering intensity in superfluid $^4$He at saturated vapor pressure and $T = 1.3$ K on a logarithmic scale as a function of wave vector $Q$ and energy, from Ref. [1]. The line shows the dispersion of the quasiparticles, corresponding to phonons at small wave vectors $Q$ and rotons near the local minimum at $Q_R \approx 1.93$ Å$^{-1}$ and energy $\Delta/k_B \approx 8.6$ K.

*Exotic quasiparticles*

A striking example of the novelty associated with emergence is the existence of quasiparticles that do not have the same quantum numbers or statistics as the constituent particles.[174] This is most common in systems of reduced dimensionality (i.e., one or two spatial dimensions) or with topological order. This shows a breakdown of the adiabatic continuity seen in Landau's Fermi liquid theory.

Some of these quasiparticles may seem to violate the spin-statistics theorem from quantum field theory. For example, quasiparticles with half-integer spin may be bosons. The spin-statistics theorem concerns systems of non-interacting particles with Lorentz invariance. Then particles with integer (half-integer) spin obey Bose-Einstein (Fermi-Dirac) statistics. The theorem does not apply to condensed matter systems as they are comprised of interacting particles and the quasiparticles can have dispersion relations that do not exhibit Lorentz invariance.

*Fractional Quantum Hall Effect (FQHE) state*s. They occur in systems that are composed of electrons (holes) which have charge -e (+e), obey Fermi-Dirac statistics, and are confined to move in only two dimensions in very high magnetic fields. However, the charge of the quasiparticles can be a fraction of the charge on a single electron, and they do not act as fermions, but have fractional statistics, and are called anyons. An alternative representation of the quasiparticles are as composite fermions.[175] They are a fermion to which is attached an integer number of magnetic flux quanta. In this picture FQHE states with fraction 1/(2n+1) can be viewed as the n-th integer quantum Hall state.

*Quantum spin chains.* These one-dimensional systems of localised spins that couple antiferromagnetically to their nearest neighbours are discussed in more detail below. For spin-1/2, there is no long-range order, and the elementary excitations are spinons, which are spin-1/2 excitations that obey semion statistics (i.e., exchange of two particles produces a phase change of $\pi/2$). This contrasts with three-dimensional antiferromagnets which exhibit long-range Neel order, and the quasiparticles are magnons: bosons with spin-1.

*Luttinger liquids.* For a system of interacting spinless fermions in one dimension the quasiparticles in the metallic state are bosons that can be identified with particle-hole excitations with a linear dispersion relation. For interacting spinful fermion systems, in the metallic state, spin-charge separation occurs. There are two types of bosonic excitations that can move independently of one another and with different velocities. One carries charge and the other spin. It is like the electron (which has charge and spin) has split in two!

*Majorana fermions.* In quantum field theory Majorana proposed in 1937 a fermion that could be its own anti-particle, but such an elementary particle has never been observed. *Toy models* have been key to proposals that Majorana's vision might be realised in condensed matter. Kitaev studied a one-dimensional lattice model for a topological superconductor and showed that the boundary states were Majorana fermions.[176] It was then suggested this might be realised in hybrid semiconductor-superconductor quantum wire structures. Several experimental groups subsequently claimed to have observed signatures of Majorana's. However, these claims are controversial since there are alternative more mundane explanations for the experimental observations and the cases reported may suffer from selection bias.[177] Kitaev also studied a toy spin model on the honeycomb lattice with highly anisotropic Heisenberg interactions and showed that it was exactly soluble with a spin liquid ground state and Majorana fermion quasiparticles.[178] In spite of extensive theoretical and experimental work it is an outstanding question as to whether Kitaev's model is relevant any real material.[179]

*Parafermions.* These occur in conformal field theory (CFT) description of the two-dimensional three-state ferromagnetic Potts model.[180] The parafermions are the $Z_3$ analogue of the chiral fermions that occur at the critical point of the transverse field Ising chain. Possible experimental realisations have been discussed by Alicea and Fendley.[181]

*A quasiparticle conjecture.* Over time it has been found that for almost every model Hamiltonian with physically realistic interactions there is a way of rewriting the system (performing a unitary transformation) so it can be viewed as a system of weakly interacting quasiparticles, even when the bare interactions are strong. This raises the question as to whether it is possible to prove some general (existence) theorem along such lines.

### 8.3 Superconductivity

Superconductivity is a poster child for emergence.

*Novelty.* Distinct properties of the superconducting state include zero resistivity, the Meissner effect, and the Josephson effect. The normal metallic state does not exhibit these properties. At low temperatures, solid tin exhibits the property of superconductivity. However, a single atom of tin is not a superconductor. A small number of tin atoms has an energy gap due to pairing interactions, but not bulk superconductivity.

There is more than one superconducting state of matter. The order parameter may have the same symmetry as a non-trivial representation of the crystal symmetry, and it can have spin singlet or triplet symmetry. For example, cuprate superconductors have an order parameter with d-wave symmetry. Type II superconductors in a magnetic field have a Abrikosov vortex lattice, another distinct state of matter.

*Unpredictability.* Even though the underlying laws describing the interactions between electrons in a crystal have been known for one hundred years, the discovery of superconductivity in many specific materials was not predicted. Even after the BCS theory was worked out in 1957, the discovery of superconductivity in intermetallic compounds, cuprates, organic charge transfer salts, fullerenes, and heavy fermion compounds was not predicted. Even with advances in electronic structure theory for weakly correlated electon systems prediction of new superconducting materials is rare.[182]

*Order and structure.* In the superconducting state the electrons become ordered in a particular way. The motion of the electrons relative to one another is not independent but correlated. Long range order is reflected in the generalised rigidity which is responsible for the zero resistivity. Properties of individual atoms (e.g., NMR chemical shifts) are different in vacuum, in the metallic state, and superconducting state.

*Universality.* Properties of superconductivity such as zero electrical resistance, the expulsion of magnetic fields, quantisation of magnetic flux, and the Josephson effects are universal. The existence and description of these properties is independent of the chemical and structural details of the material in which the superconductivity is observed. This is why Ginzburg-Landau theory works so well. In BCS theory, the temperature dependences of thermodynamic and transport properties are given by universal functions of $T/T_c$ where $T_c$ is the transition temperature. Experimental data is consistent with this for a wide range of superconducting materials, particularly elemental metals for which the electron-phonon coupling is weak.

*Modularity at the mesoscale.* Emergent entities include Cooper pairs and vortices. There are two associated emergent length scales, typically much larger than the microscopic scales defined by the interatomic spacing or the Fermi wavelength of electrons. The first scale is the coherence length; it is associated with the energy cost of spatial variations in the order parameter. It defines the extent of the proximity effect where the surface of a non-superconducting metal can become superconducting when it is in electrical contact with a superconductor. The coherence length turns out to be of the order of the size of the Cooper pairs in BCS theory. The second length scale is the magnetic penetration depth (also known as the London length) that determines the extent that an external magnetic field can penetrate the surface of a superconductor. It is determined by the superfluid density. The relative size of the coherence length and the penetration depth determines whether an Abrikosov vortex lattice is stable in a large magnetic field.

*Quasiparticles.* The elementary excitations are Bogoliubov quasiparticles. They are qualitatively different to the particle and hole excitations in a normal metal. They are a coherent superposition of a particle and hole excitation (relative to the Fermi sea), have zero charge and only exist above the energy gap. The mixed particle-hole character of the quasiparticles is reflected in the phenomenon of Andreev reflection.

*Singularities.* Superconductivity is a non-perturbative phenomenon. In BCS theory the transition temperature, $T_c$, and the excitation energy gap are a non-analytic function of the electron-phonon coupling constant $\lambda$.

$$T_c \sim exp(-1/\lambda)$$

A singular structure is also evident in the properties of the current-current correlation function. Interchange of the limits of zero wavevector and zero frequency do not commute; this is intimately connected with the non-zero superfluid density.[183]

*Effective theories.* These are illustrated in the Figure below. The many-particle Schrodinger equation describes electrons and atomic nuclei interacting with one another. Many-body theory can be used to justify considering the electrons as a jellium liquid of non-interacting fermions interacting with phonons. Bardeen, Pines, and Frohlich showed that for that system there is an effective interaction between fermions that is attractive. The BCS theory includes a truncated version of this attractive interaction. Gorkov showed that Ginzburg-Landau theory could be derived from BCS theory. The London equations can be derived from Ginzburg-Landau theory. The Josephson equations only include the phase of order parameter to describe a pair of coupled superconductors. The full quantum version of the Josephson equations, such as those that describe the Cooper pair boxes and transmons[184] used in quantum computing, can be derived from BCS theory using a functional integral approach.[185]

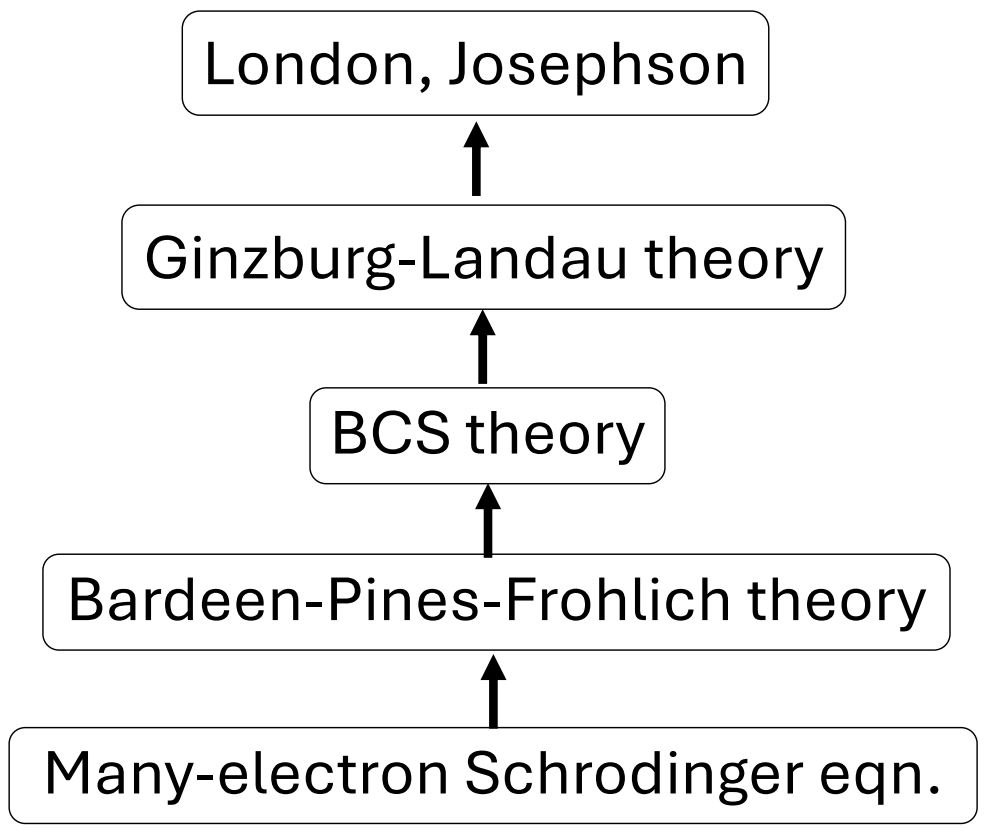

Figure 4: A hierarchy of effective theories for superconductivity.

The historical development of theories mostly went in the downwards direction in the Figure. London preceded Ginzburg-Landau which preceded BCS theory. Today for specific materials where superconductivity is known to be due to electron-phonon coupling and the electron liquid is weakly correlated one can now work upwards using computational methods such as Density Functional Theory (DFT) for Superconductors or the Eliashberg theory with input parameters calculated from DFT-based methods. However, this has had debatable success.[186] The superconducting transition temperatures calculated typically vary with the approximations used in the DFT, such as the choice of functional and basis set, and often differ from experimental results by the order of 50 percent. This illustrates how hard prediction is for emergent phenomena.

*Potential and pitfalls of mean-field theory.* Mean-field approximations and theories can provide a useful guide as what emergent properties are possible and as a starting point to map out properties such as phase diagrams. For some systems and properties, they work incredibly well and for others they fail spectacularly and are misleading. Ginzburg-Landau theory and BCS theory are both mean-field theories. For three-dimensional superconductors they work extremely well. However, in two dimensions as long-range order and breaking of a continuous symmetry cannot occur and the physics associated with the Berezinskii-Kosterlitz-Thouless transition occurs. Nevertheless, the Ginzburg-Landau theory provides the background to understand the justification for the XY model and the presence of vortices to proceed. Similarly, the BCS theory fails for strongly correlated electron systems, but a version of the BCS theory does give a surprisingly good description of the superconducting state.

*Cross-fertilisation of fields.* Cooper pairing also occurs in uncharged superfluids such as occur in liquid $^3$He and neutron stars. Concepts and methods developed for the theory of superconductivity bore fruit in other sub-fields of physics including nuclear physics, elementary particles, and astrophysics. Considering the matter fields (associated with the electrons) coupled to electromagnetic fields (a U(1) gauge theory) the matter fields can be integrated out to give a theory in which the photon has mass. This is a different perspective on the Meissner effect in which the magnitude of an external magnetic field decays exponentially as it penetrates a superconductor. This idea of a massless gauge field acquiring a mass due to spontaneous symmetry breaking was central to steps towards the Standard Model made by Nambu[187] and by Weinberg.

### 8.4 Kondo effect

The system consists of a single magnetic impurity in a metal with an antiferromagnetic coupling between the spin of the impurity and all the spins of the electrons in the metal. In real materials the system is a dilute solution of magnetic atoms in a non-magnetic metal.

*Novelty.*
The ground state is qualitatively different from the high temperature (disordered) state or if the magnetic interaction is not present. The ground state is a spin singlet and has no net magnetic moment. In contrast, the disordered state of a Kondo system is degenerate and has a net magnetic moment, seen by a Curie magnetic susceptibility at high temperatures.

*Emergent energy scale.*
The Kondo temperature $T_K$ is much smaller than the "bare" energies in the system: the antiferromagnetic exchange coupling J and the Fermi energy of the metal. Experimentally, the scale of $T_K$ is evident from the temperature at which there is a minimum in the resistivity and the magnetic susceptibility deviates from Curie-like behaviour. In multi-channel problems there can be a hierarchy of energy scales, corresponding to separate screening of orbital and spin degrees of freedom.

*Singularities.*
The Kondo effect is non-perturbative. A perturbation theory in powers of the interaction J gives expressions for physical quantities such as the resistivity that diverge as the temperature goes to zero. This is inconsistent with experiment which show physical quantities are finite and vary smoothly as the temperature decreases. This was known as the "Kondo problem". The non-perturbative behaviour is also seen in that the calculated Kondo temperature as a function of J has an essential singularity as a function of J. The system exhibits asymptotic freedom (borrowing terminology from QCD) in that as the energy scale decreases the effective coupling increases, becoming infinite at zero energy.

*Discontinuity*
There is no phase transition, i.e., all physical quantities are smooth functions of temperature. This shows that novelty does not necessarily imply discontinuity.

*Effective theories*
The Kondo model Hamiltonian can be derived from the Anderson impurity model that allows charge fluctuations on the impurity atom. The Schrieffer-Wolff transformation can be used to integrate out charge degrees of freedom in the Anderson model and obtain the Kondo model as an effective theory. In the continuum limit the Kondo model can be mapped to boundary conformal theory. This approach has been particularly fruitful for investigating multi-channel Kondo problems (where there are orbital degrees of freedom) some of which have a non-Fermi liquid ground state.[188] In some of these models there are hierarchies of multiple energy scales.

*Universality*

The temperature (T) and magnetic field (B) dependence of the contribution of the magnetic impurity to thermodynamic and transport properties are universal functions of $T/T_K$ and $B/T_K$. Experimental data for a wide range of impurity atoms and metals with diverse values for J and band structure parameters collapse onto single theoretical curves. The Sommerfeld-Wilson ratio of the magnetic susceptibility to specific heat coefficient has the universal value 2, compared to the value of 1 for Sommerfeld and Bloch models for weakly correlated Fermi liquids. The Kondo effect has also been observed in artificial systems including semiconductor quantum dots, and magnetic atoms and molecules on metallic surfaces. The experimental data for the T and B dependence of transport properties collapse onto the same universal scaling functions as for magnetic impurities in metals.

## 8.5 Quantum spin chains

*Novelty*

With antiferromagnetic nearest-neighbour interactions the spin-1 chain has properties qualitatively different from a three-dimensional system. The ground state has no long-range order and there is a finite energy gap to the lowest energy spin triplet excitation. There is topological order and there are edge excitations with spin-1/2. The latter is an example of fractionalisation of the quantum numbers of the quasi-particles as the system is composed of constituents with spin 1.

Chains of non-integer spins can be described by an effective theory which is the non-linear sigma model including a topological term in the action. This term changes the physics significantly from integer spin systems. There is no symmetry breaking and quasi-long-range order: spin correlations exhibit power law decay. There are gapless spin-1/2 excitations in the bulk and these are semions.

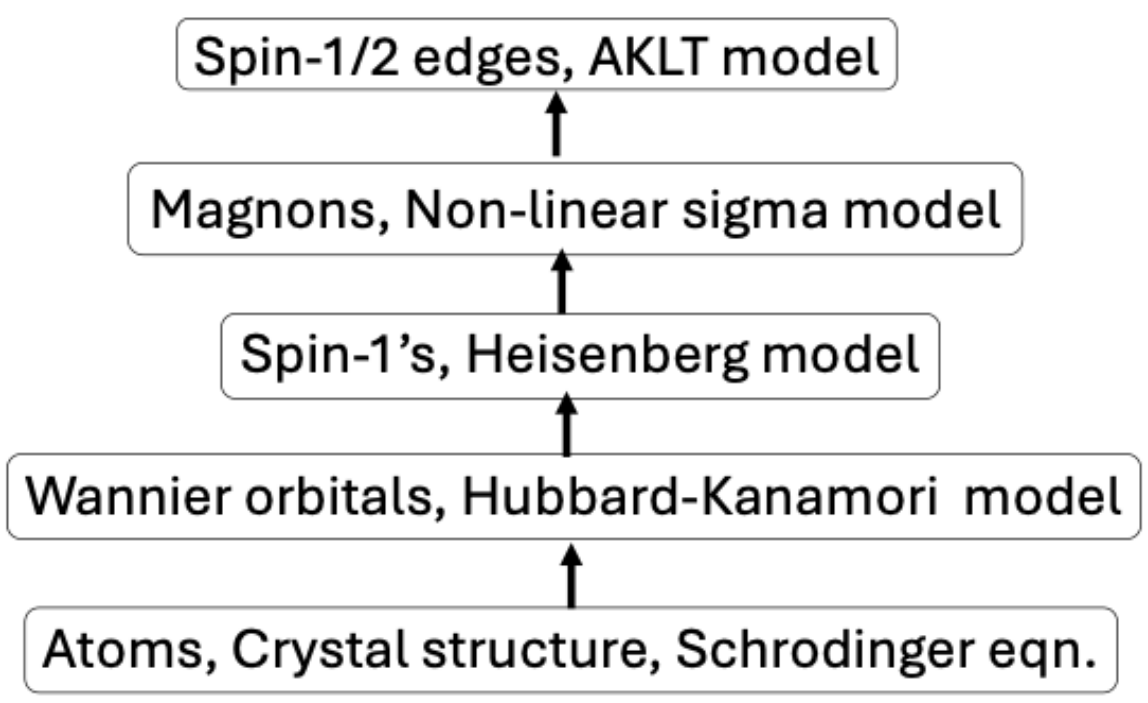

Figure 5. Hierarchy of objects and effective theories associated with a chain of atoms with a localised spin-1.

*Toy model*

The AKLT model is exactly soluble and captures key phenomena seen in spin-1 chains: the energy gap, topological order, short-range spin correlations, and spin-1/2 edge excitations.

### 8.6 Topological insulators

*Novelty*
A topological insulator is a distinct state of electronic matter with no spontaneous symmetry breaking. It is a time reversal invariant and has an electronic band gap in the bulk. Transport of charge and spin occurs by gapless edge states. This state of matter is associated with a novel topological invariant, which distinguishes it from an ordinary insulator. This invariant is analogous to the Chern number used to classify states in the integer quantum Hall effect.[189]

*Toy models*
Haldane highlighted in his Nobel Lecture in 2016, that toy models played a central role in the discovery of topological insulators in real materials. In 1988, Haldane published "Model for a Quantum Hall Effect without Landau Levels: Condensed-Matter Realization of the "Parity Anomaly"".[190] The model was a tight-binding model for spinless fermions on a honeycomb net with alternating magnetic flux through plaquettes. He noted a phase transition as the magnetic flux was varied. Haldane concluded his paper with the comment, "the particular model presented here is unlikely to be directly physically realizable." Haldane's paper received little attention until 2005. Following the fabrication of graphene, Kane and Mele, published a paper, "Spin quantum Hall effect in graphene,"[102] suggesting how the state proposed by Haldane could occur. However, their estimate of the spin-orbit coupling coupling needed was several orders of magnitude larger than reality. Nevertheless, the two-dimensional topological insulator state was subsequently proposed by Bernevig, Hughes, and Zhang[192] to exist in quantum wells of mercury telluride sandwiched between cadmium telluride, and then observed.[193]

### 8.7 Quantised properties and metrology

There are two situations associated with states of matter being referred to as "quantum." One situation is when properties are determined by the quantum statistics of particles, i.e., the system is composed of particles that obey Fermi-Dirac or Bose-Einstein statistics. The second situation is when a macroscopic property is quantized with values determined by Planck's constant. I now discuss each of these with respect to emergence.

*A. Quantum statistics*
For a system of non-interacting[j] fermions and bosons at high temperatures the properties of the system are those of a classical ideal gas. As the temperature decreases there is a smooth crossover to low-temperature properties that are qualitatively different for fermions, bosons, and classical particles. This crossover occurs around a temperature, known as the degeneracy temperature, that is dependent on the particle density and Planck's constant.

---

[j] Similar behaviour occurs for weakly interacting particles. Non-interacting particles obeying quantum statistics can be viewed as interacting in the sense that the statistics provides a constraint concerning two particles being in the same quantum state (energy level) and so could be viewed as an attractive interaction for bosons and an infinitely repulsive interaction for fermions.

Many of the properties resulting from quantum statistics also occur in systems of strongly interacting particles and this is central to the concept of Landau's Fermi liquid and viewing liquid $^4He$ as a boson liquid. If liquid $^3He$ and the electron liquid in elemental metals are viewed as a gas of non-interacting fermions, the degeneracy temperature is about 1 K and 1000 K, respectively. Thermodynamic properties are qualitatively different above and below the degeneracy temperature. Low-temperature properties can have values that differ by orders of magnitude from classical values and have a different temperature dependence. In contrast to a classical ideal gas, a fermion gas has a non-zero pressure at zero temperature and its magnitude is determined by Planck's constant. This degeneracy pressure is responsible for the gravitational stability of white dwarf and neutron stars.

These properties of systems of particles can be viewed as emergent properties, in the sense of novelty, as they are qualitatively different from high-temperature properties. However, they involve a crossover as a function of temperature and so are not associated with discontinuity. They also are not associated with unpredictability as they are straight-forward to calculate from a knowledge of microscopic properties.

*B. Quantised macroscopic properties*

These provide a more dramatic illustration of emergence. Here I consider four specific systems: superconducting cylinders, rotating superfluids, Josephson junctions, and the integer Quantum Hall effect. All these systems have a macroscopic property that is observed to have the following features.

As an external parameter is varied the quantity defining a macroscopic property varies in a step-like manner with discrete values on the steps. This contrasts to the smooth linear variation seen when the material is not condensed into the quantum state of matter.
The value on the steps is an integer multiple of some specific parameter.
This parameter (unit of quantisation) only depends on Planck's constant $h$ and other fundamental constants.
The unit of quantisation does not depend on details of the material, such as chemical composition, or details of the device, such as its geometrical dimensions.
The quantisation has been observed in diverse materials and devices.
Explanation of the quantisation involves topology.

*Superconducting cylinders.* A hollow cylinder of a metal is placed in a magnetic field parallel to the axis of the cylinder. In the metallic state the magnetic flux enclosed by the cylinder increases linearly with the magnitude of the external magnetic field. In the superconducting state, the flux is quantized in units of the magnetic flux quantum, $\Phi_0 = h/2e$ where $e$ is the charge on an electron. It is also found that in a type II superconductor the vortices that occur in the presence of an external magnetic field enclose a magnetic flux equal to $\Phi_0$.

*Rotating superfluids.* When a cylinder containing a normal fluid is rotated about an axis passing down the centre of the cylinder the fluid rotates with a circulation proportional to the speed of rotation and the diameter of the cylinder. In contrast, in a superfluid, as the speed of rotation is varied the circulation is quantised in units of $h/M$ where $M$ is the mass of one atom in the fluid. This quantity is also the circulation around a single vortex in the superfluid.

*Josephson junctions.* In the metallic state the current passing through a junction increases linearly with the voltage applied across the junction. In the superconducting state the AC Josephson effect occurs. If a beam of microwaves of constant frequency is incident on the junction, jumps occur in the current when the voltage is an integer multiple of $h/2e$. The quantisation is observed to better than one part in a million (ppm).

*Integer Quantum Hall effect.* In a normal conductor the Hall resistance increases linearly with the external magnetic field for small magnetic fields. In contrast, in a two-dimensional conductor at high magnetic fields the Hall resistance is quantized in units of $h/2e^2$. The quantisation is observed to better than one part in ten million. Reflecting universality, the observed value of the Hall resistance for each of the plateaus is independent of many details, including the temperature, the amount of disorder in the material, the chemical composition of system (silicon versus gallium arsenide), or whether the charge carriers are electrons or holes.

Macroscopic quantum effects are also seen in SQUIDs (Superconducting Quantum Interference Devices). They exhibit quantum interference phenomena analogous to the double-slit experiment. The electrical current passing through the SQUID has a periodicity defined by the ratio of the magnetic flux inside the current loop of the SQUID and the quantum of magnetic flux.

The precision of the quantisation provides a means to accurately determine fundamental constants. Indeed, the title of the paper announcing the discovery of the integer quantum Hall effect was, "New Method for High-Accuracy Determination of the Fine-Structure Constant Based on Quantized Hall Resistance."[194] It is astonishing that a macroscopic measurement of a property of a macroscopic system, such as the electrical Hall resistance, can determine fundamental constants that are normally associated with the microscale and properties of atomic systems.

Laughlin and Pines claimed that the quantisation phenomena described above reflect organizing principles associated with emergent phenomena, and their universality supports their claim of the unpredictability of emergent properties.[67],[k]

<u>*Quantum states of matter and metrology*</u>

---

[k] They stated "These things are clearly true, yet they cannot be deduced by direct calculation from the Theory of Everything, for exact results cannot be predicted by approximate calculations. This point is still not understood by many professional physicists, who find it easier to believe that a deductive link exists and has only to be discovered than to face the truth that there is no link. But it is true nonetheless. Experiments of this kind work because there are higher organizing principles in nature that make them work. The Josephson quantum is exact because of the principle of continuous symmetry breaking (16). The quantum Hall effect is exact because of localization (17). Neither of these things can be deduced from microscopics, and both are transcendent, in that they would continue to be true and to lead to exact results even if the Theory of Everything were changed. Thus the existence of these effects is profoundly important, for it shows us that for at least some fundamental things in nature the Theory of Everything is irrelevant. P.W. Anderson's famous and apt description of this state of affairs is ''more is different'' (2). The emergent physical phenomena regulated by higher organizing principles have a property, namely their insensitivity to microscopics, that is directly relevant to the broad question of what is knowable in the deepest sense of the term."

The universality of these macroscopic quantum effects has practical applications in metrology, the study of measurement and the associated units and standards. In 1990 new international standards were defined for the units of voltage and electrical resistance, based on the quantum Hall effect and the AC Josephson effect, respectively.

Prior to 1990 the standard used to define one volt was based on a particular type of electrical battery, known as a Weston cell. The new standard using the AC Josephson effect allowed voltages to be defined with a precision of better than one part per billion. This change was motivated not only by improved precision, but also improved portability, reproducibility, and flexibility. The old voltage standard involved a specific material and device and required making duplicate copies of the standard Weston cell. In contrast, the Josephson voltage standard is independent of the specific materials used and the details of the device.

Prior to 1990 the international standard for the ohm was defined by the electrical resistance of a column of liquid mercury with constant cross-sectional area, 106.3 cm long, a mass of 14.4521 grams and a temperature 0 °C. Like the Josephson voltage standard, the quantum Hall resistance standard has the advantage of precision, portability, reliability, reproducibility, and independence of platform. The independence of the new voltage and resistance standards from the platform used reflects the fact that the Josephson and quantum Hall effects have the universality characteristic of emergent phenomena.

## 9. Classical condensed matter

### 9.1 Continuous phase transitions in two dimensions

In two dimensions the phase transition that occurs for superfluids, superconductors, and planar classical magnets is qualitatively different from that which occurs in higher dimensions. Known as the Berezinskii-Kosterlitz-Thouless transition, it involves several unique emergent phenomena.

*Novelty*
The low-temperature state does not exhibit long-range-order or spontaneous symmetry breaking. Instead, the order parameter has power law correlations, below a temperature $T_{BKT}$. Hence, it is qualitatively different from the high-temperature disordered state in which the correlations decay exponentially. It is a distinct state of matter,[l] with properties that are intermediate between the low- and high-temperature states normally associated with phase transitions. The power law correlations in the low-temperature state are similar to those at a conventional critical point, which decay in powers of the critical exponent $\eta$. However, here the phase diagram can be viewed as having a line of critical points, consisting of all the temperatures below $T_{BKT}$. Along this line the critical exponent $\eta$ varies continuously with a value that depends on interaction strength. In contrast, at conventional critical points $\eta$ has a fixed value determined by the universality class. For example, for the Ising model in two dimensions $\eta=1/4$, independent of the interaction strength.

[l] Sometimes it is stated that the low-temperature state has topological order, but I am not really sure what that means.

At the critical point the U(1) symmetry of the model is extended to an emergent SU(2) symmetry.

The mechanism of the BKT phase transition is qualitatively different from that for conventional phase transitions. It is driven by the unbinding of vortex and anti-vortex pairs by thermal fluctuations. In contrast, conventional phase transitions are driven by thermal fluctuations in the magnitude of the order parameter.

*Discontinuity*
There is a discontinuity in the stiffness of the order parameter at the transition temperature, $T_{BKT}$. The specific heat capacity is a continuous function of temperature, in contrast to the singularity that occurs for conventional phase transitions.

*Toy model*
A classical Heisenberg model for a planar spin, also known as the XY model, on a two-dimensional square lattice captures the essential physics.

*Modularity at the mesoscale*
The quasiparticles of the system that are relevant to understanding the BKT phase transition and the low-temperature state are not magnons (for magnets) or phonons (for superfluids), but vortices, i.e., topological defects. These entities usually occur on the mesoscale. The relevant effective theory is not a non-linear sigma model. Thermal excitation of vortex-antivortex pairs determine the temperature dependence of physical properties and the transition at $T_{BKT}$. There is an *effective interaction* between a vortex and anti-vortex that is attractive and a logarithmic function of their spatial separation, analogous to a two-dimensional Coulomb gas.

*Universality*
The BKT transition occurs in diverse range of two-dimensional models and materials including superfluids, superconductors, ferromagnets, arrays of Josephson junctions, and the Coulomb gas. The ratio of the discontinuity in the order parameter stiffness at $T_{BKT}$, to $T_{BKT}$ has a universal value, independent of the coupling strength.

The renormalisation group (RG) equations associated with the transition are the same as those for a multitude of other systems, both classical and quantum. The classical two-dimensional systems include the Coulomb gas, Villain model, $Z_n$ model for large n, solid-on-solid model, eight vertex model, interacting dimer model, and the Ashkin-Teller model. The RG equations also describe the classical Ising chain with $1/r^2$ interactions. Quantum models with the same RG equations include the anisotropic Kondo model, spin boson model, XXZ antiferromagnetic Heisenberg spin chain, and the sine-Gordon quantum field theory in 1+1 dimensions. In other words, all these models are in the same universality class.

*Singularity*
The correlation length of the order parameter is a non-analytic function of the temperature.

$$\xi(T) \sim exp(-1/|T - T_{BKT}|)$$

This essential singularity is related to the non-perturbative nature of the corresponding quantum models.

*Two-dimensional crystals*

Related physics is relevant to the solidification of two-dimensional liquids, and the associated theory was developed by Halperin, Nelson, and Young. The relevant toy model is not the classical XY model as it is necessary to include the effect of the discrete rotational symmetry of the lattice of the solid. The low-temperature state exhibits discrete rotational, but not spatial, symmetry breaking, with power law spatial correlations. This state does not directly melt into a liquid, but into a distinct state of matter, the hexatic phase. It has short-range spatial order and quasi-long-range orientational (sixfold) order. The phase transitions are driven by topological defects, disclinations and dislocations.

*Predictability*
The BKT transition, the quasi-ordered low temperature state, and the hexatic phase were all predicted theoretically before they were observed experimentally. This is unusual for emergent phenomena but shows that unpredictability is not equivalent to novelty.

### 9.2 Spin ices

Spin ices are magnetic materials in which geometrically frustrated magnetic interactions between the spins prevent long-range magnetic order and lead to a residual entropy similar to that in ice (solid water). Spin ices provide a beautiful example of many aspects of emergence, including how surprising new entities can emerge at the mesoscale.

*Novelty*
Spin ices are composed of individual spins on a lattice. The system exhibits properties that the individual spins and the high-temperature state do not have. Spin ices exhibit a novel state of matter, dubbed the magnetic Coulomb phase. There is no long-range spin order, but there are power-law (dipolar) correlations that fall off as the inverse cube of distance. Novel entities include spin defects reminiscent of magnetic monopoles and Dirac strings. The novel properties can be understood in terms of an emergent gauge field. The gauge theory predicts that the spin correlation function (in momentum space) has a particular singular form exhibiting pinch points [also known as bow ties], which are seen in spin-polarised neutron scattering experiments.[195], [196]

*Toy models*
Classical models such as the Ising or Heisenberg models with antiferromagnetic nearest-neighbour interactions on the pyrochlore lattice exhibit the emergent physics associated with spin ices: absence of long-range order, residual entropy, ice type rules for local order, and long-range dipolar spin correlations exhibiting pinch points. These toy models can be used to derive the gauge theories that describe emergent properties such as monopoles and Dirac strings.

Actual materials that exhibit spin ice physics such as dysprosium titanate ($Dy_2Ti_2O_7$) and holmium titanate ($Ho_2Ti_2O_7$) are more complicated than these toy models. They involve quantum spins, ferromagnetic interactions, spin-orbit coupling, crystal fields, complex crystal structure and dipolar magnetic interactions. Henley states[197] that these materials "are well approximated as having nothing but (long-ranged) dipolar spin interactions, rather than nearest-neighbour ones. Although this model is clearly related to the “Coulomb phase,” I feel it is largely an independent paradigm with its own concepts that are different from the (entropic) Coulomb phase..." In different words, the classical Heisenberg models are toy models that illustrate what is possible but should not be viewed as effective theories for the actual materials exhibiting spin ice physics.

*Effective theory*
Gauge fields described by equations analogous to electrostatics and magnetostatics in Maxwell's theory electromagnetism are emergent in coarse-grained descriptions of spin ices. I briefly describe how this arises.

Consider an antiferromagnetic Heisenberg model on a bipartite lattice where on each lattice site there is a tetrahedron. The "ice rules" require that two spins on each tetrahedron point in and two out. Define a field **L**(i) on each lattice site i which is the sum of all the spins on the tetrahedron. The magnetic field **B**(**r**) is a coarse graining of the field **L**(i). The ice rules and local conservation of flux require that

$$\nabla \cdot B = 0$$

The ground state of this model is infinitely degenerate with a residual entropy $S_0$. The emergent "magnetic" field [which it should be stressed is not a physical magnetic field] allows the presence of monopoles [magnetic charges]. These correspond to defects that do not satisfy the local ice rules in the spin system. It is argued that the total free energy of the system is

$$\frac{F(\{\boldsymbol{B}(\boldsymbol{r})\})}{T} = S_0 + \frac{K}{2} \int d^3\boldsymbol{r} |\boldsymbol{B}(\boldsymbol{r})|^2$$

K is the "stiffness" or "magnetic permeability" associated with the gauge field. It is entirely of entropic origin, just like the elasticity of rubber.[m]

A local spin flip produces a pair of oppositely charged monopoles. The monopoles are deconfined in that they can move freely through the lattice. They are joined together by a Dirac string. This contrasts with real magnetism where there are no magnetic charges, only magnetic dipoles; one can view magnetic charges as confined within dipoles.

There are only short-range (nearest neighbour) direct interactions between the spins. However, these act together to produce a long-range interaction between the monopoles (which are deviations from local spin order). This *effective interaction* between the two monopoles [charges] has the same form as Coulomb's law.

$$V(\vec{r_1}, \vec{r_2}) = \frac{1}{K|\vec{r_1} - \vec{r_2}|}$$

This is why this state of matter is called the Coulomb phase.

*Universality*
The novel properties of spin ice occur for both quantum and classical systems, Ising and Heisenberg spins, and for a range of frustrated lattices. Similar physics occurs with water ice, magnetism, and charge order.

---

[m] Aside: I am curious to see a calculation of K from a microscopic model and an estimate from experiment. Henley points out that in water ice the entropic elasticity makes a contribution to the dielectric constant and this "has been long known." The references given[198,199] calculate the physical dielectric constant and the entropic contribution to K is not clear.

*Modularity at the mesoscale*
The system can be understood as a set of weakly interacting modular units at different scales. These include the tetrahedra of spins, the magnetic monopoles, and the Dirac strings. However, it should be noted that Henley states that Dirac strings are "a nebulous and not very helpful notion when applied to the Coulomb phase proper (with its smallish polarisation), for the string's path is not well defined... It is only in an ordered phase... that the Dirac string has a clear meaning."[197]

The measured temperature dependence of the specific heat of $Dy_2Ti_2O_7$ is consistent with that calculated from Debye-Hückel theory for deconfined charges interacting by Coulomb's law.[195]

*Unpredictability*
Most new states of matter are not predicted theoretically. They are discovered by experimentalists, often by serendipity. Spin ice and the magnetic Coulomb phase is an exception.

*Sexy magnetic monopoles or boring old electrical charges?*
In the discussion above the "magnetic field" B(r) might be replaced with an "electric field" E(r). Then the spin defects are just analogous to electrical charges and the "Dirac strings" become like a polymer chain with opposite electrical charges at its two ends. This is not as sexy. On the other hand, it can be argued that the emergent gauge field is "magnetic". It describes spin defects and these are associated with a local magnetic moment. Furthermore, the long-range dipolar correlations (with associated pinch points) of the gauge field are detected by magnetic neutron scattering.

*Emergent gauge fields in quantum many-body systems?*
The analogy of spin ices with classical magnetostatics raises the hope that there may be condensed matter systems that exhibit emergent classical electromagnetism including emergent photons.[200] Recently, Smith et al., [201] argued that quantum spin ices, including $Ce_2Zr_2O_7$, may have emergent "photons" and spinons.

The existence of emergent gauge fields in quantum states of matter has been investigated extensively by Xiao-Gang Wen. He has shown how certain mean-field treatments of frustrated antiferromagnetic quantum Heisenberg models (with quantum spin liquid ground states) and doped Mott insulators (such as high $T_c$ cuprate superconductors) lead to emergent gauge fields. As fascinating as this is, there is no definitive evidence for these emergent gauge fields. They just provide appealing theoretical descriptions. This contrasts with the emergent gauge fields for spin ice, where the pinch points in the neutron scattering spectrum are viewed as a "smoking gun." Based on his success at constructing these emergent gauge fields Wen has promoted the provocative idea that the gauge fields and fermions that are considered fundamental in the standard model of particle physics may be emergent entities.[171,202]

In models for the metallic state of the cuprate superconductors, Sachdev has argued that there are emergent gauge fields.[203] These are required in a 'fractionalized Fermi liquid' state. Although it has electron-like quasi-particles near a Fermi surface, the Luttinger theorem for the volume of the Fermi surface is violated.

I mention two more definitive examples of an emergent gauge field in condensed matter. One is that associated with the Berry curvature that causes the anomalous Hall effect in metallic ferromagnets. The gauge potential is defined in terms of electronic Bloch states and the Berry curvature is only non-zero in when time-reversal symmetry is broken.[204]

The second example concerns systems with quasiparticles that are Weyl fermions, i.e., gapless particles with a definite chirality. In the A phase of superfluid $^3$He, there are topologically stable nodes on the Fermi surface at the points defined by the vector parallel to the direction of the orbital angular momentum of the order parameter. Analogous to electromagnetism the vector acts as a gauge potential and its spatial or time-dependent variations of the order parameter (such as occur in vortices) define "magnetic" and "electric" fields leading to a chiral anomaly.[205]

### 9.3 Soft matter

Soft materials and states of matter that are "squishy" include polymers, membranes, foams, colloids, gels, and liquid crystals. Soft matter is ubiquitous in biological materials. This field has provided new insights into the physics of biological systems and of the science and technology of food. Seminal contributions made by Pierre de Gennes were recognized with a Nobel Prize in 1991. He began his career working on superconductivity and went on to develop a unified framework to understand soft matter, introducing ideas from condensed matter physics such as effective theories, order parameters, scaling, renormalisation, and universality.

McLeish identified six characteristics of soft matter.[206] I list them as they relate to characteristics of emergence.

1. Thermal motion

Soft matter systems exhibit large local spatial rearrangements of their microscopic constituents under thermal agitation. In contrast, "hard" materials experience only small distortions due to thermal motion. These large rearrangements lead to emergent length scales.

2. Structure on intermediate length scales

There are basic units ("fundamental" structures), typically involving a very large number (hundreds to thousands) of atoms, that are key to understand soft matter behaviour. These basic units are neither macroscopic nor microscopic (in the atomic sense), but rather mesoscopic. The relevant scales range from several nanometres to a micrometer. An example of these length scales is those associated with (topological) defects in liquid crystals. Another example is the persistence length associated with polymers[n] rather than the shorter scale of atoms and monomer units. Solutions of amphiphilic molecules form nanostructures such as micelles, vesicles, and tubules. Modularity at the mesoscale is key to theory development.

3. Slow dynamics

[n] It quantifies the bending stiffness of a polymer molecule. For parts of the polymer shorter than the persistence length, the molecule behaves like a rigid rod. For parts that are much longer than the persistence length the polymer dynamics can be described in terms of a three-dimensional random walk.

The mesoscopic length scales and complex structures lead to phenomena occurring on time scales of the order of seconds or minutes.

4. Universality

The same physical properties can arise from materials with quite different underlying chemistries. Soft matter theories that describe liquid crystals, polymers, colloids, and gels are independent of chemical details and so describe a wide range of materials.

5. Common experimental techniques

The dominant tools are microscopy, scattering (light, x-rays, neutrons), and rheometers which measure mechanical properties such as viscosity (rheology). These techniques match the relevant time and distance scales. These are typically much larger than those associated with hard materials and so require complementary techniques and instrumentation.

6. Multi-disciplinarity

Soft matter is studied by physicists, chemists, engineers, and biologists.

*Qualitative difference*
New states of matter are seen such as the sponge phase of microemulsions, ferroelectric smectics. Solutions of amphiphilic molecules form a rich array of phases through *self-assembly,* reflecting ordering of underlying micelles, vesicles, or tubules.

*Discontinuities* are seen in phase transitions including polymer and gel collapse.

*Effective interactions* describe interactions between modular structures. Volume exclusion leads to entropic attraction between large objects. Entropic elasticity, with a magnitude proportional to temperature, is relevant to rubber and polymers. The hydrophobic interaction drives self-assembly in aqueous solutions of amphiphilic molecules.

Examples of *effective theories* include the freely-jointed chain for polymers and the Maier-Saupe theory for liquid crystals.

A study of the turbulence of the fluid flow in a system of swimming bacteria is illustrative. Wensink et al.[207] found that a qualitative and quantitative description of observations of flow patterns, energy spectra, and velocity structure functions was given by a toy model of self-propelled rods (similar to that proposed by for flocking) and a minimal continuum model for incompressible flow. For the toy model they presented a *phase diagram* as a function of the volume fraction of the fluid occupied by rods and aspect ratio of the rods. There were six distinct phases: dilute state, jamming, swarming, bionematic, turbulent, and laned. The turbulent state occurs for high filling fractions and intermediate aspect ratios, covering typical values for bacteria.

Regarding this work, McLeish highlighted the importance of the identification of the relevant mesoscopic scale and the power of toy models and effective theories in the following beautiful commentary.[208]

"Individual 'bacteria' are represented in this simulation by simple rod-like structures that possess just the two properties of mutual repulsion, and the exertion of a constant swimming force along their own length. The rest is simply calculation of the consequences. No more detailed account than this is taken of the complexities within a bacterium. It is somewhat astonishing that a model of the intermediate elemental structures, on such parsimonious lines, is able to reproduce the complex features of the emergent flow structure. Impossible to deduce inductively the salient features of the underlying physics from the fluid flow alone—creative imagination and a theoretical scalpel are required: the first to create a sufficient model of reality at the underlying and unseen scale; the second to whittle away at its rough and over-ornate edges until what is left is the streamlined and necessary model. To 'understand' the turbulent fluid is to have identified the scale and structure of its origins. To look too closely is to be confused with unnecessary small detail, too coarsely and there is simply an echo of unexplained patterns."

### 9.4 Spin glasses

These are alloys of a non-magnetic metal (e.g., Cu) and a magnetic atom (e.g., Cr). They can be viewed as a large concentration of magnetic impurity atoms located in random positions inside the host metal. The magnetic moments of two impurities, separated by a distance R, interact with one another via the RKKY interaction, which has a magnitude that varies as $1/R^3$ and a sign that rapidly oscillates as a function of R. Given the random location of the spins this means that the interspin interactions are random and can be of either sign. In different words, this is a system with quenched disorder and frustrated interactions.

*Novelty*
For impurity concentrations larger than a critical value, as the temperature is lowered the material undergoes a transition to *a distinct state of matter*, below a low temperature $T_{sg}$. There is no long-range magnetic order or net magnetisation. In this state the direction of the local magnetic moments are frozen, albeit in random directions. It is not in an equilibrium state and exhibits relaxation of magnetic properties on macroscopic time scales, i.e., glassy behaviour. This means there are long-lived metastable states.

*Discontinuities*
The temperature dependence of thermodynamic properties are qualitatively different from regular magnetic phase transitions. The specific heat capacity is a continuous function of temperature, with a peak around the temperature $T_{sg}$, that has a magnitude proportional to the concentration of magnetic impurities. At extremely low magnetic fields (of the order of $10^{-4}$ Tesla) the magnetic susceptibility has a sharp cusp at $T_{sg}$, which becomes a broad peak for larger fields. This nonlinearity has been interpreted as a singularity in the third-order magnetic susceptibility.

Discovery of the spin glass state raised two fundamental questions about this state of matter. What is the order parameter? What symmetry is broken? Following the proposal of the toy models discussed below, these answers were answered definitively by Parisi[209] and highlight how dramatically different this state of matter is.

*Toy models*
The Edwards-Anderson model is a Ising model where the inter-spin interactions $J_{ij}$ have random values assigned by a Gaussian probability distribution with average zero. A similar but slightly different model that is more amenable to analytic treatment is the Sherrington-Kirkpatrick model. In it the inter-spin interactions are infinite-range and the mean-field theory is exact. These toy models capture essential ingredients to understanding spin glasses including the rugged energy landscape, replica symmetry breaking, and ultrametricity.

*Replica symmetry breaking*

Theoretical calculations of the properties of disordered systems can be performed by using the replica trick which is based on the mathematical identity.

$$\ln Z = \lim_{n \to 0} \frac{Z^n - 1}{n}$$

One first considers n identical replicas of the system, calculates the partition function Z and then takes the limit of n tending to zero, making use of the mathematical identity.

The toy models exhibit multiple local equilibria that become infinite in number in the thermodynamic limit. Furthermore, the magnitude of the typical energy barriers between local minima scales with N. This leads to the concept of a rugged energy landscape that can describe the glassy behaviour. The energy minima exhibit a hierarchical structure that can be described in terms of the mathematical concept of ultrametricity.

In the spin glass state, the breaking of replica symmetry is related to the existence of many pure thermodynamic states (i.e., local minima). The order parameter and breaking of replica symmetry can be understood as follows. A replica $\alpha$ can be viewed as one realization of the system. $q_{\alpha\beta}$ is a measure of the average overlap of the two solutions $\alpha$ and $\beta$. The order parameter is a function q(x) that is defined in terms of a probability distribution $P(q_{\alpha\beta})$ and varies between 0 and 1. q(x) is non-uniform in the glassy phase.

*Inter-disciplinarity*

The underlying models, physical ideas and mathematical techniques are relevant and useful for a wide range of complex systems.[210] This led to new insights into problems in neuroscience, the folding of proteins, [211] biological evolution, computer science, and optimization in applied mathematics. Examples of the latter include the travelling salesman problem and simulated annealing. I now discuss how spin glass ideas inspired Hopfield's work on associative memory and neural networks.

## 10. Nuclear physics

Atomic nuclei are complex quantum many-body systems. The table below summarises how nuclear physics is concerned with phenomena that occur at a range of length and number scales. As the atomic number A, number of protons Z, and number of neutrons (A-Z) in a nucleus increases qualitatively different behaviour occurs. The nature of the ground state and low-lying excitations changes.

Large nuclei exhibit collective excitations.[212] Nuclear electric quadrupole moments are sometimes more than an order of magnitude larger than can be attributed to a single proton. Rotational band structures reflect the possibility of nuclear shapes including prolate, oblate, and triaxial deformations from a sphere. The moments of inertia associated with the rotational states are markedly smaller than the moments for rigid rotation expected for uncorrelated single-particle motion.

| Scale, Number of nucleons | Entities | Effective interactions | Effective Theory | Phenomena |
|---|---|---|---|---|
| $10^3$ - $10^{54}$ | Nuclear matter | Fermi degeneracy pressure | Equation of state | Neutron stars, superfluidity, quark deconfinement transition |
| 100 | Heavy nuclei | Surface tension | Liquid drop model | Fission, fusion, shape vibrations |
| 8-300 | α-structures | α-α | Cluster models | α-decay, nucleosynthesis |
| 4-300 | Nucleon pairs | | Interacting boson model | Shape phase transitions |
| 3-300 | Nucleus, nucleon quasiparticles | Nuclear mean-field | Shell model | Magic numbers |
| 1-300 | Nucleons, neutrinos | Weak nuclear force | Fermi V-A theory | Beta decay |
| 2 | Protons, Neutrons, Pions | Strong nuclear force, Pion exchange | Yukawa, Chiral field theory | |
| 1 | Quarks, Gluons | Gluon exchange | QCD | Quark confinement, Asymptotic freedom |

Table 2. Hierarchies in nuclear physics. Moving from the bottom level to the second top level, relevant length scales increase from less than a femtometre to several femtometres. Table entries are not exhaustive but examples.

At each level of the hierarchy there are effective interactions that are described by effective theories. Some big questions in the field concern how the effective theories that operate at each level are related to the levels above and below. The best-known are the shell model, the Bohr-Mottelson theory of non-spherical nuclei, and the liquid drop model. Here I also describe the Interacting Boson Model (IBM), which generalises and provides a microscopic basis for the Bohr-Mottelson theory. Other effective theories include chiral perturbation theory, Weinberg's theory for nucleon-pion interactions, and Wigner's random matrix theory.

The challenge in the 1950's was to reconcile the liquid drop model and the nuclear shell model. This led to discovery of collective rotations and shape deformations. Since the 1980s a major challenge is to show how the strong nuclear force between two nucleons can be derived from Quantum Chromodynamics (QCD). Figure 6 illustrates how the attractive interaction between a neutron and a proton can be understood in terms of creation and destruction of a down quark-antiquark pair. This program has had limited success, e.g., in calculating from QCD the strength of the three-nucleon interaction or the parameters in chiral effective theory.

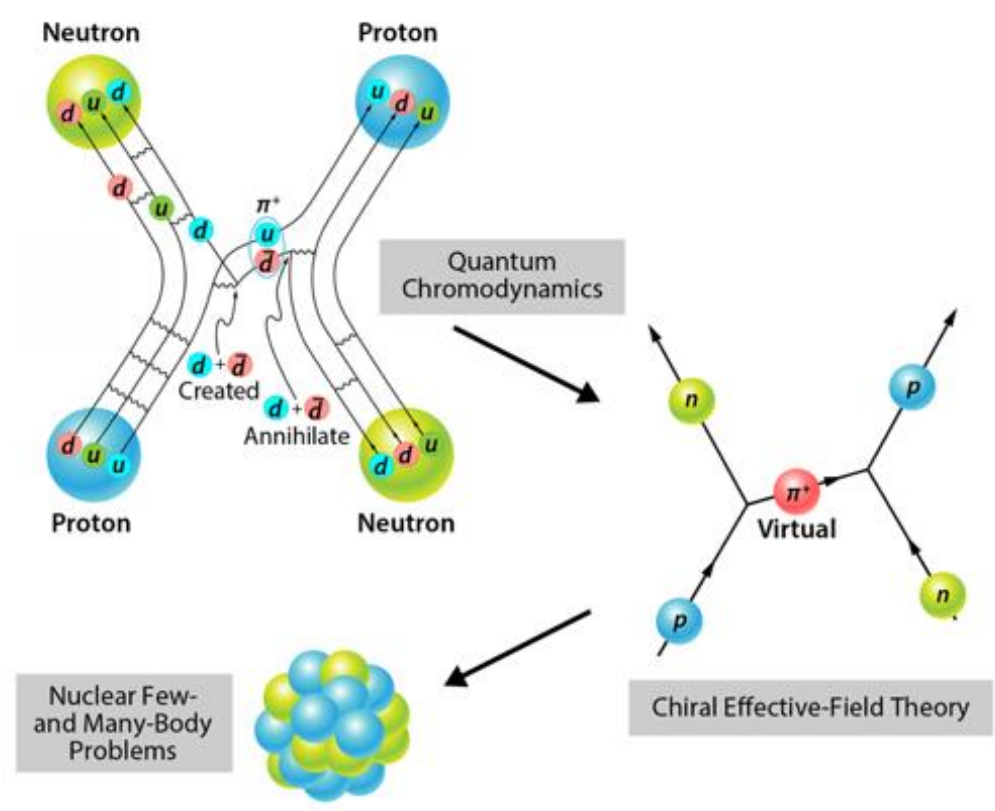

Figure 6. Effective interactions in nuclear physics.[213] Proton-neutron scattering can be described in terms of virtual exchange of a pion or in terms of creation and annihilation of quarks described by Quantum ChromoDynamics (QCD).

An outstanding problem concerns the equation of state for nuclear matter, such as found in neutron stars. A challenge is to learn more about this from the neutron star mergers that are detected in gravitational wave astronomy.[214]

*Effective theories*

The centrality of emergence in nuclear physics can be illustrated by considering the different effective theories that have been proposed to describe phenomena at different scales.

*Liquid drop model*

A nucleus is treated as a drop of incompressible liquid with an energy E(A,Z). This energy function has separate terms corresponding to the volume, surface, Coulomb interaction, nucleon asymmetry, and pairing. Parameters in the model are determined by fitting the observed binding energies of a wide range of nuclei. The theory can be used to explain spontaneous fission and estimate the energy barrier to it.

*Shell model*

The shell model has similarities to microscopic models in atomic physics. A major achievement is it explains the origins of magic numbers, i.e., that nuclei with atomic numbers 2, 8, 20, 28, 50, 82, and 126 are particularly stable because they have closed shells. Other nuclei can then be described theoretically as an inert closed shell plus valence nucleons that interact with a mean-field potential due to the core nuclei and then with one another via effective interactions.

Landau's Fermi liquid theory provides a basis for the nuclear shell model which assumes that nucleons can be described in terms of weakly interacting quasiparticles moving in an average potential from the other nucleons.[215] The BCS theory of superconductivity, was adapted to describe the pairing of nucleons, leading to energy differences between nuclei with odd and even numbers of nucleons. This helped explain why the moments of inertia for rotational bands were smaller than for independent nucleons. The shell model is the basis of many-body calculations that include interactions between nucleons. This means considering Slater determinants (configurations) which describe different occupations of the orbitals associated with the nuclear mean-field potential. A measure of the complexity of this many-body system is that a J=2+ state of $^{154}$Sm can contain contributions from more than $10^{14}$ configurations involving just a single nucleon moved to an excited orbital. This highlights the necessity of simpler theories such as those considered below and makes their successes even more impressive.

*Cluster structures*

For certain nuclei there are hints in shell model calculations of internal structures that consist of alpha particles. These are particularly evident in excited states near the threshold for alpha particle emission. The excited state of the second $J=0^+$ state of $^{12}$C has a structure of three alpha particles. This state was made famous by Fred Hoyle as the fine-tuning of its' energy is necessary for the stellar nucleosynthesis of carbon. This internal structure is consistent with trends seen in binding energies. These hints are also seen in a toy model where all the nucleons move in a potential which is an anisotropic harmonic oscillator. This has motivated effective theories where alpha particles are quasiparticles.[216]

Isolated nuclei reside near a quantum phase transition between a nuclear liquid and a Bose-Einstein condensate of alpha clusters.[217] In theoretical models, changing the range and locality of the nuclear force, one can traverse from one phase to the other. In different words, if the density of everyday nuclear matter is reduced by a factor of three, the nucleons condensed in alpha particles.

*Bohr-Mottelson model*

This model bridges the liquid drop and shell models and describes collective effects in heavy nuclei such as shape vibrations and large electric quadrupole moments. The effective Hamiltonian can be written in terms of the kinetic energy and potential energy associated with two nuclear deformation parameters $\beta$ and $\gamma$. The first term of the Hamiltonian is the kinetic energy of the $\beta$-vibrations. The second term describes the rotational energy and the kinetic energy of the $\gamma$-vibrations and contains the Casimir operator of the SO(5) Lie group. The last term is the potential energy as a function of both deformation parameters $\beta$ and $\gamma$.

*Interacting Boson Model (IBM)*

The IBM explains the rapid changes as a function of the number of nucleons observed in the low-lying excitation spectrum of nuclei containing even numbers of protons and of neutrons. $R_{4/2}$ is the ratio of the energies of the $J=4^+$ state to that of the $2^+$ state, relative to the ground state. As the system changes from that of nucleus with close to a magic number of nuclei to an almost half-filled shell and ellipsoidal shape, $R_{4/2}$ changes from less than 2 to about 3.3. In between, the ratio is 2.0 and the nucleus exhibits spherical vibrations. Associated with this rapid change is a large increase in B(E2), the strength of the quadrupole transition between the $2^+$ state and the ground state.

The IBM is surprisingly simple and successful. It is an example of the power of toy models and effective theories. It illustrates the importance of quasiparticles, builds on the stability of closed shells and neglects many degrees of freedom. It describes even-even nuclei, i.e., nuclei with an even number of protons and an even number of neutrons. The basic entities in the theory are pairs of nucleons, which are either in an s-wave or a d-wave state. There are five d-wave states (corresponding to the 2J+1 possible states of total angular momentum with J=2). Each state is represented by a boson creation operator and so the Hilbert space is six-dimensional. If the states are degenerate [which they are not] the model has U(6) symmetry.

The IBM can describe transitions in the shape of nuclei[218] and can be justified from the shell model with certain assumptions.[219] The IBM Hamiltonian is written in terms of the most general possible combinations of the boson operators. This has a surprisingly simple form, involving only four parameters,

$$H = \epsilon n_d - \kappa\vec{Q}\cdot\vec{Q} - \kappa'\vec{L}\cdot\vec{L} + \kappa''\vec{P}\cdot\vec{P},$$

where $n_d$ is the number of d-bosons, $L$ is the boson angular momentum operator, and $\vec{Q}$ and $\vec{P}$ roughly correspond to quadrupole and pairing interactions between bosons, respectively.

For a given nucleus the four parameters can be fixed from experiment, and in principle calculated from the shell model. The Hamiltonian can be written in a form that gives physical insight, connects to other nuclear models and is amenable to a group theoretical analysis that makes calculation and understanding of the energy spectrum relatively simple.

Central to the group theoretical analysis are subalgebra chains which connect U(6) to O(3), rotational symmetry. The three distinct chains of broken symmetry are associated with the subalgebras U(5), SU(3), and O(6). The former two connect to an anharmonic vibrator and to a quadrupole deformed rotator, respectively. This provides an elegant connection to Bohr and Mottelson's model of collective quadrupole surface excitations and Elliott's SU(3) model. For each of the three subalgebras the Hamiltonian can be written in terms of Casimir operators of the corresponding subalgebra and so are analytically soluble.

A major achievement of the model, and which was unanticipated, was that the O(6) limit provided a qualitative and quantitative description of a dozen of the lowest-lying excitations of the $^{196}$Pb nucleus.[220] It described the ordering and spacing of the levels, their quantum numbers, and the intensity of gamma ray transitions between them.

For all even-even nuclei in the nucleotide chart estimates of the parameter values for the IBM model have been made. The different parameter regimes of the model can be related to the different subalgebra chains.

Nuclear theorists have suggested that the model and the associated nuclei exhibit quantum phase transitions as the model parameters and nuclei numbers change.[218] However, these are crossovers not true phase transitions. The nucleus has a finite number of degrees of freedom and there is no spontaneous symmetry breaking: the ground state has J=0. On the other hand, the low-lying excitation spectrum does reflect the crossover. Perhaps, this is analogous to the "tower of states" first introduced by Anderson to describe how an antiferromagnetic Heisenberg model on a finite lattice exhibits signatures of incipient spontaneous symmetry breaking which is prevented by quantum tunneling between degenerate ground states.[221,222]

Taking the limit of an infinite number of bosons (nucleon pairs) gives a semi-classical model, allowing connection with the Bohr-Mottelson model. For example, in the SU(3) case, taking the limit where the quantum numbers $\lambda,\mu$ become large they tend to the nuclear deformation parameters $\beta$ and $\gamma$ that appear the Bohr-Mottelson model.

*Random matrix theory*

Wigner proposed this could describe the statistical distribution of energy level spacings in heavy nuclei. This effective theory makes no assumptions about the details of interactions between nucleons, except that the Hamiltonian matrix has unitary symmetry. Universality is evident in that it can describe experimental data for a wide range of heavy nuclei. Furthermore, random matrix theory can also describe aspects of quantum chaos and zeros of the zeta function relevant to number theory. It has also been applied to ecology and finance, as discussed later.

*Skyrmion model*

Skyrme considered a classical field theory defined in terms of an SU(2) matrix describing three pion fields and dynamics defined by a non-linear sigma model Lagrangian.[223] Solutions with topological charge A were identified with a nucleus with atomic number A. It is remarkable that nucleons emerge purely from a pion field theory. Witten showed that Skyrme's model could be regarded as the low-energy effective field theory of Quantum ChromoDynamics (QCD) in the limit of a large number of quark colours.[224]

*Chiral effective field theory*

Weinberg proposed a theory to describe the interaction of nucleons and mesons[225]. It reproduced results from current algebra calculations. Chiral perturbation theory is a systematic derivation of Weinberg's theory that ensures the theory is consistent with the symmetries of QCD.[226] As discussed below, developing the theory led Weinberg to change his views about effective theories and renormalizability as a criteria for theory selection.[225]

## 11. Elementary particles and fields

The systems of interest can be viewed as being composed of many interacting components because quantum fields have components with a continuous range of momenta. Even the "vacuum" is composed of many fluctuating fields, and an elementary particle interacts with these fluctuating fields. Elementary particles can be viewed as excitations of the vacuum, just as in quantum condensed matter quasiparticles can be viewed as elementary excitations of the ground state.

### Big questions

Big questions in the field include unification, quantum gravity, the hierarchy problem, baryon asymmetry, and connections to big-bang cosmology, including the origins of dark matter and dark energy, and the value of the cosmological constant.

The hierarchy problem is that measured values of some masses and coupling constants in the Standard Model are many orders of magnitude different from the "bare" values used in the Lagrangian. For example, why is the weak force $10^{24}$ times stronger than gravity? In different words, why is the Higgs mass parameter $10^{16}$ times smaller than the Planck mass?

There are two aspects to the problem of the cosmological constant. First, the measured value is so small, 120 orders of magnitude smaller than estimates based on the quantum vacuum energy. Second, the measured value seems to be finely tuned (to 120 significant figures!) to the value of the mass energy.

*Scales and stratum*

| Scale, Energy | Entities | Effective interactions | Effective Theory | Phenomena |
|---|---|---|---|---|
| | Electrons, photons | Coulomb | Quantum electrodynamics (QED) | Vacuum polarisation, e-e+ creation, anomalous magnetic moments |
| 1 MeV | Nucleons and pions | Strong nuclear force | Chiral EFT | |
| 200 MeV | Light quarks, gluons | | Heavy quark EFT | Quark confinement, chiral symmetry breaking |
| 2 GeV | Heavy quarks | | QCD | |
| 100 GeV | W and Z gauge bosons | Weak nuclear force | Weinberg-Salam | Beta decay, Neutral currents |
| 125 GeV | Higgs boson, leptons | | Standard model | Electroweak symmetry breaking, lepton mass |
| $10^{16}$ GeV | | SU(5) | GUT | Proton decay |
| $10^{19}$ GeV | Strings, space-time foam | | Quantum gravity | Hawking radiation |

Table 3. The energy scales and associated stratum associated with elementary particles and fields. EFT is Effective Field Theory.

As far as is known there is no qualitatively new physics between 100 GeV and $10^{16}$ GeV, a range covering 14 orders of magnitude, and beyond the range of current experiments. This absence is sometimes referred to as the "desert".

*Novelty*

At each of the stratum shown in Table 3 new entities emerge and there are effective theories describing their interactions. As in condensed matter physics, a unifying concept is *spontaneous symmetry breaking,* introduced by Nambu, drawing on analogues with superconductivity.[227] It leads to new particles and effective interactions. Some massless particles acquire mass. For example, the W and Z gauge bosons in the electroweak theory.

When the relevant quantum field theories are considered at non-zero temperature spontaneous symmetry breaking is associated with phase transitions at temperatures that

would have occurred in the very early universe.[121221421821421821922422722 8] For example, electroweak symmetry breaking occurs at a temperature of about $10^{15}$ K, which was the temperature of the universe about $10^{-12}$ sec after the big bang.

*Anomalies* refer to where a quantum field theory does not have the same symmetry as the classical Lagrangian on which the theory is based. A famous example is the Adler–Bell–Jackiw anomaly, which explains the decay rate of neutral mesons, and describes the non-conservation of axial currents. They are conserved in the classical Lagrangian but not in the quantum theory.

*Dimensional transmutation* concerns how the quantum fluctuations associated with interactions can lead to the emergence of energy scales. For example, consider a classical field theory in which the coupling constant describing the interactions is a dimensionless constant. In the corresponding quantum field theory, there may be logarithmic *divergences* in one-loop diagrams of perturbation theory. This implies that this strength of the interactions depends on $\Lambda$, the typical energy scale of the properties of interest. From a renormalization group perspective this is referred to as the "running" of the coupling constant and is described by the beta function. $\Lambda$ is sometimes referred to as the cutoff (range of validity) for the perturbation theory. Depending on the sign of the beta function, $\Lambda$ may be a high energy (UV) or low energy (IR) scale. Note that $\Lambda$ is not predicted by the theory but is a parameter that must be fixed by experiment, and it specifies the bounds of validity of the theory. *Universality* is reflected in the theory being valid over a wide range of energy scales and the beta function defining the relationship between the corresponding theories.

*Modularity at the mesoscale*

Composite particles such as nucleons and mesons can be viewed as elementary excitations of the vacuum for Quantum ChromoDynamics (QCD). These particles interact with one another via effective interactions that are much weaker than the interactions in the underlying theory (Figure 6). Specifically, the force between nucleons is short range and much weaker than the force between the quarks of which they are composed. The composite particles have masses (energies) intermediate between the IR and UV cutoffs of the theory. In that sense they occur at the mesoscale.

In some theories there are soliton or instanton solutions the classical equations of motion. Coleman referred to these as “classical lumps”, [229] and identified their quantum descendants with elementary particles.

In the QCD vacuum there are centre vortices in the SU(3) gauge field. These are thick two-dimensional objects imbedded in four-dimensional spacetime. They determine the properties of quark confinement and dynamical symmetry breaking. It has become possible to visualise centre vortices in recent lattice QCD simulations.[230]

*Effective theories*

Over the past fifty years there has been a significant shift in perspective concerning effective theories. Originally, QED, the Standard Model, and quantum field theory in general, were considered to be fundamental. Renormalisability of a theory was considered to be a criteria for the validity of that theory, arguably following the work of 't Hooft and Veltman showing that Yang-Mills theory was renormalisable. In 2016, Steven Weinberg described how his perspective changed, following his development in 1979 of chiral effective theory for nucleons.[225]

"Non-renormalizable theories, I realized, are just as renormalizable as renormalizable theories… For me in 1979, the answer involved a radical reconsideration of the nature of quantum field theory… Perhaps the most important lesson from chiral dynamics was that we should keep an open mind about renormalizability. The renormalizable Standard Model of elementary particles may itself be just the first term in an effective field theory that contains every possible interaction allowed by Lorentz invariance and the SU (3) × SU (2) × U (1) gauge symmetry, only with the non-renormalizable terms suppressed by negative powers of some very large mass M... confirmation of neutrino oscillations lends support to the view of the Standard Model as an effective field theory, with M somewhere in the neighborhood of $10^{16}$ GeV.

The Standard Model is now considered an effective field theory. Effective theories have been reviewed by Georgi[231] and Levi.[74] The associated history and philosophy has been reviewed by Hartmann[232] and by Crowther.[233]

There are two distinct approaches to finding effective theories at a particular scale, referred to as bottom-up and top-down approaches. However, as noted earlier these terms are used in the opposite sense to in condensed matter physics, chemistry, and biology. To avoid confusion, I will here refer to them as low-energy and high-energy approaches.

*High-energy* approaches require having a theory at a higher energy scale and then integrating out the high energy degrees of freedom (fields and particles) to find an effective theory for the lower energy scale. This is what Wilson did in his approach to critical phenomena. The renormalisation group provides a technique to compute the effective interactions at lower energies. In quantum field theory a successful example of the high-energy approach is the electroweak theory of Weinberg and Salam. In the low energy limit, it produced quantum electrodynamics and the theory of the nuclear weak interactions.

*Low-energy* approaches can be done without knowing the higher energy theory. Sometimes the Lagrangian for the effective theory can be written down based on symmetry considerations and phenomenology. An example is Fermi's theory of beta decay and the weak interactions.

As for condensed matter, the historical progression of theory development has always been from low-energy theories to high-energy theories. In different words, the low-energy theory has been developed, understood, and tested first. Later, high-energy theories have been developed that have been shown to reduce to the low-energy theory in that limit. This requirement has guided the development of the high-energy theory.

## Toy models

Toy models have played a central role in understanding quantum field theory, as these models show what is possible and aid the development of concepts. When first proposed it was known that many of these toy models did not describe known physical forces or particles. Many are not in the 3+1 dimensions of the real world.

Sidney Coleman was a master at using toy models to illuminate key aspects of quantum field theory.[229] He discussed models in 1+1 dimensions including the sine-Gordon, Gross-Neveu, and Thirring models. They illustrate how theories with simple Lagrangians can have rich properties including spontaneous symmetry breaking, singularities (non-perturbative behaviour), and a rich particle spectrum. The Gross-Neveu model exhibits asymptotic freedom, spontaneous symmetry breaking, and dimensional transmutation. The Sine-Gordon model has a ground state and excitation spectrum that changes qualitatively as the coupling constant increases. The classical theory exhibits solitons and there are analogous entities in the quantum theory. Although the theory is bosonic these excitations can have fermionic character.

I now consider a few other examples of toy models: the Anderson model for massive gauge bosons, the decoupling theorem, scalar electrodynamics, and lattice gauge theory.

*The Nambu-Jona-Lasinio model*

Nambu showed the importance of spontaneously broken symmetries in quantum field theory, guided by an analogy with the theory of superconductivity. Following the work of Bogoliubov and Anderson, he showed in 1960 that in the BCS theory "gauge invariance, the energy gap, and the collective excitations are logically related to each other" and that in case of the strong interactions "we have only to replace them by (chiral) $\gamma_5$ invariance, baryon mass, and the mesons."[227] This connection was worked out explicitly in two papers in 1961, written with Jona-Lasinio. In the first paper,[187] they acknowledged, "that the model treated here is not realistic enough to be compared with the actual nucleon problem. Our purpose was to show that a new possibility exists for field theory to be richer and more complex than has been hitherto envisaged, …"

The model consists of a massless fermion field with a quartic interaction that has chiral invariance, i.e., unchanged by global gauge transformations associated with the $\gamma_5$ matrix. At the mean-field level, this symmetry is broken. Excitations include massless bosons (associated with the symmetry breaking and like those found earlier by Goldstone) and bound fermion pairs. It was conjectured that these excitations could be analogues of mesons and baryons, respectively. The model was proposed before quarks and QCD. Now, the fermion degrees of freedom would be identified with quarks, and the model illustrates the dynamical generation of quark masses. When generalised to include SU(2) or SU(3) symmetry the model is considered to be an effective field theory for QCD,[234] similar to the chiral effective theory used in nuclear physics.

*Anderson's model for massive gauge bosons*

In 1963 Anderson argued that the case of superconductivity was relevant to how massless gauge bosons could acquire mass.[235] It is a toy model in that it is not Lorentz invariant or a non-abelian gauge theory. In a neutral superfluid the U(1) gauge symmetry is broken and there are associated massless collective modes, and they are density oscillations. Bogoliubov had shown that these modes can also ensure that physical quantities are gauge invariant. In a superconductor, which is a charged superfluid, the Coulomb interaction (which is itself mediated by massless gauge bosons, i.e., photons) causes the density oscillations to have mass (i.e., non-zero frequency at long wavelengths). These are plasmons. Anderson's toy model provided justification for the belief that symmetry breaking could lead to gauge bosons acquiring mass, which turned out to be the case for the non-Abelian gauge theory for the electroweak interaction. Anderson’s work inspired Higgs work on the Higgs field and its associated boson.

*Scalar electrodynamics*

This consists of a scalar massless boson field (a meson) interacting with an Abelian gauge field [i.e., electromagnetism]. It was studied by Sidney Coleman and Erik Weinberg in 1973.[236] They showed how radiative corrections beyond the semi-classical approximation lead to spontaneous symmetry breaking. In different words, the interactions (and associated quantum fluctuations) lead to the system having qualitatively different properties. In the broken symmetry state the excitations (effective fields) are a massive vector boson and a massive scalar boson. The model also exhibited *dimensional transmutation*. In fact, this paper is the origin of that term. This work went against the view of quantum field theory only being perturbative and promoted the importance of effective field theory. It illustrates violation of adiabatic continuity, or in different words, non-perturbative behaviour. The excitations (fields) in the effective theory are not the same as those in the original classical Lagrangian. They also extended their analysis to non-Abelian gauge theories.

*The decoupling theorem*

In 1975, Appelquist and Carazzone[74] considered a set of massless gauge fields coupled to a set of massive spin-1/2 fields, and showed that the low-energy theory was simply a gauge theory with renormalised interactions. Thus, the gauge fields decoupled from the massive fields. They called this the decoupling theorem and argued it was true for a wide range of theories. This concept is central to the emergence of effective field theories (EFTs) and a hierarchy of scales, as occurs in unified theories.[231] In its simplest case, this theorem demonstrates that for two coupled systems with different energy scales $m_1$ and $m_2$ (with $m_2 > m_1$) and described by a renormalisable theory, there is always a renormalisation condition according to which the effects of the physics at scale $m_2$ can be effectively included in the theory with the smaller scale m1 by changing the parameters of the corresponding theory. The decoupling theorem implies the existence of an EFT at scale $m_1$. However, the EFT will cease to be applicable once the energy gets close to $m_2$.

*Unification and GUTs*

In 1974, Georgi and Glashow proposed a Grand Unified Theory (GUT) to unify electromagnetism with the weak and strong nuclear forces. The symmetry group SU(5) can be broken down to SU(2) x U(1) x SU(3) where the three subgroups correspond to the weak, electromagnetic, and strong nuclear force respectively. This symmetry breaking can occur through the Higgs mechanism acting at a high energy scale of order $10^{17}$ GeV. The decoupling theorem can be used to show that the strong interaction decouples from the electroweak interaction and is described by QCD. Below this high energy scale there are effectively three independent coupling constants, corresponding to the three forces, and obeying their own scaling relations. At laboratory scales they differ by many orders of magnitude but become comparable at the unification scale. In different words, the standard model emerges as a low-energy effective theory following symmetry breaking. Although this GUT has aesthetic appeal it is unlikely to be correct as it predicts the existence of magnetic monopoles and the spontaneous decay of protons. Neither have been observed. The proton's half-life is constrained to be at least $1.67\times10^{34}$ years, whereas most predictions of the GUT give a value that is one to three orders of magnitude smaller.

*Lattice gauge theory*

Lattice gauge theory was arguably a toy model when first proposed by Wilson in 1974. He treated space-time as a discrete lattice, purely to make analysis more tractable. Borrowing insights and techniques from lattice models in statistical mechanics, Wilson then argued for quark confinement, showing that the confining potential was linear with distance.

Earlier, in 1971 Wegner had proposed a $Z_2$ gauge theory in the context of generalised Ising models in statistical mechanics to show how a phase transition was possible without a local order parameter, i.e., without symmetry breaking. Later, it was shown that this phase transition is similar to the confinement-deconfinement phase transition that occurs in QCD.[237] In condensed matter, Wegner's work provided a toy model to illustrate the possibility of a quantum spin liquid.[238]

*Lattice QCD*

The discrete nature of lattice gauge theory means it is amenable to numerical simulation. It is not necessary to have the continuum limit of real spacetime because of universality. Due to increases in computational power over the past fifty years and innovations in algorithms, lattice QCD can be used to calculate properties of nucleons and mesons, such as mass and decay rates, with impressive accuracy. The mass of three mesons is typically used to fix the mass of the light and strange quarks, and the length scale. The mass of nine other particles, including, the nucleon, is calculated with an uncertainty of less than one per cent, and in agreement with experimental values.[239,240] An indication that this is a strong coupling problem is that about 95 per cent of the mass of nucleons comes from the interactions. Only about 5 per cent is from the rest mass of the constituent quarks. Like the Ising model, lattice QCD shows how  because of universality toy models can sometimes even give an excellent description of experimental data.

*Phase diagram of QCD*

QCD has a rich phase diagram as a function of temperature versus baryon density.[241] This spans the range from the "low" densities and temperature of everyday nuclear matter to neutron stars, and to the high temperatures of the early universe and relativistic heavy ion colliders. There is a first-order phase transition between quark confinement and the deconfinement of the quark-gluon plasma. An open questions is whether there is a critical point at low density and high temperature.[242] Colour superconductivity has been proposed to exist at high densities and low temperatures near the deconfinement transition.[243] It is possible that this transition occurs inside the dense cores of neutron stars.[244]

*Beyond the Standard Model*

The Standard Model (SM) is now widely considered to be an effective theory.[245,246] An approach to finding physics beyond the SM is to write down a Lagrangian with SM as the leading term and next term is in powers of $E^{-6}$ where E is an energy scale associated with a higher-energy theory from which the SM emerges. This provides a framework to look for small deviations from the SM at currently accessible experimentally energies and to propose new experiments.

## 12. Quantum gravity

Einstein's theory of General Relativity successfully describes gravity at large scales of length and mass. In contrast, quantum theory describes small scales of length and mass. Emergence is central, often only implicitly, to most attempts to unify the two theories. Crowther has given a comprehensive discussion, from a philosophical perspective, emphasising the central role of effective field theory.[233] Wuthrich and Huggett have discussed how spacetime in emergent in three different theories of quantum gravity: causal sets, loop quantum gravity, and string theory.[247]

Before considering specific examples, it is useful to make some distinctions. First, a quantum theory of gravity is not necessarily the same as a theory to unify gravity with the three other forces described by the Standard Model. Whether the two problems are inextricable is unknown.

Second, there are two distinct possibilities on how classical gravity might emerge from a quantum theory. In Einstein's theory of General Relativity, space-time and gravity are intertwined. Consequently, the two possibilities are as follows.

Space-time is not emergent. Classical General Relativity emerges from an underlying quantum field theory describing fields at small length scales, probably comparable to the Planck length. Space-time emerges from some underlying granular structure. In some limit, classical gravity emerges with the space-time continuum.

Third, there are "bottom-up" and "top-down" approaches to discovering how classical gravity emerges from an underlying quantum theory, as was emphasised by Hu[248] and Crowther.[249] As discussed earlier about conflicting terminology, here I will refer to low-energy and high-energy approaches.

Finally, there is the possibility that quantum theory itself is emergent, as discussed in the earlier section on the quantum measurement problem. Some proposals of Emergent Quantum Mechanics (EQM) attempt to also include gravity.[160]

I now mention several different approaches to quantum gravity and for each point out how they fit into the distinctions above. Universality presents a problem for the high-energy approaches highlighting the problem of protectorates. General Relativity has been shown to be the low-energy limit of several very different underlying theories. This shows that just deriving General Relativity from a higher energy theory is not a sufficient condition for finding the correct quantum theory of gravity. Crowther[249] stated that, "there is very little that ties the emergent theory to the one that it emerges from… the high-energy theory is severely underdetermined."

*Gravitons and semi-classical theory*

A simple low-energy approach is to start with classical General Relativity and consider gravitational waves as the normal modes of oscillation of the space-time continuum. They have a linear dispersion relation and move with the speed of light. They are analogous to sound waves in an elastic medium and electromagnetic waves in free space. Semi-classical quantisation of gravitational waves leads to gravitons which are a massless spin-2 field. They are the analogue of phonons in a crystal or photons in the electromagnetic vacuum. However, this reveals nothing about an underlying quantum theory, just as phonons with a linear dispersion relation reveal nothing about the underlying crystal structure.

On the other hand, one can start with a massless spin-2 quantum field and consider how it scatters off massive particles. In the 1960s, Weinberg showed that gauge invariance of the scattering amplitudes implied the equivalence principle (inertial and gravitational mass are identical) and the Einstein field equations. In a sense, this is a high-energy approach, as it is a derivation of General Relativity from an underlying quantum theory. In passing, I mention Weinberg used a similar approach to derive charge conservation and Maxwell's equations of classical electromagnetism, and classical Yang-Mills theory for non-abelian gauge fields.[250] Weinberg pointed out that this could go against his reductionist claim that in the hierarchy of the sciences the arrows of the explanation always point down, stating that "sometimes it isn't so clear which way the arrows of explanation point… Which is more fundamental, general relativity or the existence of particles of mass zero and spin two?"[20]

More recently, Weinberg discussed General Relativity as an effective field theory[225]

> "... we should not despair of applying quantum field theory to gravitation just because there is no renormalizable theory of the metric tensor that is invariant under general coordinate transformations. It increasingly seems apparent that the Einstein–Hilbert Lagrangian √gR is just the least suppressed term in the Lagrangian of an effective field theory containing every possible generally covariant function of the metric and its derivatives..."

This is a low-energy approach that was first explored by Sakharov in 1967,[251] and experienced a renewal of interest around the turn of the century.[252] Weinberg then went on to discuss a high-energy approach:

> "it is usually assumed that in the quantum theory of gravitation, when $\Lambda$ reaches some very high energy, of the order of $10^{15}$ to $10^{18}$ GeV, the appropriate degrees of freedom are no longer the metric and the Standard Model fields, but something very different, perhaps strings... But maybe not..."

*String theory*

Versions of string theory from the 1980s aimed to unify all four forces. They were formulated in terms of nine spatial dimensions and a large internal symmetry group, such as SO(32), where supersymmetric strings were the fundamental units. In the low energy limit, vibrations of the strings are identified with elementary particles in four-dimensional space-time. A particle with mass zero and spin two appears as an immediate consequence of the symmetries of the string theory. Hence, this was originally claimed to be a quantum theory of gravity. However, subsequent developments have found that there are many alternative string theories and it is not possible to formulate the theory in terms of a unique vacuum.

*AdS-CFT correspondence*

In the context of string theory, this correspondence conjectures a connection (a dual relation) between a theory of gravity in Anti-deSitter space-time (AdS) and a conformal field theory (CFT) for a supersymmetric Yang-Mills gauge theory. It is argued that it illustrates a holographic principle for quantum gravity as the gauge theory is defined on the boundary of the space-time and completely determines the theory of gravity. String theorists interpret the duality as implying that space-time is emergent.[253,254] Crowther disagrees:[233]

> "Even though the relationship between the two theories is symmetric, and the duality is considered to be exact, many authors take the AdS/CFT duality as implying that the gauge theory on the boundary is fundamental, while the higher-dimensional bulk spacetime is emergent from it… The fallacious line of reasoning that leads to the suggestion that the gravitational theory is emergent is that an exact formulation of it has not been found, whereas we do have an exact formulation of the gauge theory, and so (the thought is that) the gauge theory is more fundamental.
>
> Instead, an appropriate interpretation of the duality would be one that treats the dual theories as being on equal footing. Although the duality apparently features two (very) different theories, we should view them as describing the same physical quantities, just using different concepts… Describing one theory as emergent from the other, on this picture, is misguided."

In 2018, Witten argued that AdS-CFT suggests that gauge symmetries are emergent.[255] However, I cannot follow his argument.

Seiberg reviewed different approaches, within the string theory community, that lead to spacetime being emergent.[253] An example of a *toy model* is a matrix model for quantum mechanics (which can be viewed as a zero-dimensional field theory). Perturbation expansions can be viewed as discretised two-dimensional surfaces. In a large N limit, two-dimensional space and general covariance (the starting point for general relativity) both emerge. Thus, this shows how both two-dimensional gravity and spacetime can be emergent. However, this type of emergence is distinct from how low-energy theories emerge. Seiberg also notes that there are no examples of toy models where time (which is associated with locality and causality) is emergent.

*Loop quantum gravity*

This is a high-energy approach where both space-time and gravity emerge together from a granular structure, sometimes referred to as "spin foam" or a "spin network", and has been reviewed by Rovelli.[256] The starting point is Ashtekar's demonstration that General Relativity can be described using the phase space of an SU(2) Yang-Mills theory. A boundary in four-dimensional space-time can be decomposed into cells and this can be used to define a dual graph (lattice) Γ. The gravitational field on this discretised boundary is represented by the Hilbert space of a lattice SU(2) Yang-Mills theory. The quantum numbers used to define a basis for this Hilbert space are the graph Γ,  the "spin" [SU(2) quantum number] associated with the face of each cell, and the volumes of the cells. The Planck length limits the size of the cells. In the limit of the continuum and then of large spin, or visa verse, one obtains General Relativity.

*Quantum thermodynamics of event horizons*

A low-energy approach to quantum gravity was taken by Padmanabhan.[257] He emphasises Boltzmann's insight: "matter can only store and transfer heat because of internal degrees of freedom". In other words, if something has a temperature and entropy then it must have a microstructure. He does this by considering the connection between event horizons in General Relativity and the temperature of the thermal radiation associated with them. He frames his research as attempting to estimate Avogadro's number for space-time.

The temperature and entropy associated with event horizons has been calculated for the following specific space-times:

i. For accelerating frames of reference (Rindler space-time) there is an event horizon which exhibits Unruh radiation with a temperature that was calculated by Fulling, Davies and Unruh.

ii. The black hole horizon in the Schwarschild metric has the temperature of Hawking radiation.

iii. The cosmological horizon in deSitter space is associated with a temperature proportional to the Hubble constant H, as discussed in detail by Gibbons and Hawking.[258]

Padmanabhan considers the number of degrees of freedom on the boundary of the event horizon, $N_s$, and in the bulk, $N_b$. He argues for the holographic principle that $N_s = N_b$. On the boundary surface, there is one degree of freedom associated with every Planck area, $N_s = A/L_p^2$, where $L_p$ is the Planck length and A is the surface area, which is related to the entropy of the horizon, as first discussed by Bekenstein and Hawking. In the bulk, classical equipartition of energy is assumed so the bulk energy $E = N_b\, k\, T/2$.

Padmanabhan gives an alternative perspective on cosmology through a novel derivation of the dynamic equations for the scale factor R(t) in the Friedmann-Robertson-Walker metric of the universe in General Relativity. His starting point is a simple argument leading to

$$\frac{dV}{dt} = L_p^2(N_s - N_b)$$

.

V is the Hubble volume, $4\pi/3H^3$, where H is the Hubble constant, and $L_p$ is the Planck length. The right-hand side is zero for the deSitter universe, which is predicted to be the asymptotic state of our current universe.

He presents an argument that the cosmological constant is related to the Planck length, leading to the expression

$$\Lambda L_P^2 = 3\exp(-24\pi^2\mu)$$

where μ is of order unity and gives a value consistent with observation.

## 13. Complexity theory

Complexity and emergence are often used interchangeably, as discussed earlier. Since the 1980's a field known as complexity theory or complexity science has developed[259,260] with associated conferences, journals, and institutions. Jensen has written an academic monograph that states what defines the field is the emphasis on networks.[6] Popular books have been published by Gleick,[87] Holland,[22] Mitchell,[261] Parisi,[262] Waldrop,[263] and Coveney and Highfield.[264] Topics I have already discussed include chaos theory and pattern formation.

Below I briefly discuss some other relevant topics. Others will feature in discussions of biology, economics, and computer science.

*Networks*

These are also known as graphs: large collections of vertices connected by edges. Different forms of networks that have been studied extensively include scale-free and small-world networks and they have been reviewed by Albert and Barabasi,[265] Newman[85,266,267] and Strogatz.[268] These networks have attracted considerable interest because they occur in many biological and social systems. Scale-free networks have the property that P(k), the probability of a node being connected to k other nodes scales like an inverse power of k. This means there is a high probability of the network having hubs, i.e., particular nodes that are connected to a large number of other nodes. Such hubs represent *modularity at the mesoscale.* Statistical methods have been developed to identify such hubs and their associated communities in real networks.[269] This may reveal overlapping community structure.[269]

*Power laws*

Examples include fractals, Zipf's law and the Pareto distribution. Power laws have been reviewed by Newman[85] and are the subject of a popular book by West.[270] Their origin has been contested.

*Cellular automata*

Conway's Game of Life is a popular and widely studied version of cellular automata. It is based on four simple rules for the evolution of a two-dimensional grid of squares that can either be "dead" or "alive." Distinct complex patterns can emerge, including still lifes, oscillators, and spaceships. Blundell argued that the Game of Life[271] illustrated emergence, considering the "scattering" of different objects off one another could lead to the creation and destruction of objects such as spaceships, "Canada geese", and "pulsars".

Wolfram performed an exhaustive study of one-dimensional automata where bits are updated based only on the state of their two nearest neighbours. He found[272] that the 256 different automata of this type, and their behaviour, falls into four universality classes: uniformity, periodic time dependence, fractal, and complex non-repetitive patterns. Wolfram speculated that the latter may perform universal computation. He also introduced the concept of computational irreducible complexity, associated with Rule 110. Calculation of the final outcome of its evolution cannot be simplified, but only determined by going through every step of the computation.

*"Order at the edge of chaos."* This concept has been heavily promoted[263] but is contentious. In the context of evolution, Kauffman conjectured that when a biological system must perform computations to survive, natural selection will favour systems near a phase transition between chaos and order. The concept has also been invoked to understand and improve international aid programs.[273]

Packard considered how the rules of cellular automata (CA) might evolve according to a genetic algorithm which is characterised by a parameter, lambda, defined by Langdon. According to Mitchell et al.[274] Packard made "two hypotheses: (1) CA rules able to perform complex computations are most likely to be found near "critical" lambda values, which have been claimed to correlate with a phase transition between ordered and chaotic behavioural regimes for CA; (2) When CA rules are evolved to perform a complex computation, evolution will tend to select rules with lambda values close to the critical values.” These have been contested by Mitchell et al.[274] Crutchfield pointed out[275] that the observation that in several specific models that the intrinsic computational capacity is maximised at phase transition led to the conjecture that there is a universal interdependence of randomness and structure. This led to the hope that there was a single universal complexity-entropy function. However, this turned out not to be the case.

*Self-reproducing machines*
Until the 1950’s it was assumed that a machine could not reproduce itself and that this was the fundamental difference between machines and living systems. However, von Neumann[276] showed that this was incorrect. In the process, he invented the concept of a cellular automaton. The one he studied was two-dimensional and each cell could in one of 29 different states. Self-replication is an emergent property. The parts of the machine do not have this property. The whole machine only has this property if it has enough parts and the list of instructions is long enough. Quantitative change produces qualitative change. Brenner argued[277] that von Neumann’s work, together with that of Turing on universal computation, central to biological theory.

*Self-organized criticality*

This is another heavily promoted concept that has been claimed to be an organising principle for complex systems[99] but has remained contentious.

Bouchaud reviewed[278] the relevance of the idea of self-organised criticality for finance and economics: "The seminal idea of Per Bak is to think of model parameters themselves as dynamical variables, in such a way that the system spontaneously evolves towards the critical point, or at least visits its neighbourhood frequently enough." A key property of systems exhibiting criticality is power laws in the probability distribution of a property.

Next, he reviewed the critical branching transition, a toy model that describes diverse systems including “sand pile avalanches, brain activity, epidemic propagation, default/bankruptcy waves, word of mouth, ..." This illustrates universality. The model involves the parameter $R_0$ which became famous during the COVID-19 pandemic. $R_0$ is the average number of uninfected people who become infected due to contact with an infected individual. For sand piles $R_0$ is the average number of grains that start rolling in response to a single rolling grain. When $R_0 < 1$, a single unstable grain will on average only dislodge $(1 - R_0)^{-1}$ grains. Very large avalanches have an exponentially small probability. As $R_0$ approaches 1, the probability P(S) of an avalanche involving a large number of grains S is given by [279]

$$\mathrm{P}(S) S^{-3/2} exp \propto (-(1 - R_0)^2 S)$$

When $R_0 = 1$ the distribution of avalanche sizes is a scale-free, power-law distribution $1/S^{3/2}$, with infinite mean. Bouchaud observed that "most avalanches are of small size, although some can be very large. In other words, the system looks stable, but occasionally goes haywire with no apparent cause."

## 14. Quantitative measures of causal emergence

The relationship between emergence and causation is contentious and has a long history. Most discussions are qualitative. Presented with a new system how does one identify the microscopic and macroscopic scales that may be most useful for understanding and describing the system? Can Pearl's seminal ideas about causality[280] be implemented in a practical way for understanding emergence?

Broadly speaking, a weakness of discussions of emergence and causality is that it hard to define these concepts in a rigorous and quantitative manner that makes them amenable to empirical testing, with respect to theoretical models and to experimental data. Fortunately, in the past decade there have been some specific proposals to address this issue, mostly using information theory. A helpful recent review is by Yuan et al.[281]

> "Two primary challenges take precedence in understanding emergence from a causal perspective. The first is establishing a quantitative definition of emergence, whereas the second involves identifying emergent behaviors or phenomena through data analysis. To address the first challenge, two prominent quantitative theories of emergence have emerged in the past decade. The first is Erik Hoel et al.'s theory of causal emergence [19],[282] whereas the second is Fernando E. Rosas et al.'s theory of emergence based on partial information decomposition [24].[283]
>
> Hoel et al.'s theory of causal emergence specifically addresses complex systems that are modeled using Markov chains. It employs the concept of effective information (EI) to quantify the extent of causal influence within Markov chains and enables comparisons of EI values across different scales [19,25]. Causal emergence is defined by the difference in the EI values between the macro-level and micro-level."

One perspective on causal emergence is that it occurs when the dynamics of a system at the macro-level is described more efficiently by macro-variables than by the dynamics of variables from the micro-level.

Klein et al.[284] used Hoel's information theoretic measures of causal emergence to analyse protein interaction networks (interactomes) in over 1800 species, containing more than eight million protein–protein interactions, at different scales. They showed the emergence of 'macroscales' that are associated with lower noise and uncertainty. The nodes in the macroscale description of the network are more resilient than those in less coarse-grained descriptions. Greater causal emergence (i.e., a stronger macroscale description) was generally seen in multi-cellular organisms compared to single-cell organisms. The authors quantified casual emergence in terms of mutual information (between large and small scales) and effective information (a measure of the certainty in the connectivity of a network). Ball[56] (pp. 218-220) gives an account of this work in terms of the emergence of multicellularity in biological evolution. He introduced (pp. 225-7) the term *causal spreading*, arguing that over the history of evolution the locus of causation has changed.

Returning to the review of Yuan et al., they stated[281]

> "However, in Hoel's theory of causal emergence, it is essential to establish a coarse-graining strategy beforehand. Alternatively, the strategy can be derived by maximizing the effective information (EI) [19]. However, this task becomes challenging for large-scale systems due to the computational complexity involved. To address these problems, Rosas et al.[283] introduced a new quantitative definition of causal emergence [24] that does not depend on coarse-graining methods, drawing from partial information decomposition (PID)-related theory. PID is an approach developed by Williams et al., which seeks to decompose the mutual information between a target and source variables into non-overlapping information atoms: unique, redundant, and synergistic information [29]…"

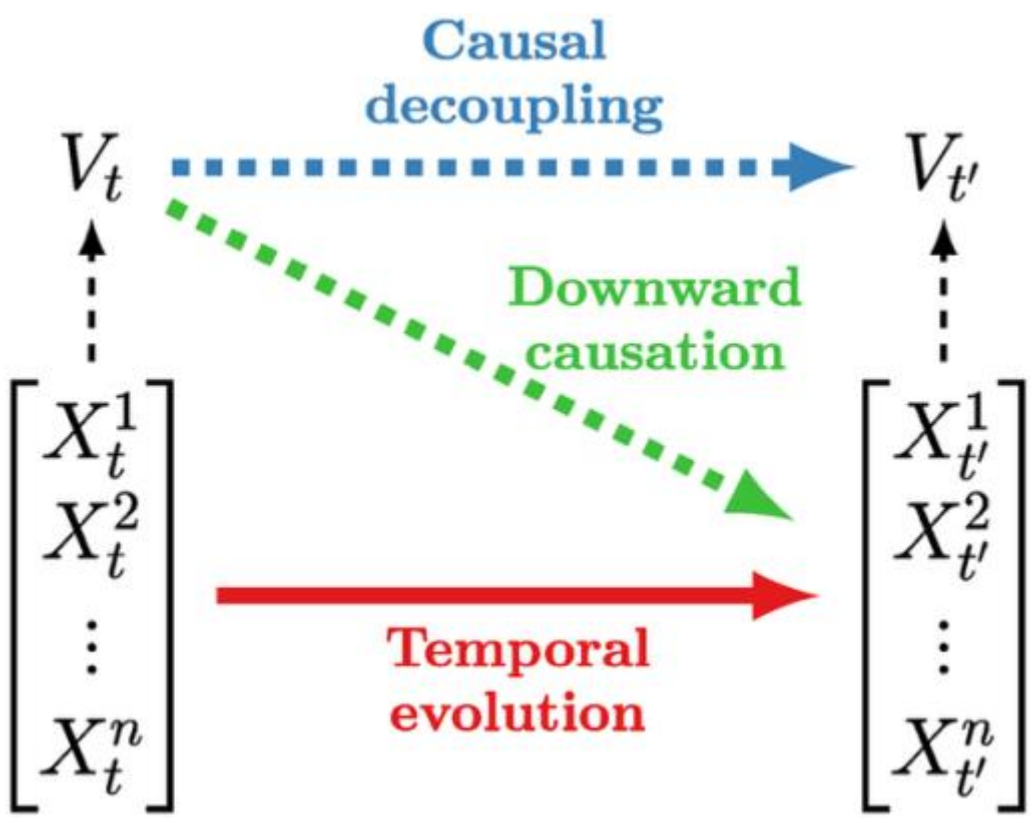

Figure. Diagram of causally emergent relationships. Taken from Rosas et al.[283] $X_t^j$ (j=1,…,n) are microscopic variables that define a Markov chain. $V_t$ is a macroscopic variable that is completely determined by the microscopic variables. Rosas et al. stated "Causally emergent features have predictive power beyond individual components. Downward causation takes place when that predictive power refers to individual elements; causal decoupling when it refers to itself or other high-order features."

Rosas et al.[283] applied the method to specific systems including Conway's Game of Life, Reynolds' flocking model, and neural activity as measured by electrocorticography. More recently it was used to describe emergence in computer science[46] including identification of modular structures. Calculations were performed for specific examples, including Ehrenfest's urn model for diffusion, the Ising model with Glauber dynamics, a Hopfield neural network model for associative memory.

On the second challenge of understanding emergence from a causal perspective, Yuan et al. stated:

> The second challenge pertains to the identification of emergence from data. In an effort to address this issue, Rosas et al. derived a numerical method [24]. However, it is important to acknowledge that this method offers only a sufficient condition for emergence and is an approximate approach. Another limitation is that a coarse-grained macro-state variable should be given beforehand to apply this method.

Sas et al.[285] recently reference empirical applications of this framework to "the study of gene regulatory networks [22], the dynamics of the human brain [23], the internal dynamics of reservoir computing [24], and the formation of useful internal representations in machine learning [25]."

Yuan et al.[281] also discussed two significant connections between causal emergence and machine learning. First, machine learning can be used to improve calculations of causal emergence. Second, causal emergence measures can be used to better understand how machine learning works and improve it.

The work described above built on earlier work by Crutchfield who claimed[286] that the identification of emergence and hierarchies could be made operational, stating that "different scales are delineated by a succession of divergences in statistical complexity at lower levels." More recently Rupe and Crutchfield have reported progress[287] towards identifying emergent self-organisation in a system.

Although this work on quantitative measures of emergence based on information theory represents significant progress, there are many open problems. Examples include the extension to non-Markovian systems and the development of computationally feasible methods for large systems.[285] The latter is particularly important in physical systems where spontaneous symmetry breaking occurs, as this only happens in the thermodynamic limit of an infinite system.

There is an unrecognised similarity between the work described above and techniques recently developed to characterise phase transitions in statistical mechanics models such as the Ising model and classical dimer models.[288,289] Coarse graining (CG) is optimised by maximising the Real-Space Mutual Information (RSMI) between a spatial block and its distant environment. In general, maximising mutual information is notoriously hard but can be done using state-of-the-art machine learning algorithms. Gokmen et al.[289] have developed an algorithm that they claimed "can, unsupervised, construct order parameters, locate phase transitions, and identify spatial correlations and symmetries for complex and large-dimensional real-space data." Furthermore, the optimal CG explicitly identifies the scaling operators associated with the critical point. The classical dimer model provides a stringent test as "the relevant low-energy degrees of freedom are profoundly different from the microscopic building blocks of the theory and change qualitatively throughout the phase diagram." In other words, the emergent entities (quasiparticles such as vortices associated with the height field which is described by a sine-Gordon classical field theory) are different from the basic constituents of the system (dimers).

There may be significant potential for fruitful interaction between these methods being developed independently by two different scientific communities.

## 15. Molecular structure

Simple molecules (such as water, carbon dioxide, methane, and benzene) have a unique structure defined by fixed bond lengths and angles. In other words, there is a well-defined geometric structure that gives the relative locations of the centre of atomic nuclei in the molecule. This structure is a classical entity that emerges from the interactions between the electrons and nuclei of the constituent atoms. Chemists draw pictures of structures and use ball and stick models to represent them. The concept of potential energy surfaces can illuminate the relative stability of different structures and the mechanisms and kinetics of chemical reactions.

Most philosophical debates about the emergence of molecular structure centre around the issue of irreducibility. Specifically, can the existence of structures be predicted from quantum theory without assuming their existence or invoking classical concepts? I will argue that the answer is yes, contrary to much of the philosophical literature, which relies heavily on the widespread use of the Born-Oppenheimer approximation in quantum chemistry calculations. However, the fact that these arguments for irreducibility are weak does not mean that emergence (defined in terms of novelty) is not central to chemistry.

*Scales.*
Energy (or equivalently time) provides the most natural scale to consider emergence associated with molecular structure. Transitions between electronic, vibrational, and rotational quantum states are associated with energy changes typically of the order of 1, 0.01, and 0.0001 eV, respectively. The Figure below shows a schematic of the hierarchy of energy levels associated with a typical small molecule. There also smaller energy scales associated with the spin states of atomic nuclei in the molecule.

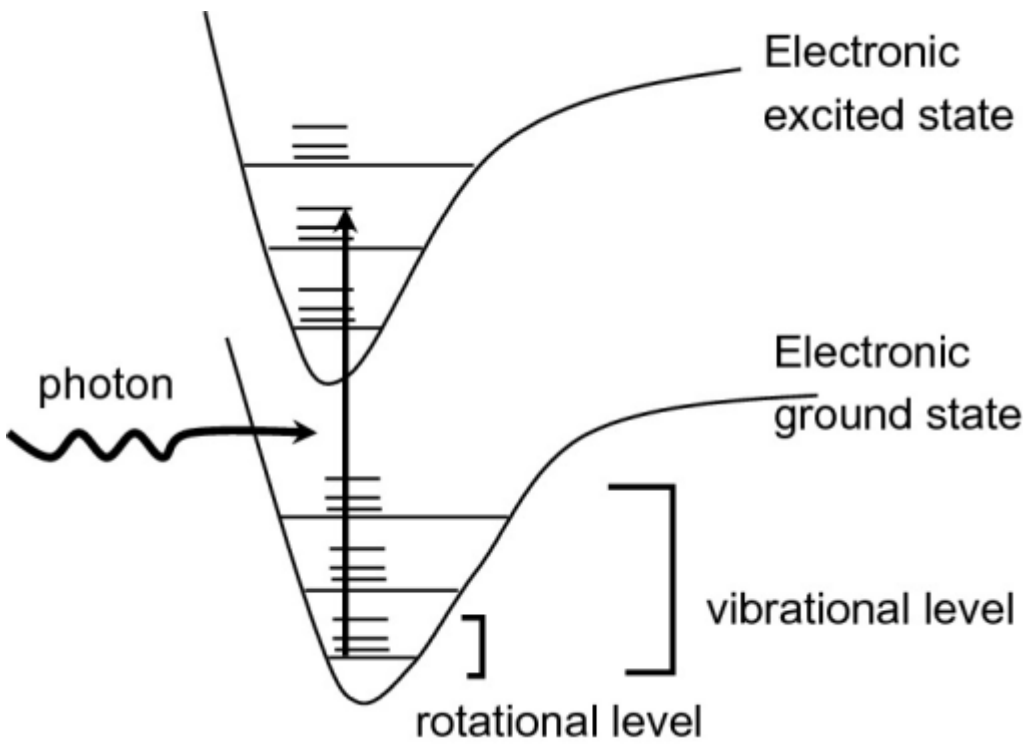

Figure. Schematic of quantum states associated with the electronic ground and excited state associated with a typical molecule. For the ground and an excited electronic state, the curves are the potential energy surfaces associated with a vibrational co-ordinate. Taken from Ref. [290]

The separation of energy scales can also be seen in the temperature dependence of the specific heat capacity of molecules in the gaseous phase. Transitions between plateaus occur when the thermal energy is comparable to the energy level spacing associated with a particular degree of freedom. (See for example, Figure 1.3 in Schroeder, *Thermal Physics*.)

*Novelty.*
Is molecular structure emergent? In other words, does the system of interacting atoms have a property that its components do not? Each of the atoms that compose the molecule are spherically symmetric. There is no geometry associated with them, whereas the atoms in the molecule are in a particular spatial relationship to one another. If the constituent atoms are isolated, they do not have the vibrational modes that the molecule does. Hence, molecular structure can be viewed as an emergent property, when defined in terms of novelty.

<u>*Isomers*</u>

Isomers are often given as an example of an emergent phenomenon. In Anderson's original "More is Different" article, he discussed the chirality of sugar molecules as an example of symmetry breaking.

*Structural isomers* occur when a specific collection of atoms (chemical formula) can have more than one molecular structure. An example is $C_3H_4$ which can be realised in three different structures: propadiene, cyclopropene, and propyne. Each structure has different chemical and physical properties. Aggregates of each molecule can have different properties, such as boiling and melting points, because the intermolecular interactions are different. Some isomers are more stable than others. They may be able to interconvert, but sometimes do not on laboratory time scales.

*Stereoisomers* are equally stable structures and are related by a chiral (mirror) symmetry. They differ physically in that they rotate the plane of polarisation of incident light in opposite directions. The two isomers, known as enantiomers, differ chemically in how they react with other chiral molecules. The simplest example is HFClBr, which has a tetrahedral structure. Chirality is central to molecular biology. Proteins are made of amino acids, and in nature they all have the L-form (left-handed chirality). Most forms of DNA involve double helices with right-handed chirality. The chirality of drug molecules matters, as tragically found with thalidomide in the 1950s.

In 1927 Hund suggested[291] that the observed optical activity of chiral molecules presented a paradox for quantum theory. Given that the Hamiltonian for the molecules was invariant under inversion, the ground state should be an equal superposition of states with left and right chirality, and have no net optical activity.

*Quantum theory of molecular structure*

Consider a molecule that contains $N_e$ electrons and $N_n$ atomic nuclei (ions). The full quantum-mechanical Hamiltonian for the system is

$$\begin{aligned}\hat{H} \quad &= \quad -\sum_{i=1}^{N_a} \frac{\hbar^2}{2M_i}\nabla^2_{\mathbf{R}_i} - \sum_{i=1}^{N_e} \frac{\hbar^2}{2m_e}\nabla^2_{\mathbf{r}_i} + \sum_{i=1}^{N_a}\sum_{j>i} \frac{Z_i Z_j e^2}{4\pi\epsilon_0 \left|\mathbf{R}_i - \mathbf{R}_j\right|} \\ &- \quad \sum_{i=1}^{N_a}\sum_{j=1}^{N_e} \frac{Z_i e^2}{4\pi\epsilon_0 \left|\mathbf{R}_i - \mathbf{r}_j\right|} + \sum_{i=1}^{N_e}\sum_{j>i} \frac{e^2}{4\pi\epsilon_0 \left|\mathbf{r}_i - \mathbf{r}_j\right|}\end{aligned}$$

where e is the electronic charge, $\mathbf{r}_j$ is the position of the j-th electron, $Z_i$ and $M_i$ are the charge and mass, respectively, of the i'th ion with position co-ordinate $\mathbf{R}_j$.
The challenge is to find the eigenstates and eigenvalues of the Hamiltonian,

$$\hat{H}\Psi_\alpha = \mathcal{E}_\alpha \Psi_\alpha$$

This Hamiltonian was dubbed "The Theory of Everything" by Laughlin and Pines because if the solution (i.e., eigenstates and eigenvalues) could be found it would describe almost all of chemistry and materials science.[67]

This Hamiltonian treats the electrons and nuclei on an equal footing.

For isomers the Hamiltonian is identical. However, as will be discussed below, that does not preclude solutions to the Hamiltonian that can describe isomers.

The Hamiltonian has global translational and rotational symmetry, where all the particles undergo the same rotation or translation. In contrast, molecular structures may have discrete rotational symmetries. However, this is not necessarily a problem, as an eigenstate of a quantum problem can transform according to a non-trivial irreducible representation of the symmetry of the Hamiltonian. For example, except the s-orbitals all the orbitals of the hydrogen atom are spatially anisotropic.

The Hamiltonian does not contain any terms involving the spins of the particles. However, the spins appear in the eigenstates and different spin states can have different energies. The electrons are identical particles and so have permutation symmetry. They are fermions with spin-1/2 and so any eigenstate must be antisymmetric under the exchange of two electrons. The energies associated with this exchange, can be of order eV and are crucial to the formation of chemical bonds and the stability of molecular structures.

If two or more atoms in the molecule are identical, then any exact eigenstate must be consistent with permutation symmetry. If a nucleus is composed of an even (odd) number of nucleons, then it is a boson (fermion) with integer (half-integer) spin, and eigenstates must be symmetric (antisymmetric) under exchange of identical nuclei. However, the corresponding exchange energies are relatively small (because the quantum delocalisation of the nuclei is small) and consequently most practical calculations of the eigenstates do not make this requirement of the eigenstates. To get a sense of scale, consider the case of the water hexamer, for which calculations and measurements[292] showed that the energies associated with nuclear exchange are of the order of nano eV. Nevertheless, if the electrons and nuclei are treated on a completely equal footing, the exchange of identical nuclei should be taken into account. Although this is challenging, it has been done recently, as discussed below.

Even if the electrons and nuclei are treated on an equal footing, in a sense the Hamiltonian still mixes quantum and classical descriptions. The Hamiltonian treats the ions as point particles with a charge. However, nuclei have a finite size and are composed of nucleons. The interaction between particles is assumed to be the classical Coulomb interaction. However, a complete quantum description would treat the interaction as due to exchange of virtual photons and described by quantum electrodynamics. Nevertheless, these classical elements have a strong justification because for chemical phenomena the quantum corrections only lead to very small quantitative differences. In a sense, the Hamiltonian can be viewed as an *effective theory*, valid at the scales relevant to chemistry.

*Full quantum solutions of the Hamiltonian*

In most of computational quantum chemistry the Hamiltonian is solved in the Born-Oppenheimer *approximation*, which will be introduced and discussed later. This is a source of some confusion and contention in philosophical discussions about the emergence of molecular structure.

Due to advances in methodology and computational power over the past few decades it has become possible in *practise* to solve the full quantum Hamiltonian for small molecules.[293–295] There are three levels of complication associated with this: quantum nuclear motion, rotational symmetry, and some nuclei being identical particles. There are also two challenges: first, finding the eigenstates and second, deducing molecular structure from the eigenstates.

To begin I consider the simplest case which ignores the complications associated with rotational symmetry or identical nuclei. This provides some insight and undermines some objections in the philosophical literature.

The ground state eigenfunction can be written as

$$\Psi_0(s, \mathbf{r}, \mathbf{R})$$

where **r** and **R** are $3N_e$ and $3N_a$ -dimensional vectors, respectively. Note this function will have a complicated structure as it will depend on the spin states of all the electrons, denoted by s.

A probability distribution for the positions of the nuclei is given by

$$\rho(\mathbf{R}) = \sum_s \int d^{3N_e}\mathbf{r}\, |\Psi_0(s, \mathbf{r}, \mathbf{R})|^2$$

where the sum is over all the electron spin degrees of freedom.

For many molecules, but not all, this probability distribution will have a *unique* global maximum at the co-ordinates $\mathbf{R}_0$. This set of co-ordinates defines the geometry of the molecular structure. The physics underlying the existence of well-defined maxima is that the mass of the nuclei is much larger than the electrons and the zero-point motion of the nuclei are much smaller than the separation of the nuclei in the molecular structure.

Note that the nuclear probability distribution is regularly measured in scattering experiments (using X-rays, neutrons, or electrons) and its maxima are used to determine the structures of molecules and crystals. The Debye-Waller factor is a measure of the width of the probability distribution. At low temperatures it is determined by quantum zero-point motion of the nuclei. In other words, it is well established experimentally that classical molecular structures approximate a fluctuating quantum structure.

Not every molecule will have a probability distribution with a unique maximum. An example is ammonia. As discussed further below, it has two maxima; each represents an umbrella geometry, and they are related by an inversion symmetry. The ground state wavefunction of the whole system is a superposition of two quantum states, each being associated with one of the two umbrella geometries, and the electronic and nuclear degrees of freedom are entangled with one another.

*A general quantum definition of molecular structure*

Lang et al.[295] have recently overcome the challenges mentioned above to determine molecular structure in a manner that treats the electrons and nuclei on an equal footing with respect to quantum theory. They have considered both rotational symmetry and nuclear permutation symmetry and given a general definition of molecular structure involving nuclear probability densities calculated from the full wavefunction. They have explicitly performed these calculations for $D_3^+$, (where D is deuterium). The result is the molecule has the same triangular structure that is observed experimentally and calculated using the Born-Oppenheimer approximation. This work is significant because it explicitly shows that molecular structure can be predicted *in practice*, not just *in principle*, from quantum theory.

*Born-Oppenheimer approximation*

In 1927 Born and Oppenheimer introduced an approximation to allow solution of the full quantum equations.[296] Without it, much of theoretical chemistry and solid state physics would be incredibly difficult in practice. The approximation is based on the separation of time and energy scales associated with electronic and nuclear motion. It leads to the concept of *potential energy surfaces* for electronic states. They define an *effective theory* for the dynamics of the atomic nuclei in a molecule or solid.

The full Hamiltonian (given above) can be denoted by

$$\hat{H} = -\sum_{i=1}^{N_a} \frac{\hbar^2}{2M_i} \nabla^2_{\mathbf{R}_i} + \hat{H}_e(\mathbf{R})$$

where the first term is the kinetic energy operator for the nuclei. In the Born-Oppenheimer approximation (BOA) the full wavefunction is written as a product of a nuclear wavefunction and an electronic wavefunction.

$$\Psi(s, \mathbf{r}, \mathbf{R}) = \Phi(\mathbf{R})\Psi_e(s, \mathbf{r}, \mathbf{R})$$

Substituting this in the eigenvalue equation for the full Hamiltonian leads to separate eigenvalue equations for the electronic and nuclear wavefunctions, assuming terms depending on gradients with respect to **R** of the electronic terms can be neglected.

$$\hat{H}_e(\mathbf{R})\Psi_e(s, \mathbf{r}, \mathbf{R}) = E_e(\mathbf{R})\Psi_e(s, \mathbf{r}, R)$$

$$(-\sum_{i=1}^{N_a} \frac{\hbar^2}{2M_i} \nabla^2_{\mathbf{R}_i} + E_e(\mathbf{R}))\Phi(\mathbf{R}) = \mathcal{E}\Phi(\mathbf{R})$$

In the first equation the nuclear co-ordinates appear as parameters not as operators. This is central to the philosophical debates. The second equation can be viewed as an *effective Hamiltonian* for the nuclear degrees of freedom. The function $E_e(\mathbf{R})$ defines the potential energy surface of the molecule.

In the BOA the nuclear probability distribution defined above is

$$\rho(\mathbf{R}) = |\Phi(\mathbf{R})|^2$$

As discussed earlier, the structure of many molecules can be defined in terms of the value of **R** at which the probability is maximum.

I make four points about the BOA that are relevant to philosophical debates about whether molecular structure is predictable in a logically consistent manner from quantum theory.

i. The BOA does treat the nuclear degrees of freedom quantum mechanically. They are described by the nuclear wavefunction $\Phi(\mathbf{R})$, which is determined by the second eigenvalue equation. Consequently, the BOA does not violate Heisenberg's uncertainty principle, contrary to some claims in the philosophy literature.[297,298]

ii. The BOA is not *ad hoc*. Corrections to it can be calculated and have been for many molecules.[296] These corrections are typically small, being of order $(m_e/M_i)^{1/2}$. Exceptions, such as near conical intersections (where the potential energy surfaces for two electronic states touch) are well-known and well-studied.

iii. For most small molecules, the results of BOA calculations compare favourably with wave-functions obtained from solutions of the full quantum Hamiltonian. When there are differences, they are largely small quantitative differences. When the differences are qualitative, they have largely been anticipated from knowledge of the limitations of the BOA.

iv. The Born-Oppenheimer approximation is an example of a general approach to quantum mechanics problems, discussed by Migdal.[299] Consider a system composed of two subsystems that have dynamics on two vastly different time scales, termed fast and slow. The effects of the fast system on the slow system can be treating by adding a potential energy term to the Hamiltonian operator of the slow system.

*Singularity.* The Born-Oppenheimer approximation can be viewed as an asymptotic expansion in powers of $(m_e/M)^{1/4}$, where $m_e$ is the mass of an electron and M is a typical mass of an atomic nucleus in the molecule. Primas[60,300] and Bishop[11] claim that as a result quantum chemistry is associated with a different ontology to quantum theory. In other words, molecular structures are classical objects and not present in quantum theory.

*Quantum justification for classical discussions of potential energy surfaces*

In computational quantum chemistry the BOA is used to determine potential energy surfaces (PES) for electronic states of molecules. It is standard practise to identify local minima on a PES with molecular structures. Molecular binding energies are identified with the difference in energy between these minima and the energies of the isolated atoms of which the molecule is composed. A chemical reaction between two molecules A and B to produce C can be understood in terms of the PES for the composite system consisting of all the atoms in A and B. The dynamics of the chemical reaction can be described in terms of a path on the PES that goes from the local minimum associated with A and B infinitely far apart to the minimum associated with the structure C. The path will pass through a saddle point on the PES and this is identified with a transition state in the chemical reaction and its energy determines the activation energy for the chemical reaction. The local curvature of the PES near a minima can be used to determine force constants for harmonic motion and the associated vibrational frequencies of a molecule.

This picture is a completely classical one and so may motivate a claim that chemists mix classical and quantum concepts and calculations in an *ad-hoc* manner. However, this is unfair because a quantum description of the nuclear dynamics can be given in terms of quantum wave-packets, consistent with the Heisenberg uncertainty principle, and whose dynamics is defined by the quantum equation for nuclear motion that is given by the BOA. Furthermore, the dynamics of the centre of a wave-packet is given by classical equations of motion (Ehrenfest's theorem). Thus, the classical language used by chemists can be viewed as a justified and compact version of a quantum description.[301]

*Structural isomers.* Isomers are associated with different local minima on the electronic ground state potential energy surface for a given combination of atoms. For a specific example, see the Figure below. To understand a quantum description of isomers, consider a reaction coordinate associated with an isomerisation reaction (i.e. conversion of one isomer to the other). There are three energy scales of relevance: the energy difference between the ground state energy of the two isomers, the magnitude of the barrier height (activation energy), and the quantum zero-point energy associated with vibrations in the direction of the reaction coordinate. Denote these energies as dE, $E_b$, and $E_{zp}$, respectively. If dE ~ $E_{zp} << E_b$ then the nuclear probability density $\rho_0(\mathbf{R})$ for the vibrational ground state will have two local maxima, corresponding to the geometries of the two isomers. If dE >> $E_{zp}$ then then the nuclear probability density $\rho_0(\mathbf{R})$ for the vibrational ground state will have only one local maxima, corresponding to the lower energy isomer geometry. However, the geometry of the higher energy isomer can be found as a local maximum in the nuclear probability density $\rho_i(\mathbf{R})$ for one of the excited vibrational ground states.

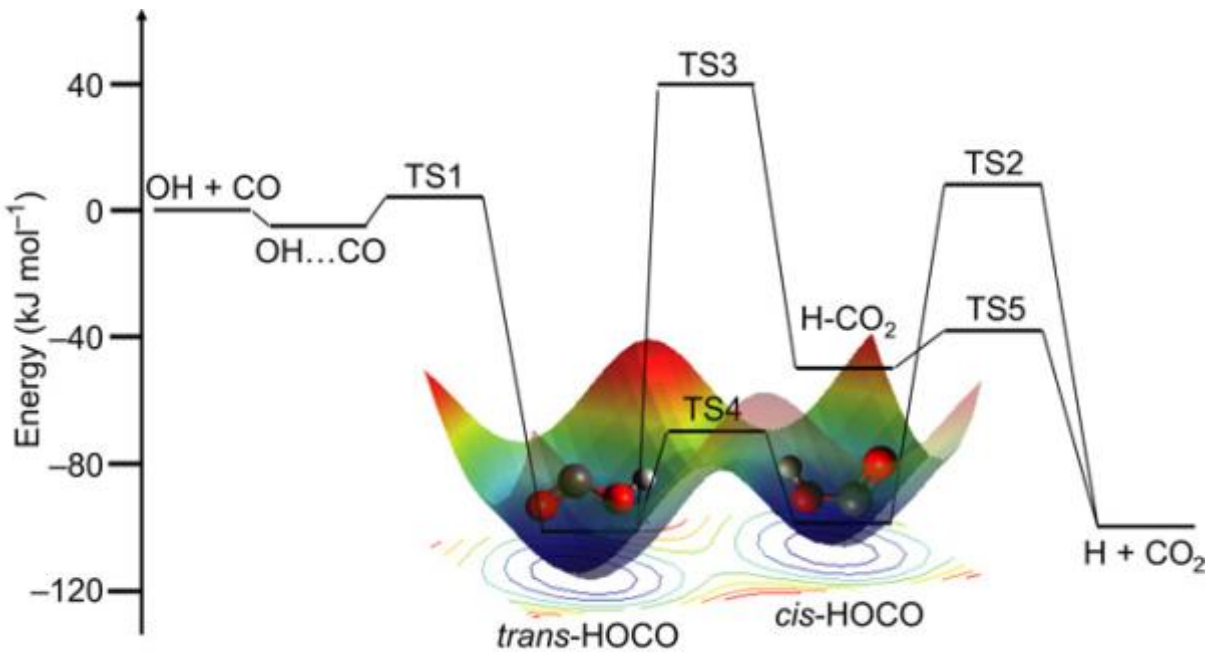

Figure. Potential energy surface associated with the two structural isomers of HOCO. TSn denote different transition states associated with the chemical reaction OH + CO -> H + $CO_2$. Taken from Bui et al.[302]

*Molecular structure of ammonia*

Ammonia is different to many molecules in several respects. First, the probability density $\rho_0(\mathbf{R})$ has two distinct maxima and they have equal probability. These correspond to two distinct molecular structures. Second, quantum tunnelling between these structures can occur and has been observed at low temperatures in dilute gases. The two structures have an umbrella structure and are related by an inversion. (See the Figure below). The example of ammonia was discussed by Anderson at the beginning of his seminal *More is Different* article[25] and sets the stage for his discussion of how symmetry breaking leads to well-defined molecular structures in large chiral molecules. Ammonia often features in discussions about the emergence of molecular structure and the reduction of chemistry to physics.

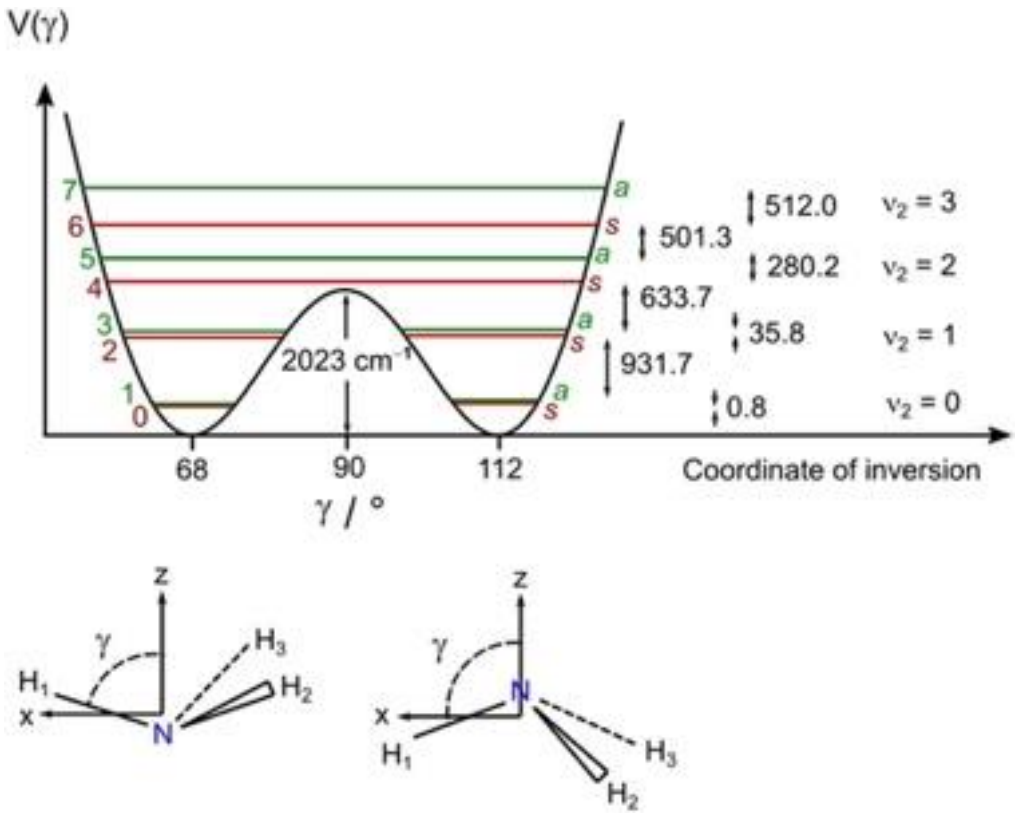

Figure. Potential energy surface and vibrational energy levels associated with pyramidal inversion of ammonia. Below the barrier the levels are split by quantum tunnelling. Energies are given in $cm^{-1}$. Taken from Nguyen et al.[303]

Systems such as ammonia and stereoisomers have two possible molecular structures of equal energy that are related by an inversion symmetry. They can be modelled in terms of a two-state Hamiltonian that quantum mechanically treats electronic and nuclear degrees of freedom on an equal footing.[304] In the adiabatic limit, the potential energy surface associated with the electronic ground state is described by a symmetric double well potential, when the electron-nuclear coupling is sufficiently strong. But the following discussion does not assume the adiabatic limit. In general, low-lying states can be written in terms of diabatic electronic states[304]

$$\Psi_{\pm,m}(\mathbf{r},\mathbf{R}) = \frac{1}{\sqrt{2}}\left(\Psi_d^L(\mathbf{r},\mathbf{R})\Phi_{L,m}(\mathbf{R}) \pm \Psi_d^R(\mathbf{r},\mathbf{R})\Phi_{R,m}(\mathbf{R})\right)$$

where the +,- quantum number is associated with the inversion symmetry, d denotes a diabatic electronic state, L and R, refer to the chirality associated with the two molecular structures, and $\Phi_{L,m}(\mathbf{R})$ is the m-th vibrational state associated with the L structure. For these states the quantum entanglement of the electronic and nuclear degrees of freedom has been calculated for a range of model parameters.[304]

For an isolated ammonia molecule quantum tunnelling back and forth between the two structures can occur. The four lowest vibrational states are quantum superpositions of the two molecular structures, as shown in the Figure above. The electronic dipole moment of the two terms in the superposition states point in opposite direction and as a result the dipole moment of the whole state is zero.

Coherent tunnelling between the two structures does occur in a dilute gas of ammonia at low temperature. The quantum transition between the two lowest states is the basis of the maser, the forerunner of the laser. At higher temperatures and in denser gases the tunnelling and quantum superpositions are washed out by decoherence and the nuclear wavefunction "collapses" onto a definite structure. This is why in most situations, ammonia has a definite structure, just like most molecules.

To illustrate how relatively rare the case of ammonia is, where coherent tunnelling between structures occurs, I contrast it to some other specific molecules. Phosphine ($PH_3$) is isoelectronic to ammonia, but the energy barrier for inversion is about 1.5 eV, whereas for ammonia is about 0.22 eV. Since, the tunnel splitting decreases approximately exponentially with the barrier height and width, the splitting is orders of magnitude smaller in $PH_3$ and has never been observed. For the ground state it has been estimated to be $10^{-8}$ $cm^{-1}$,[305] compared to 0.8 $cm^{-1}$ for ammonia. Sahu et al.[306] calculated the tunnel splittings for more than ten chiral molecules, including the family $H_2X_2$ (X=O,S,Se,Te,Po). Except for O, the splitting is smaller than $10^{-6}$ $cm^{-1}$. For water hexamers, tunnel splittings have been both measured and calculated[292] and the largest ones are about 3 x $10^{-5}$ $cm^{-1}$. These very small splittings mean it requires on weak environmental interactions to destroy any possible quantum coherence between the two possible molecular structures.

On the one hand, the decoherence problem is not mysterious and can be understood in terms of the spin-boson model, where a two-state system is coupled to infinite bath of harmonic oscillators (representing the environment).[307] For chemical systems the parameters in the model can be well-characterised.[308] When the coupling is sufficiently strong or the tunnel splitting is sufficiently small the system is localised in one of the two quantum states and cannot tunnel to the other. In most chemistry contexts, these conditions are easily realised. On the other hand, the connection between decoherence, definite outcomes, and quantum measurement problem is contested.

*Philosophical perspectives*

In philosophical discussions of emergence and reductionism in chemistry, molecular structure has received significant attention and debate.[11,309,60,300,310] . Sometimes molecular structure is used to justify strong emergence, i.e., that molecular cannot be predicted, even in principle, solely from quantum theory.

Primas was one of the first to claim that molecular structure could not be reduced to quantum physics. Consider the following statement.[60]

> "From a physical point of view the crucial point of a Born–Oppenheimer description is not a simplification of the mathematical problem, but the replacement of the basic theory by a related but *qualitatively new one*…
>
> [the structure of benzene] *does not exist* in a full quantum-theoretical description since *electrons and nuclei are entangled by Einstein-Podolsky–Rosen correlations*. The concept of molecular structure does not follow from first principles - all molecules with the same empirical formula have the same Schrödinger equation, *so that, at this level, the shape of a molecule as the main feature of molecular chemistry is simply not in evidence*. In a quantum theoretical description the molecular shape emerges by abstracting from *the actually existing Einstein-Podolsky–Rosen correlations* between the electrons and the nuclei. Historically, the structure concept has been introduced into quantum chemistry by the so called Born-Oppenheimer approximation. But this terminology is misleading since *the main issue is not an approximation, but the breaking of a holistic symmetry*."

I have italicised claims that are contestable and that I will discuss further below.

Cartwright has given philosophical arguments as to why chemistry cannot be reduced to physics. In this context, she claimed[311] (pp. 106-7)

> "The typical method of quantum-mechanical treatment of molecules begins with the Born–Oppenheimer approximation…

> This approximation treats the atomic nucleus as a classical particle. But this fundamentally violates quantum mechanics which, following the Heisenberg uncertainty principle, maintains that we cannot have a simultaneous assignment of fixed positions and fixed momenta. The approximations that provide the reduction violate the very theory that the chemistry is being reduced to… the success of quantum chemistry relies fundamentally on assumptions that belong to classical chemistry."

This claim that the BOA violates the Heisenberg uncertainty principle was rebutted above and discussed in more detail by Huggett et al.[298] Nevertheless, Lombardi et al. are not satisfied by the rebuttal.[312]

Hendry[310] claimed molecular structure is evidence of strong emergence and/or downward causation. In a similar spirit, Ellis[313] and Drossel[314] argued that crystal structures in solid state physics are evidence of strong emergence. The arguments of Hendry have been criticised by Seifert.[315] The arguments centre around the fact that the molecular structure is a classical entity and a concept that is imposed, whereas a logically self-consistent approach would treat both electrons and nuclei quantum mechanically. It is claimed that the existence of molecular structures is assumed and not derived in quantum chemistry calculations as they assume the Born-Oppenheimer approximation (BOA).

Scerri[297] responded to these arguments claiming that chemistry (particularly the concept of molecular structure) is irreducible to quantum physics. He claimed these arguments are not valid because they misunderstand the role of the BOA. It does not violate the uncertainty principle and in practice chemists can and do perform non-BOA approximations. Woolley has responded[316] to some of Scerri's claims.

Fortin et al.[317] rejected the view that decoherence can explain isomerism, as decoherence does not resolve issues associated with the quantum measurement problem.

Franklin and Seifert[318] claimed "the problem of molecular structure just is the quantum measurement problem." This is debatable. Most molecular structures can be understood in terms of the nuclear probability density having a unique maxima and decoherence then is not relevant. Decoherence and the collapse of the nuclear wavefunction is only relevant in systems such as ammonia and stereoisomers in which there are molecular structures with equal energy and separated by high energy barriers.

I now respond to some of the contestable claims of Primas.

"*electrons and nuclei are entangled by Einstein-Podolsky–Rosen correlations"*

It is possible to quantify and calculate the amount of entanglement between electrons and nuclei in a specific quantum state.[304,319,320] In an EPR state the entanglement is maximal. In the BOA wavefunction entanglement is present, but is absent in the crude BOA.[320] The entanglement has been estimated in benzene and is very small.[304] The only molecules where the entanglement may be significant is for those, such as isomers, where there are two degenerate molecular structures and the overlap of the associated nuclear wavefunctions is small (i.e., the tunnel splitting is small). But then, in most chemical situations decoherence will wash out this entanglement.

> "all molecules with the same empirical formula have the same Schrödinger equation, *so that, at this level, the shape of a molecule as the main feature of molecular chemistry is simply not in evidence*."

This is the problem of isomers. It is resolved because isomers are present in the *solution* to the Schrödinger equation, as I argued earlier.

> *"*the crucial point of a Born–Oppenheimer description is not a simplification of the mathematical problem, but the replacement of the basic theory by a related but *qualitatively new one… the main issue is not an approximation, but the breaking of a holistic symmetry*."

This seems subjective to me. I see the BOA as just a well-justified approximation. The electrons and nuclei are treated differently because they are. They have vastly different masses and this affects how they can be treated in any solution to the full Hamiltonian. Nevertheless, the BOA is a quantum theory and the nuclei are described by a wavefunction. As discussed earlier, for small molecules in practise it is no longer necessary to use the BOA and the electrons and nuclei can be treated on an equal footing. Molecular structure is present in solutions to the full Schrödinger equation.

I wonder if the objection to use of the BOA is any different to the use of approximations in other theories? For example, consider theoretical treatments of the motion of planets in the solar system. The effects of all the planets are not treated on an equal footing. The effect of other planets on a planet of interest are treated perturbatively.

In conclusion, the arguments that molecular structure is evidence of strong emergence are weak. Some confusion may partly arise from misinterpreting the significance of the following valid observations.

i. Molecular structures were first conjectured before quantum theory was proposed.
ii. The BOA was proposed to explain molecular structure from quantum theory.
iii. Today, almost all calculations of molecular structure assume BOA.
iv. Chemists talk about molecular structures largely in classical not quantum terms.

However, the scientific reality is that for small molecules their structure, can be understood, described, and calculated in purely quantum terms. Yet, that understanding provides a strong justification for the validity of the BOA and for the convenience of using classical language to describe molecular structure.

I stress that the weakness of the arguments for the strong emergence of molecular structure, does not mean that an emergent perspective is not relevant to chemistry. Molecular structure is emergent, when defined in terms of *novelty*. This leads to *effective theories* defined in terms of potential energy surfaces. Furthermore, as the next section argues chemistry exhibits novel properties, concepts, and theories that are not present in physics.

## 16. Chemistry

It is important to be clear what the system is. Most of chemistry is not really about isolated molecules. Most chemical reactions happen in an environment, often a solvent. Then the system is the chemicals of interest and the solvent. For example, when it is stated that HCl is an acid, this is not a reference to isolated HCl molecules but a solution of HCl in water which dissociates into $H^+$ and $Cl^-$ ions. Chemical properties such as reactivity can change significantly depending on whether a compound is in the solid, liquid, or gas state, or on a solid surface or dissolved in a solvent.

*Scales*

Relevant scales include the total numbers of atoms in a compound, which can range from two to millions, the total number of electrons, and the number of different chemical elements in the compound. As the number atoms and electrons increases so does the dimensionality of the Hilbert space of the corresponding quantum system.

The time scales for processes, which range from molecular vibrations to chemical reactions, can vary from femtoseconds to days. Relevant energy scales, corresponding to different effective interactions, can vary from tens of eV (strong covalent bonds) to microwave energies of 0.1 meV (molecular rotational energy level transitions).

*Novelty*

All chemical compounds are composed of a discrete number of atoms, usually of different type. For example, acetic acid, denoted $CH_3COOH$, is composed of carbon, oxygen, and hydrogen atoms. A compound usually has chemical and physical properties that its constituent atoms do not have. Chemistry is all about transformation. Reactants combine to produce products, e.g., A + B -> C. The product C may have chemical or physical properties that the reactants A and B did not have.

Chemistry involves concepts that do not appear in physics. Hoffmann argued[321] against a reductionist perspective.

> "Scientists have bought the reductionist mode of thinking as their guiding ideology… Yet this philosophy bears so little relationship to the reality within which scientists themselves operated... There are vertical and horizontal ways of understanding.there are concepts in chemistry which are not reducible to physics… ideas such as aromaticity, acidity and basicity, the concept of a functional group,… They cannot be mathematicisized,… But they are of fantastic utility to our science."

*Diversity*

Chemistry is a wonderland of diversity as it puts chemical elements in a multitude of different arrangements that produce a plethora of phenomena. For example, much of organic chemistry involves only three different atoms: carbon, oxygen, and hydrogen.

*Irreducibility*

In 1929 Paul A. M. Dirac claimed that "the underlying physical laws necessary for the mathematical theory of ... the whole of chemistry are thus completely known, and the difficulty is only that the exact application of these laws leads to equations much too complicated to be soluble." This sentence of Dirac's is often cited in discussions of emergence because it raises the question of whether chemistry is reducible to quantum physics.

*Qualitative difference with increase in molecular size*

Consider the following series with varying chemical properties: formic acid ($CH_2O_2$), acetic acid ($C_2H_4O_2$), propionic acid ($C_3H_6O_2$), butyric acid ($C_4H_8O_2$), and valerianic acid ($C_5H_{10}O_2$), whose members involve the successive addition of a $CH_2$ radical. The Marxist Friedrich Engels used these examples as evidence for Hegel's law: "The law of transformation of quantity into quality and *vice versa*", that was discussed earlier.

In 1961 Platt[322] discussed properties of large molecules that "might not have been anticipated" from properties of their chemical subgroups. Table 1 in Platt's paper listed "Properties of molecules in the 5- to 50- atom range that have no counterpart in diatomics and many triatomics." Table 2 listed "Properties of molecules in the 50- to 500- atom range and up that go beyond the properties of their chemical sub-groups." The properties listed included internal conversion (i.e., non-radiative decay of excited electronic states), formation of micelles for hydrocarbon chains with more than ten carbons, the helix-coil transition in polymers, chromatographic or molecular sorting properties of polyelectrolytes such as those in ion-exchange resins, and the contractility of long chains.

Platt also discussed the problem of molecular self-replication. As discussed earlier, until the 1950's it was assumed that a machine could not reproduce itself and this was the fundamental difference between machines and living systems. However, von Neumann showed that a machine with enough parts and a sufficiently long list of instructions can reproduce itself. Platt pointed out that this suggested there is a threshold for autocatalysis: "this threshold marks an essentially discontinuous change in properties, and that molecules larger than this size differ from all smaller ones in a property of central importance for biology." Thus, self-replication is an emergent property. A modification of this idea has been pursued by Kauffman with regards to the origin of life,[3] that when a network of chemical reactions is sufficiently large it becomes self-replicating.

*Coarse graining in multi-scale modelling*

Computational chemists now routinely study systems containing as many as billions of atoms. Dynamical simulations on such large systems are restricted to treating the atoms classically moving in effective potentials that known as "force fields". These are all parameterised and some are determined by electronic structure calculations using the Born-Oppenheimer approximation.

Given the large number of atoms, the goals of computational efficiency and physical insight have led to the development of a range of coarse-graining methods. These have recently been reviewed by Jin et al.[72] Basically, sub-groups of atoms are replaced by "blobs" that interact with one another via a new effective potential.

Consider a system, defined by a Fine-Grained (FG) set of n coordinates $\mathbf{r}^n$, and interacting with one another via a potential $u_{FG}(\mathbf{r}^n)$. Coarse-Graining involves defining a set of $N < n$, collective coordinates,

$$M_I(r^n) = \sum_{i=1}^{n} c_{Ii} r_i$$

where I=1,…,N. The *effective interaction* between the entities defined by these collective co-ordinates is given by

$$U_{\text{CG}}(\mathbf{R}^N) = -k_\text{B}T \ln \int d\mathbf{r}^n \delta(\mathbf{M}(\mathbf{r}^n) - \mathbf{R}^N) \times \exp\left(-\frac{u_{\text{FG}}(\mathbf{r}^n)}{k_\text{B}T}\right) + (\text{const.}) \quad (11)$$

In chemistry terminology this effective interaction is a Potential of Mean Force (PMF) and can be temperature dependent. The big challenge is to decide how to choose the collective coordinates $\mathbf{M}_I(\mathbf{r}^n)$. There are a plethora of different methods for making this choice. They contain varying amounts of physical and chemical insight, statistical criteria, and brute force computation.[72] One approach[323] is based on Renormalisation Group methods from statistical physics. In physics language, all of these coarse-graining procedures correspond to what is referred to as integrating out degrees of freedom to derive effective fields and interactions.

*Self-assembly*

Molecular self-assembly is of interest in chemistry, materials science, and biology. This is a form of spontaneous ordering or *self-organisation* and involves *modular structures*. New structures emerge resulting from weak interactions between molecular components. Examples of molecular self-assembly include the formation of molecular crystals, colloids, lipid bilayers, phase-separated polymers, self-assembled monolayers, protein folding, and the association of a ligand with a receptor. The *effective interactions* between molecules are generally weak (i.e., comparable to thermal energies) and non-covalent (van der Waals, electrostatic, hydrophobic, or hydrogen bonds).

Concepts and techniques for molecular self-assembly have been extended to mesoscale and macroscale components.[324] Millimetre-scale components either float at an interface between two fluids or are suspended in fluid or similar density. The *effective interactions* are capillary forces that result from minimise interfacial free energy by minimizing interfacial areas.

*Computational quantum chemistry*

There is a plethora of acronyms associated with different methods (Huckel, PPP, HF, PT2, CC, CAS-SCF, VB, …) used to calculate properties of molecules. These methods are all approximations to the electronic many-body Schrodinger equation in the Born-Oppenheimer approximation. Each of these methods can be viewed as solving an effective Hamiltonian or toy model as they restrict the solution to lie in a particular sub-space for the full many-electron Hilbert space.

*Water*

Much of chemistry and all of biochemistry happens in systems where water is the solvent. Despite the simplicity of the $H_2O$ molecule, collections of water molecules exhibits rich and diverse structures. For example, bulk water has 18 different stable solid states, depending on pressure and temperature. Solid water can also exist in a metastable glass state. Snowflakes exhibit rich structures with a complex phase diagram as a function of temperature and humidity. Clathrate hydrates are solid states of water with diverse cage-like structures that are stable at negative pressure[325] and can trap gaseous molecules in the cages.

The hydrophobic interaction is an *effective interaction*. It determines micelle formation, immiscibility of oil and water mixtures, the local structure around some solutes, and the folding of proteins.

Water is a rich source of t*oy models* including monatomic water,[326] the Sceats-Rice random network model for liquid water[327], and the ice-type (six-vertex) models that are exactly soluble in two dimensions.[328]

## 17. Protein folding

There are two distinct protein folding problems. One problem is to predict the final stable structure (native state) of a protein given a knowledge of the complete amino acid sequence of the protein. The second problem is to explain the dynamics of how an unfolded protein (i.e., a polymer chain) folds so rapidly, efficiently, and reversibly into its native state. Arguably, the first problem was recently solved using the AI-based programs Alphafold.[329] This was developed at Google DeepMind and recognised by the Nobel Prize in Chemistry in 2024, shared by Demis Hassabis and John Jumper. Nassar et al. recently argued[330] that the second challenge has largely been solved, at least at the qualitative level, using a range of techniques drawing on statistical mechanics.

Proteins are a distinct *state of matter.* Globular proteins are tightly packed with a density comparable to a crystal but without the spatial regularity found in crystals. The native state is thermodynamically stable, in contrast to the globule state of synthetic polymers which is often glassy and metastable, with a structure that depends on the preparation history.

For a given amino acid sequence, the native folded state of the protein is emergent. It has a structure, properties, and function that the individual amino acids do not, nor does the unfolded polymer chain. Examples of functions are chemical catalysis, DNA transcription, signal transduction, and photosynthesis.

Protein folding is an example of *self-organisation.* A key question is how the order of the folded state arises from the disorder (random configuration) of the unfolded state.

There are *hierarchies* of structures, length scales, and time scales associated with the folding. The hierarchy of structures include primary, secondary, tertiary, and ternary structures. The primary structure is the amino acid sequence in the heteropolymer which contains up to twenty different amino acids. Secondary structures include alpha-helices, beta-sheets and disordered regions. The tertiary structure is the native folded state. An example of a ternary structure is in hemoglobin which consists of four myoglobin units in a particular geometric arrangement.

The hierarchy of time scales varies over more than seventeen orders of magnitude, including folding ($10^{-8}$ to $10^{3}$ sec), helix-coil transitions (microseconds), hinge motion (nanoseconds), and bond vibrations (10 femtoseconds).

Folding exhibits a hierarchy of processes, nucleation of ordered regions from the disordered polymer, and the growth and coalescence of these regions, until the native conformation is reached. These processes can be understood in terms of modularity and cooperativity.[331] Frauenfelder argued that the energy landscape associated with folding was rugged and had a hierarchical structure.[332]

*Modularity* is reflected in the concept of foldons, segments of the polymer chain that can fold into stable structures independently. The protein folding process can be viewed as the stepwise assembly of foldons into the native structure of the protein.[333]

*Discontinuities.* The folding-unfolding transition [denaturation] is a sharp thermodynamic transition, similar to a first-order phase transition. This sharpness reflects the cooperative nature of the transition. There is a well-defined enthalpy and entropy change associated with the transition.

*Universality.* Proteins exhibit "mutational plasticity", i.e., native structures tolerant to many mutations (changes in individual amino acids). Aspects of the folding process such as its speed, reliability, reversibility, and modularity appear to be universal, i.e., hold for all proteins.

*Diversity with limitations.* On the one hand, there are a multitude of distinct native structures and associated biological functions. On the other hand, this diversity is much smaller than the configuration space, presumably because thermodynamic stability vastly reduces the options.

*Effective interactions.* These are subtle. Some of the weak interactions matter as the stabilisation energy of the native state is of order 40 kJ per mole, which is quite small as there are typically about 1000 amino acids in the polymer chain. Important interactions include hydrogen bonding, hydrophobic, and volume exclusion. In the folded state monomers interact with other monomers that are far apart on the chain. The subtle interplay of these competing interactions produces complex structures with small energy differences, as is often the case with emergent phenomena.

*Toy models*

I now discuss four different models that have provided insight into the folding process.

*Dill's HP polymer on a lattice*

This consists of a polymer which has only two types of monomer units and undergoes a self-avoiding walk on a lattice. H and P denote hydrophobic and polar amino acid units, denoted by red and blue circles, respectively, in the figure below. The relative simplicity of the model allows complete enumeration of all possible conformations for short chains. The model is simpler in two dimensions yet still captures essential features of the folding problem.

As the H-H attraction increases the chain undergoes a relatively sharp transition to just a few conformations that are compact and have hydrophobic cores. The model exhibits much of the universality of protein folding. Although there are twenty different amino acids in real proteins, this model divides them into two classes and still captures much of the phenomena of folding, including mutational plasticity.

*Blob-based models*

These are built around the modular structures that emerge in the folding process and give reasonable estimates of protein folding times for about 200 proteins, including their correlation with chain length.[334] The folding time $\tau_{tcs}$ is estimated to be given by the product of three factors,

$$\tau_{tcs} = \Omega_M \Omega_B \tau_p$$

where $\Omega_M$ is the number of conformations within the blob, $\Omega_B$ is the number of conformations among, the blobs, and $\tau_p$ is the time required for a single amino acid to probe its possible conformations.

*Wako-Saito-Munoz-Eaton model*

This is an Ising-like model on a chain. A short and helpful review is by Munoz.[335] A protein conformation for a chain of N residues is described by a set of Ising-like variables, $m_k$ assigned to each residue k of the protein ($m_k = 1$ for native and 0 for other conformations), where k=1,..,N. The Hamiltonian of the model is defined as:

$$H(\{m\}) = \sum_{i=1}^{N-1} \sum_{j=i+1}^{N} \varepsilon_{i,j} \mathrm{m}_{i,j}$$

and $\varepsilon_{i,j}$ is the contact energy between residues i and j in the native state. The protein state represents a set of all residue states, with 2N possible conformations, and:

$$\mathrm{m}_{i,j} = \mathrm{m}_i \, \mathrm{m}_{i+1} \cdots \mathrm{m}_j = \prod_{k=i}^{j} \mathrm{m}_k$$

An order parameter is the average value of the $m_k$ is a measure of the degree of native structure formation, and can be used as a reaction co-ordinate for folding. Note that in the model the interactions are not pairwise but involve strings of "spins" between native contacts, i.e., they are non-local.

The model has successfully predicted the folding rates for 22 proteins from their known three-dimensional structures. [336] It has predicted the presence of folding intermediates for four proteins that exhibit them. The calculations suggest that folding speed is largely determined by the distribution and strength of contacts in the native structure.

*Spin glass models*

Bryngelson and Wolynes[211]proposed a spin-glass type model to study protein folding. This sought to explain the slow folding time of many proteins, the presence of multi-exponential decay, and the possibility that the energy landscape associated with different conformations had a complex structure. Each amino acid in the protein was treated as a “spin” with of the order of ten different states (conformations). The Hamiltonian contained terms associated with the primary (single site), secondary (nearest-neighbour), and tertiary (long-range) structures (interactions). The interaction parameters were treated in the random energy approximation, following Derrida’s model for glasses. In the mean-field approximation the model had a phase diagram with four phases, corresponding to unfolded; correctly folded; folded frozen (a frozen phase in which the native structure is favoured); and misfolded frozen (a frozen phase in which the native structure is not favoured). This work led to the “principle of minimal frustration" for proteins. Naturally occurring proteins may appear to be composed of a random sequence of amino acids. However, they fold much more rapidly, reliably, and reversibly than a polymer composed of an actual random sequence of amino acids. This suggests evolution may have selected amino acid sequences to minimise frustration. Random heteropolymers and proteins are both described by rugged energy landscapes but for proteins this landscape is superimposed on a funnel that drives folding towards the native state.[337]

## 18. Biology

In the sixty years following the discovery of the structure of DNA and the associated beginnings of molecular biology, a gene-centred reductionism has been embraced by most biologists. However, over the past two decades some biologists are shifting away from this perspective,[338] advocating a more systems-based approach, with an emphasis on networks of interactions between units as much as the nature of the units (DNA, proteins, genes). Popular books by Ball,[56] Noble,[48,339] and Kauffman[3] have advocated such an emergentist perspective. Ball introduced the term of causal spreading, arguing that over the history of evolution the locus of causation has changed. Specifically, Ball discussed causal emergence where "it is preferable to match the scale of the cause to the scale of the effect", partly based on ideas of Hoel and collaborators.[282,340,341] In a similar spirit, Noble introduced the terms, biological relativity,[339] and a-mergence,[342] arguing that there is no preferred scale of causality. In other words, there is both upward causation and downwards causation. Properties at one level do not just emerge from the interactions between lower-level components. They can result from boundary conditions and constraints imposed by higher levels. For example, error correction associated with DNA copying is increased by about six orders of magnitude in a three-stage process that makes use of the cell's protein and lipid machinery.

Earlier Ernst Mayr, a prominent evolutionary biologist and philosopher of biology, discussed the role of emergence in biology,[54] the autonomy of biology as a scientific discipline,[343] and disagreed with Weinberg about the primacy of reductionism, pointing out its limits.[344] In Mayr's book, *What Makes Biology Unique? Considerations on the Autonomy of a Scientific Discipline*,[343] chapter 4 is entitled "Analysis or Reductionism." There he stated:

> "Needless to say, the workers in the more complex branches of science saw in this [reductionist] claim only a ploy of the chemists and physicists to boost the importance of their fields. As Hilary Putnam said correctly: "What [reductionism] breeds is physics worship coupled with neglect of the 'higher-level' sciences. Infatuation with what is supposedly possible in principle goes with indifference to practice and to the actual structure of practice"
>
> What is the crucial difference between the concepts analysis and reduction? The practitioner of analysis claims that the understanding of a complex system is facilitated by dissecting it into smaller parts. Students of the functions of the human body choose as their first approach its dissection into bones, muscles, nerves, and organs. They make neither of two claims made by the reductionists
> (A) that the dissection should proceed "down to the smallest parts," – i.e., atoms and elementary particles, and
> (B) that such a dissection will provide a complete explanation of the complex system.

> This reveals the nature of the fundamental difference between analysis and reduction. Analysis is continued downward only as long as it yields useful new information and it does not claim that the "smallest parts" give all the answers.
>
> ... the view that composite wholes have properties not evident in their components has been widely accepted since the middle of the nineteenth century. The principle was already enunciated by Mill, but it was Lewes (1875) who not only presented a thorough analysis of the topic but also proposed the term emergence for this phenomenon.
>
> ... emergence is characterized by three properties
> ... first, that a genuine novelty is produced – that is, some feature or process that was previously nonexistent;
> second, that the characteristics of this novelty are qualitatively, not just quantitatively, unlike anything that existed before;
> third, that it was unpredictable before its emergence, not only in practice, but in principle, even on the basis of an ideal, complete knowledge of the state of the cosmos."

Recently, Schmit and Dill made "A modest proposal for cell biophysics going forward"[73] which is to "encourage more big-picture, top-down modeling (i.e., the "forest"), in contrast to details-first, bottom-up modelling (i.e., the "trees")." In other words, they advocate an emergentist perspective and make the case for toy models, in analogue to the Ising model in condensed matter. To illustrate the need for a systems-based approach that highlights modularity, they mention, the article, "Can a biologist fix a radio?"[345] There is also an article written in a similar spirit, "Can a neuroscientist understand a microprocessor?"[346]

<u>*Scales and stratification*</u>

Biology has sub-disciplines associated with ecosystems, organisms (animals and plants), organs, cells, genes, and molecules. Each of these entities correspond to a level of self-organisation and length scale. The entities are composed of many smaller entities from lower levels. To provide an overview of modern biology, *The Economist* published a beautiful series of six brief articles in 2021, covering six different levels of life.[347]

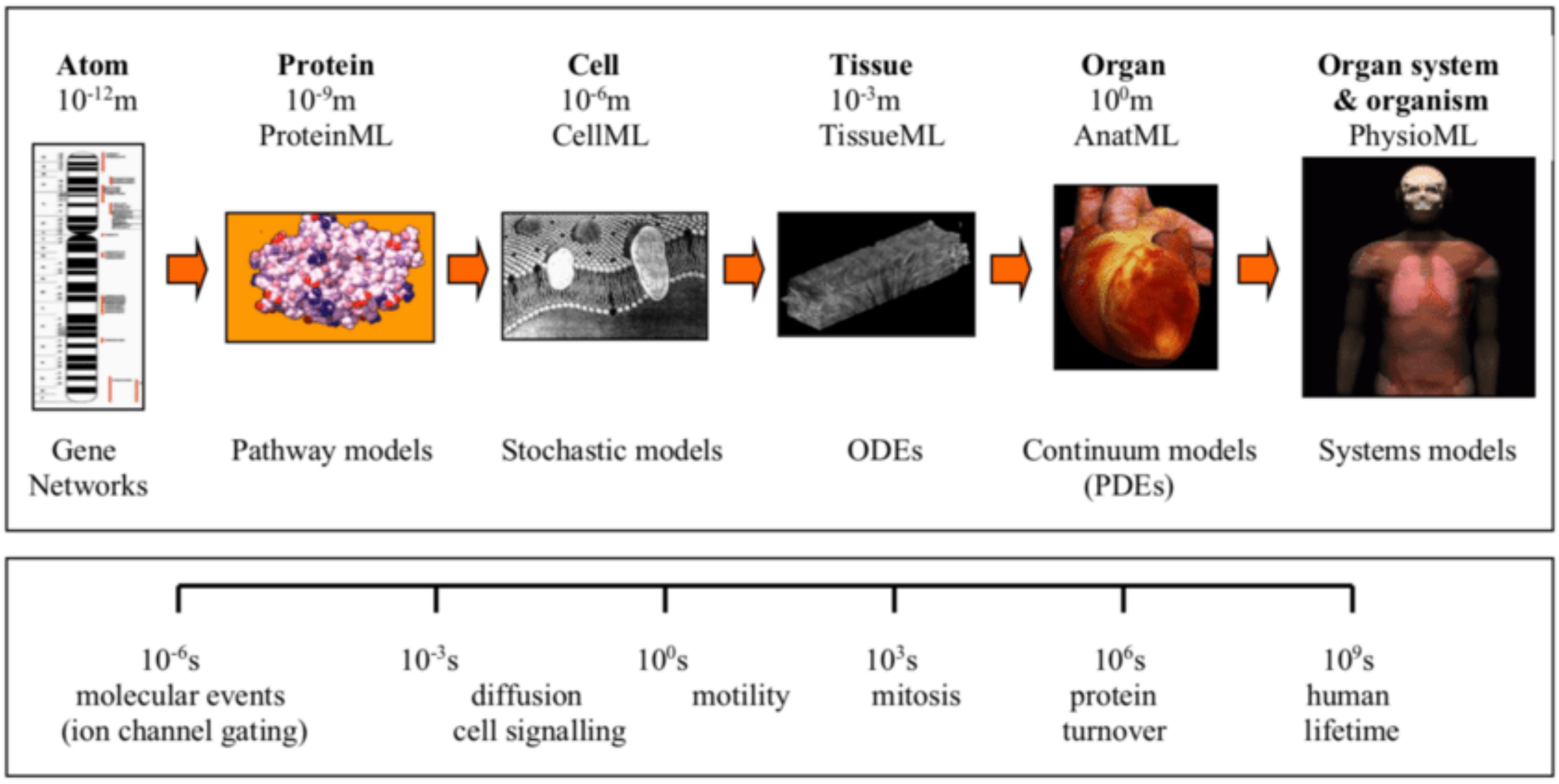

Figure 7. The stratification of biology.[348] The top line lists the emergent entity and the second line the corresponding length scale. The next line gives examples of the types of theoretical models used in one large modelling project, the IUPS Human Physiome project.[348] The bottom panel lists corresponding time scales and associated dynamical phenomena.

*Novelty: Function is emergent*

The concept of function is central to biology. DNA, proteins, cells, or organs all have a specific function, i.e., they have specific properties that produce specific outcomes in specific contexts. This contrasts with physics and chemistry: entities such as quarks, atoms, fluids, crystals, and molecules do not have a function or purpose. They only have properties. The individual base pairs in a strand of DNA or individual amino acids in a protein do not have a function. However, a linear sequence of amino acids and folded into the native form of the protein have a property that enables the protein to have a specific function. Hence, the function is emergent.

In the 1960s Polanyi argued that biological function was emergent[17] and that life was *irreducible* to chemistry and physics.[349] In the context of physicists working in molecular biology, Hopfield stated "The word "function" does not exist in physics, but physicists need to learn about it, otherwise they will be in a sandbox playing by themselves."[350]
Hartwell et al. stated:[47] "Although living systems obey the laws of physics and chemistry, the notion of function or purpose differentiates biology from other natural sciences… To describe biological functions, we need a vocabulary that contains concepts such as amplification, adaptation, robustness, insulation, error correction and coincidence detection."

Farnsworth et al.[351] made a precise definition: "A biological function is a process enacted by a biological system A at emergent level n which influences one or more processes of a system B at level n +1, of which A is a component part."The function contributes to replication and growth and can be quantified in terms of its contribution to biomass production. Farnworth et al. give an example of a specific calculation for an autocatalytic set. [As an aside, I note that they say that the system has "causal closure" showing how this term means different things to different people.]

An example of novelty is the work of Bhalla and Iyengar, who considered emergent properties of networks of biological signaling pathways within cells.[352] The found that the networks had properties that the individual pathways did not and were non-intuitive. These properties included "integration of signals across multiple time scales, generation of distinct outputs depending on input strength and duration, and self-sustaining feedback loops." This suggests that simple biochemical reactions can store information when they are coupled together appropriately.

*Modularity*

Modularity is clearly apparent at the level of DNA, proteins, genes, cells, organs, and organisms. However, there is also a "hidden" modularity present in biological networks that has only become more apparent over the past few decades.

Hartwell et al. argued[47] "for the recognition of functional 'modules' as a critical level of biological organization. Modules are composed of many types of molecules. They have discrete functions that arise from interactions among their components (proteins, DNA, RNA and small molecules), but these functions cannot easily be predicted by studying the properties of the isolated components. We believe that general 'design principles' - profoundly shaped by the constraints of evolution - govern the structure and function of modules."

Mayr argued[54] (p.19) that "two major pillars in the explanatory framework in modern biology" are the "genetic program" and "the concept of emergence – that new properties emerge at higher levels of integration which could not have been predicted from a knowledge of the lower-level components." Following Jacob's description[353] of the stratification of biological systems, he invoked the concept of "integrons", units at one level "formed by assembling integrons from the level below". They can be said to have emerged as each "integron has new characteristics and capacities not present at any lower level of integration."

*Protein interactomes*. Klein et al.[284] used information theoretic measures of causal emergence developed by Hoel[282] to analyse protein interaction networks (interactomes) in over 1800 species, containing more than eight million protein–protein interactions, at different scales. They showed the emergence of 'macroscales' that are associated with lower noise and uncertainty. The nodes in the macroscale description of the network are more resilient than those in less coarse-grained descriptions. Greater causal emergence (i.e., a stronger macroscale description) was generally seen in multi-cellular organisms compared to single-cell organisms. The authors quantified casual emergence in terms of mutual information (between large and small scales) and effective information (a measure of the certainty in the connectivity of a network). These quantitative notions of emergence have been developed more by Rosas et al.[46] in the context of computer science.

*Biomolecular condensates.* Many modular units in cells are defined by membranes. However, modular structures without membranes, consisting of compartments of proteins and nuclei acids, with unique functionality can form. These self-organised structures are known as biomolecular condensates and their formation is associated with a thermodynamic phase transition, similar to liquid-liquid transitions (i.e., when two different fluids become immiscible).[354] Schmit et al., argued that a hierarchy of interactions promotes formation of modular units and functionality. Examples of the interplay of strong and weak interactions are found in the nucleolus, SPOP/DAXX condensates, polySUMO/polySIM condensates, chromatin, and stress granules. They stated: "Strong interactions provide structural specificity needed to encode functional properties but carry the risk of kinetic arrest, while weak interactions allow the system to remain dynamic but do not restrict the conformational ensemble enough to sustain specific functional features." As an aside, I note some similarity to Granovetter's seminal work in sociology on how social cohesion emerges due to weak social links.[355]

*Physiology*

Noble's systems-based perspective of biology developed out of his work in physiology on modelling heartbeats using networks of coupled ion channels.[356] This work showed that heartbeat was not controlled by a single oscillator, but rather heartbeat was an emergent property of the feedback loops involving the ion channels. This reflects self-organisation. It has an endogenous cause not an exogenous one. This surprised many biologists.[48]

There is a hierarchy of modes of transportation seen in animals. Platt noted[322] that as mass increases there are transitions between float in air, buzz, flutter, fly, soar, and walk.

For a wide range of species scaling laws are observed where specific properties scale with a power of the body mass M of the species. An example is Kleiber's law, that the total metabolic rate, scales as $M^{3/4}$, which holds over more than five orders of magnitude of mass, for mammals, from mice to elephants. Other examples, include diameters of tree trunks and aortas scaling as $M^{3/8}$, rates of cellular metabolism and heartbeat as $M^{-1/4}$, and blood circulation time and life span as $M^{1/4}$.

Based on Euclidean geometric scaling it would be expected that the exponents would be multiples of 1/3 since the characteristic length scales as $M^{1/3}$, and the surface area scales as $M^{2/3}$. However, the observed exponents are mostly multiples of ¼. West, Brown, and Enquist explained the observed power-laws in terms of the fractal-like architecture of the hierarchical branching vascular networks that distribute resources within organisms[357],[358]. They stated: "Organisms have evolved hierarchical branching networks that terminate in size-invariant units, such as capillaries, leaves, mitochondria, and oxidase molecules. Natural selection has tended to maximize metabolic capacity, by maximizing the scaling of exchange surface areas, and internal efficiency, by minimizing the scaling of transport distances and times. These design principles are independent of detailed dynamics and explicit models and should apply to virtually all organisms." Thus, this work illustrates the importance of hierarchies, modularity, and universality. A popular book by West puts the work in the context of a wider array of scaling laws.[270]

It should be noted that Hatton et al.[359] considered a wider range of lighter eukaryotes, covering more than ten orders of magnitude of mass, and found deviations from Kleiber's law. They claimed their results were "incompatible with a metabolic basis for growth scaling and instead point to growth dynamics as foundational to biological scaling."

*Collective behaviour in animal communities*

Diverse species with a wide range of sizes exhibit collective behaviour. Birds and fish exhibit flocking (murmuration) which is collective motion characterised by cohesion, polarity, shape, and structure. Beautiful videos of flocking of starlings can be viewed on YouTube.[o] There is no leader directing the overall motion. Self-organisation occurs due to the interactions between each bird and its neighbours. Thus, the causes are endogenous. In contrast, exogenous causes are associated with the seasonal migration of flocks.

In ant colonies complex structures and functions emerge even though there is no individual directing the whole operation. In popular accounts of emergence, this example is often cited, including to illustrate how complexity can emerge from simple rules.[1,2,360] A colony is composed of several distinct classes (castes) of ant: soldiers, excavators, foragers, garbage collectors, and gardeners. Each ant has a very limited repertoire of methods to interact with other ants and their environment. Ants have poor hearing and sight. They communicate with a few signals involving touch but mostly communicate by producing trails of distinct chemicals (pheromones). Each organic molecule is identified with a specific message such as follow this trail, detection of food, presence of an enemy, or danger. For an ant colony the components are simple and the interactions between the parts are simple. Nevertheless, complex structures such as bridges, tree houses, and trail networks emerge. Trail networks can extend over distances as large as one hundred kilometres. There is no chief engineer directing the construction of these structures or a blueprint drawn up by an architect. The queen ant is not a dictator mandating that the colony must last for her lifetime, which covers many generations of worker ants.

Human crowds exhibit collective behaviour such as Mexican waves and mosh pits.[361] Groups of pedestrians can spontaneously form unidirectional lanes or stop-and-go waves and at high densities body collisions can generate turbulent seen in some crowd panics.[362] Traffic exhibits novel properties such as jams, waves, and clusters.[363] These emergent phenomena can be described by toy models which also exhibit phase transitions between the qualitatively different behaviour.

Universality is highlighted in a review by D.J.T Sumpter.[364]

> "I argue that the key to understanding collective behaviour lies in identifying the principles of the behavioural algorithms followed by individual animals and of how information flows between the animals. These principles, such as positive feedback, response thresholds and individual integrity, are repeatedly observed in very different animal societies. The future of collective behaviour research lies in classifying these principles, establishing the properties they produce at a group level and asking why they have evolved in so many different and distinct natural systems."

*Toy model for flocking*

Vicsek et al.[365] proposed a two-dimensional toy model to describe flocking. Each bird moves at a constant speed and at discrete time steps its direction is updated according to the average direction of its nearest neighbours. There is a phase transition from a disordered state (where the average bird velocity is zero) to an ordered flocking state (with non-zero average velocity) with increasing number of birds and decreasing noise. Flocking states are non-equilibrium analogues of ferromagnetic and nematic liquid crystal phases. The existence of a ferromagnetic-type ordering in the two-dimensional model may appear to violate the Mermin-Wagner theorem. However, it has been shown that the inclusion of hydrodynamics as this allows "violation" of Mermin-Wagner.[366] Detailed analysis of observational data on large flocks of starlings led to refinement of the model showing that the local interaction of each bird with its neighbours is topological.[367] This was confirmed by Bialek and collaborators[368] who used maximum entropy methods to show the data could be described by a classical Heisenberg model where the interaction is independent of density, i.e., distance to neighbours. More generally, the model describes any collection of self-propelled particles and has led to a field known as active matter[369]. This another example of universality.

---

[o] For example,
https://www.youtube.com/watch?v=eakKfY5aHmY&ab_channel=raisingmaggie

## *Evolution*

The theory of evolution explains the origin of biological diversity and levels of similarity between species. A characteristic of emergence is that *many* iterations of a simple law (natural selection of the fittest to reproduce) can produce novel, diverse and rich structures. In biological evolution many generations in a population can produce new traits and species.

Many of the most debated issues about evolution relate to the different characteristics of emergence and are briefly discussed below.

### *Scales*

Central to emergence are the ideas of "many" and of scales. The former can take two forms: a system composed of *many* interacting components, or a system that undergoes *many* iterations according to a rule that is repeated many times. For evolution both forms of "many" are relevant and have several dimensions. Evolution occurs in a population, i.e., a community of many members of a species living in a specific environment. Each member of the population has a specific genotype (many genes) which largely determines biological characteristics, from proteins to organs, defined as the phenotype. The environment also consists of many interacting species. Natural selection can act at multiple levels: on genes, cells, organisms, species, and groups of species.

Microscopic and macroscopic scales also manifest in different ways. In terms of length, the micro- and macro- scales can be defined in terms of genotypes and phenotypes, respectively. In terms of time, microevolution and macroevolution roughly correspond to directly observable timescales and geological timescales, respectively. They are associated with the emergence of new traits within a species and new species, respectively.

### *Novelty*

Development of *new* traits and species occurs over many generations, due to the repetition of the rule of natural selection.

Evolution theory uses novel concepts such as natural selection, survival of the fittest, niches, and hierarchical trees, that are not present in chemistry and physics.

### *Connecting micro- and macro- properties*

As for other systems, this is one of the great challenges of emergence. Genotypes and phenotypes are extremely well characterised. Genotype-phenotype maps seek to connect these micro- and macro- levels. A detailed understanding of how microevolution leads to macroevolution is outstanding.

*Discontinuities*

In microevolution new traits occur within a species due to (continuous) adaptation to the environment. In contrast, in macroevolution new organs and species can occur suddenly (at least on geological timescales). An example is the Cambrian explosion of new life forms. Extinctions can also represent discontinuities.

Evolution of a population occurs in response to changes in an environment. New traits, new species, and extinctions can be viewed as qualitative changes due to quantitative changes. For example, small changes in the oxygen concentration in the atmosphere is one (among many) hypothesis for the cause of the Cambrian explosion.

Using techniques from statistical physics, the transition of a species from survival to extinction can be viewed as a non-equilibrium phase transition to an absorbing state. The order parameter is the population and a toy model is directed population.[370]

*Diversity with limitations*

All species are based on the same biochemistry of DNA and proteins. Yet from these same building blocks there is an incredible diversity: more than 8 million distinct species, including more than 10,000 species of birds and more than 15,000 species of ants. Darwin said nature produces "endless forms most beautiful."

But there are limitations. For example, the number of species with more than one head, brain, heart, or liver is limited. There are many more genotypes than phenotypes.

The dominant view is that evolution is driven by random genetic mutations. Debates have arisen about how much evolution is limited (constrained) by morphology and environment. Ball stated:[56] (p. 332)

> "convergent evolution is often regarded as a sign that certain shapes or structures are ideal adaptations to particular environments for physical reasons: wings consisting of flat, thin membranes are best for flying, torpedo-shaped bodies a streamlined for efficient swimming, and so on… There is a tendency in evolutionary biology to regard natural selection as a process with an infinite palette: anything is possible so long as it doesn't break the laws of physics. But the laws of physics might impose more constraint than that, precisely because biology uses rather than merely suffers them."

*Universality*

Not all mutations produce a change in phenotype. There are neutral mutations. There are many more genotypes than phenotypes. In other words, genotype-phenotype maps are many to one.

Species that are unrelated or distantly related (in the tree of life) sometimes have traits or behaviours that are similar. Convergent evolution is the hypothesis that natural selection produced the same outcome in a different context.

*Modularity at the mesoscale*

The economist Simon pointed out that evolution can occur on much faster time scales than might be expected because of modularity.[23] According to Clune et al.[371]

> "A long-standing, open question in biology is how populations are capable of rapidly adapting to novel environments, a trait called evolvability [1]. A major contributor to evolvability is the fact that many biological entities are modular, especially the many biological processes and structures that can be modelled as networks, such as metabolic pathways, gene regulation, protein interactions and animal brains [1–7]."

Ball highlighted how domains in proteins provide functional modules that evolution uses:[56] (pp. 174-5)

> "the evolution of metazoan proteins is not so much a slow affair of letting random genetic mutations change one amino acid for another and seeing what effect it produces. Rather, it constitutes a reshuffling of already functional modules to produce multidomain molecules with new potential - a strategy much more likely to yield successful results… the "unit" of molecular evolution here is not really the base pair of DNA or the amino acid or protein, or the gene itself, by the peers at a scale intermediate between the two: the module of a domain. It seems that this shuffling, rather than the slow mutation of primary base sequences, is what has driven the evolution of animals."

Johnston et al.[372] considered an algorithmic picture of evolution that

> "suggests that symmetric structures preferentially arise not just due to natural selection but also because they require less specific information to encode and are therefore much more likely to appear as phenotypic variation through random mutations… many genotype–phenotype maps are exponentially biased toward phenotypes with low descriptional complexity. A preference for symmetry is a special case of this bias… Lower descriptional complexity also correlates with higher mutational robustness, which may aid the evolution of complex modular assemblies of multiple components."

*Self-organisation*

Complex biological structures, from proteins to organisms, have formed spontaneously due to evolution over millions of years. Their intricacy and functionality have led to claims of purpose and design. However, this is argued to be an "apparent" design, just like an economy whose self-organisation appears "as if" it is guided by an "invisible hand."[8]

Kaufman claimed[51] that self-organisation is more important than natural selection in driving evolution and this is contested.

*Unpredictability*

A contested question about evolution is the role of contingency (historical accidents) and whether the evolution of complex life forms, particularly humans, was an accident of history or inevitable.[373]

*Irreducibility*

Until recently evolutionary biology has been dominated by a reductionist gene-centric view, popularised by Dawkins. However, recent discussions about systems biology, evo-devo, and epigenetics have questioned this view. Some characterise these alternative views as form of structuralism.

*Complexity*

An algorithmic picture of evolution suggested that simplicity spontaneously emerges as many genotype-phenotype maps may be biased towards phenotypes with low descriptional complexity.[372]

*Toy models*

A key to the Modern Synthesis of evolutionary biology (that brought together Darwinian natural selection with Mendelian genetics) was the development of simple mathematical models. Mesoudi et al.[374] pointed out that "Significant advances were made in the study of biological [micro]evolution before its molecular basis was understood, in no small part through the use of simplified mathematical models, pioneered by Fisher (1930), Wright (1931), and J.B.S. Haldane (1932)..."

Nevertheless, sixty years ago objections were raised that these toy models ignored so many details, just like the objections often made in other fields of science. Mesoudi et al.[374] noted

> "…reservations about simplifying assumptions were voiced about the use of population genetic models in biology by the prominent evolutionary biologist Ernst Mayr (1963). He argued that using such models was akin to treating genetics as pulling coloured beans from a bag (coining the phrase "beanbag genetics"), ignoring complex physiological and developmental processes that lead to interactions between genes.
>
> In his classic article "A Defense of Beanbag Genetics," J.B.S. Haldane (1964)[91] countered that the simplification of reality embodied in these models is the very reason for their usefulness. Such simplification can significantly aid our understanding of processes that are too complex to be considered through verbal arguments alone, because mathematical models force their authors to specify explicitly and exactly all of their assumptions, to focus on major factors, and to generate logically sound conclusions. Indeed, such conclusions are often counterintuitive to human minds relying solely on informal verbal reasoning.
>
> Haldane (1964) provided several examples in which empirical facts follow the predictions of population genetic models in spite of their simplifying assumptions, and noted that models can often highlight the kind of data that need to be collected to evaluate a particular theory."[p]

Mayr appreciated the significance of emergence in biology but not of universality and how it can justify the use of toy models. A recent systematic justification for such mathematical models in evolutionary biology has been given by Servedio et al.,[88] who refer them as "proof-of-concept" models.

*Cross-fertilisation of fields*

Ideas from evolution stimulated the development of genetic algorithms in computer science.[1]

Drossel reviewed[377] connections between evolution and statistical physics, including a wide range of *toy models*. Examples include spin glass models that give rise to rugged landscapes for fitness and can describe hierarchal structures, comparable to Darwin's tree of life. Goldenfeld and Woese viewed evolution as a collective phenomenon far from equilibrium.[378] The toy model central to their discussion is directed percolation.

*Ecology*

Fifty years ago, Robert May argued[379] that as an ecosystems becomes more diverse (i.e., the the number of species in it increases) the less stable it becomes. He derived this counter-intuitive result using a simple model and results from Random Matrix Theory. This is an example of an emergent property: qualitative difference occurs as a system of interacting parts becomes sufficiently large. Mehta pointed out[q] that a "major deficiency of May's argument is that it does not allow for the possibility that complex ecosystems can self-organize through immigration and extinction. The simplest model that contains all these processes is the Generalized [to many species] Lotka-Volterra model (GLV)… Despite its simplicity, this equation holds many surprises, especially when the number of species is large".

*Evolution and ecology*

Two fundamental questions are the origins. of cooperation and of biodiversity. Game theory provides toy models for cooperation. Evolutionary game theory was reviewed by Frey.[380]

Mehta stated[r] that "Ecological and evolutionary dynamics are intrinsically entwined. On short timescales, ecological interactions determine the fate and impact of new mutants, while on longer timescales evolution shapes the entire community." Understanding this interplay is "one of the biggest open problems in evolution and ecology."

New experimental techniques for measuring the properties of large microbial ecosystems (including the relationship between fine-grained diversity and genetic sequencing) have stimulated significant theoretical work, including from some with a background in theoretical condensed matter physics.[381] Some recent work has provided new insights using techniques adapted from Dynamical Mean-Field Theory, which was originally developed to describe strongly correlated electron systems. One special case is when the interactions are reciprocal – how species i affects species j is identical to how species j affects species i. However, when the interactions are non-reciprocal the system can exhibit complex dynamical behaviour, including chaos.

---

[p] A full account of the debate between Mayr and Haldane and its historical and philosophical significance has been given by Rao and Nanjundiah[375] and by Dronamraju. [376]
[q] P. Mehta, "Understanding chaos and diversity in complex ecosystems – insights from statistical physics," https://www.condmatjclub.org/jccm_april_2024_03/
[r] ibid.

*Epidemology*

Disease epidemics, such as the recent Covid-19 pandemic, involve multiple length scales from the nanometre size of virion particles to neighbourhoods in which an outbreak occurs, to the thousands of kilometres that the infected may travel spreading the disease.

In the simplest epidemic models, such as the SIR [Susceptible-Infectious-Recover] model, there is a dimensionless parameter, $R_0$, known as the reproduction number, that determines how contagious the disease is. A value larger than one means an epidemic will occur. $R_0 = 1$ is a tipping point. Herd immunity occurs when a fraction of the population equal to $1-1/R_0$ is immune, e.g., due to vaccination. Herd immunity is an emergent property.[382]

More sophisticated toy models consider the network of interactions between people over which the disease can spread. In a network the degree of a node, denoted k, is equal to the number of edges connected to that node. In complex networks[383] in which there are hubs (super-spreaders), the ratio of the average value of k to the average value of $k^2$ is small, and the threshold for the phase transition to an epidemic becomes smaller. In scale-free networks,[266] which are common in both society and biology, an epidemic always occurs.

## 19. Economics

As an academic discipline, modern economics began with Adam Smith's, *The Wealth of Nations*, published in 1776. He considered how specialisation and the division of labour led to broad prosperity and the efficient allocation of resources. He asked readers to marvel at the unseen cooperation of the thousands of economic agents that enable the production, distribution, and sale of a simple woollen coat to a labourer. All the agents act locally and based on local information. There is no central control by a director who has all the necessary information. Smith said it is like there is an "invisible hand" guiding the whole process. The market self-organises. A spontaneous order emerges from all the local interactions. Information and decision-making are decentralised. A market can be adaptive without central co-ordination. Scarcities or oversupplies of parts of the process (whether materials, labour, knowledge) will often be solved without the intervention of a central coordinator.

The centrality of emergence to economics has been highlighted by Nobel Laureates Frederik Hayek, Herbert Simon, Thomas Schelling, and Paul Krugman. There perspectives will be discussed further below. Jean-Philippe Bouchaud began his career in theoretical condensed matter physics and moved into finance[384] and econophysics.[385] He is now chair of Capital Fund Management which has approximately 350 employees worldwide and manages about US$20 billion as of June 2025. He recently organised a multi-disciplinary symposium, "More is Different" at the College de France, that brought together economists, biologists, and

physicists.[84] Bouchaud's perspective and work on toy models will be considered in detail below. Didier Sornette's career has followed a similar trajectory.[386]

The *Journal of Economic Behaviour and Organisation* published a special issue in 2012 devoted to emergence in economics. The introduction to the issue by Harper and Lewis[33] highlighted how emergence has been defined differently by authors from different sub-fields (evolutionary, complexity, Austrian, spatial economics) and that a range of characteristics have been associated with emergence: novelty, unpredictability, irreducibility, self-organisation. They also noted how emergence relates to properties and processes, i.e., how self-organisation occurs.

Lewis has considered the role of emergence in the thought of the influential economist, Friedrich Hayek:[50]

> "the capacity of the price mechanism to coordinate the decision-making of a multitude of individuals, each of whom is pursuing his or her self-interest in the light of his or her own local knowledge, can be thought of as an emergent property of the market system. The power in question is emergent because it is possessed only by a particular whole – namely the free market system that is constituted by a group of people whose interactions are structured by the abstract rules of contract, tort and property law – and not by those individuals taken either in isolation or as a group whose interactions are governed and structured by some other set of rules. And that emergent whole – the people plus the relations engendered by the rule of law – cannot be eliminated from causal explanations of the coordinative properties of free markets because, if people were not so related, then their interactions would not be structured in the way required to produce an orderly outcome (Hayek, 1973: 39)."

Emergent concepts are also central to Marxist views of economics. "The law of transformation of quantity into quality and *vice versa*" due to Hegel was promoted by Friedrich Engels and was claimed to provide a scientific basis for Marxist-Leninist ideology, including claims about the qualitative difference between workers and capitalists, the instability of capitalism, and the inevitability of political revolution.[387]

Economic history is filled with bubbles and crashes. They are collective properties resulting from herding and are often not anticipated. The crashes have similarities to avalanches, where very small perturbations generate large disruptions. This has led to concepts such as Black Swans[59] (future shocks are unpredictable because of the power-law distributions associated with self-organised criticality), Dragon Kings[388] (extreme events that do not belong to the same population as the other events and so may be predictable), Minsky moments[5] (a sudden collapse of asset values marking the end of a growth phase in credit markets or business activity).

*Grand challenges and big questions*

Why do markets often work so well?
Why do markets sometimes fail spectacularly?
What are the domains of validity of neoclassical economics?

---

[5] https://www.newyorker.com/magazine/2008/02/04/the-minsky-moment

What is the relationship between microeconomics and macroeconomics?
To what extent can government policy promote economic prosperity and justice?
Can a government end an economic recession by "stimulus" spending?
How can one address problems that require collective action such as climate change, corruption and the “tragedy of the commons”?

Neoclassical economic theory often makes three assumptions: individual economic agents are rational, systems are close to equilibrium, and the system can be described by collective variables that are simply averages of the properties of individual agents. In other words, stochasticity, non-equilibrium, interactions between agents, heterogeneity, and fluctuations are often not considered. Emergence in physics suggests that each of these can lead to qualitatively new behaviour. This section will touch on work that explores when these assumptions may break down and the consequences. Bouchaud[389] and Sornette[390] have given a more detailed discussion of the limitations of neoclassical economics.

The “small shocks, large business cycle puzzle”, is a term coined by Ben Bernanke, Mark Gertler and Simon Gilchrist in a 1996 paper.[391] [Bernanke shared the 2022 Nobel Prize in Economics for his work on business cycles]. The paper begins with the following paragraph:

> “A longstanding puzzle in business cycle analysis is that large fluctuations in aggregate economic activity sometimes arise from what appear to be relatively small impulses. For example, large swings in investment spending and output have been attributed to changes in monetary policy that had very modest effects on long-term real interest rates.”

The excess volatility puzzle in financial markets was identified by Robert Shiller: The volatility "is at least five times larger than it "should" be in the absence of feedback". In the views of some, this puzzle highlights the failings of the efficient market hypothesis and the rationality of investors, two foundations of neoclassical economics. Shiller shared the 2013 Nobel Prize in Economics for this work.

According to Bouchaud,[278] “Asset prices frequently undergo large jumps for no particular reason, when financial economics asserts that only unexpected news can move prices. Volatility is an intermittent, scale-invariant process that resembles the velocity field in turbulent flows...”

The “tragedy of the commons” is a situation where a shared resource is depleted due to many individuals acting in their own self-interest. This collective outcome is detrimental to everyone, even though no single individual intended for that outcome to occur. This is an example of emergence as there is a qualitative difference between what happens at the micro- and the macro- levels. This is an example of what economists refer to as “collective action problems.”[392,393]

*System = components plus interactions*

Economic systems involve many components. Types of components include individuals (consumers, workers), professions, households, companies, public institutions, information, capital (human, financial, social), governments. Furthermore, for each type of component there may be many of that type. The numbers will depend on the economic phenomena of interest. Interactions between different components can be determined by price, geographic proximity, trust, laws, and social conventions. There are complex networks of interactions, as discussed below.

*Scales and hierarchies*

There multiple spatial scales associated with economic activity: global, regional, national, state, city, and neighbourhood. Sometimes spatial scales are emergent, as discussed by Krugman.[7] Time scales can vary from minutes to decades.

*Power laws*

A wide range of economic quantities have probability distributions that exhibit power laws covering many orders of magnitude in size. Quantities include income, wealth, firm size, and city size. They have been reviewed by Gabaix[394] and Yakovenko[395]. This means that there are "fat tails" in the probability distribution and extreme events are much more likely than in a system with a Gaussian probability distribution. Explaining the origin of these power laws remains a major challenge. One early explanation was given by Simon and his explanation has been adapted by those interested in scale-free networks.[265]

Some of these power laws exhibit *universality.* For example, Gabaix et al.[396] observed that a wide range of different financial markets exhibited fluctuations in stock price, trading volume and number of trades, with the same exponents. The universal form of the probability distribution for returns held over as many as eighty standard deviations. It described returns (both positive and negative) for periods ranging from one minute to one month, across different sizes of stocks, and a range of stock market indices. It included the most extreme events such as the 1929 and 1987 market crashes.

Schweitzer et al.[397] stated that “Economic systems are increasingly built on interdependencies, implemented through trans-national credit and investment networks, trade relations, or supply chains that have proven difficult to predict and control.” They showed that as environmental volatility increased above a critical value a network could undergo a phase transition from a state of high network efficiency to none.

In economics, different schools of thought have arisen partly because they have different answers to the following two related questions: Can economic behaviour be managed?
Is the spontaneous order that emerges in a free market always beneficial? Harper and Lewis stated[33] (p. 331) that in the Austrian school “emergent economic phenomena are typically regarded as socially beneficial in that they help to coordinate individual actions and promote the attainment of many individuals’ purposes (i.e. ends)… Kirzner (1982) argues cogently that normative claims about the socially benign character of a spontaneous order are distinct and separable from claims about the systematic nature of the social patterns generated by it.”

*Modularity at the mesoscale*

Dopfer and Potts argued[398] that emergence generates new rule systems (technologies, institutions) at the meso level by combining existing building blocks. Dopfer, Foster, and Potts,[399] argued that evolutionary economics should be understood in terms of a micro-meso-macro framework. Kharrazi et al.[400] argued modularity in economic networks is a key component (along with diversity and redundancy) to their resilience, drawing on similarities with ecological systems.

*Toy models*

*The Sante Fe Institute stock market model*

Arthur argued[401] that the standard neoclassical theory of financial markets "cannot account for actual market phenomena such as the emergence of a market psychology, price bubbles and crashes, the heavy use of technical trading (trades based on the recent history of price patterns) and random periods of high and low volatility (price variation)." This motivated development of the Sante Fe Institute stock market model.[401] The model has two phases which Arthur described[401] as follows.

> "At low rates of investors trying out new forecasts, the market behaviour collapsed into the standard neoclassical equilibrium (in which forecasts converge to ones that yield price changes that, on average, validate those forecasts). Investors became alike and trading faded away. In this case, the neoclassical outcome holds, with a cloud of random variation around it. But if our investors try out new forecasting methods at a faster and more realistic rate, the system goes through a phase transition. The market develops a rich psychology of different beliefs that change and do not converge over time; a healthy volume of trade emerges; small price bubbles and temporary crashes appear; technical trading emerges; and random periods of volatile trading and quiescence emerge. Phenomena we see in real markets emerge."

*Agent-based models*

These models can be viewed as generalised versions of the Ising model and are playing a similar role in economics and finance that the Ising model played in physics.[390] The simulation software Netlogo contains some economics models.[402] Bouchaud argued that there is a need for the phase diagrams of these models to be mapped out.[81] I would add that it would be worth exploring how these phase diagrams differ from those calculated from mean-field theories.

*Discrete choice models*

Bouchaud argued[403] that the random-field Ising model from physics can provide insight into a range of socio-economic phenomena including tipping points, hysteresis, and avalanches. I now consider some key elements used to justify the model: discrete choices, utility, incentives, noise, social interactions, and heterogeneity.

*Discrete choice*. The system consists of N agents {i=1, …, N} who make individual choices. Examples of binary choices are whether to buy a particular product, vote for a political candidate, believe a conspiracy theory, accept bribes, get vaccinated, or join a riot. For binary choices, the state of each agent i is modelled by an "Ising spin", $S_i = +1$ or -1.

*Utility.* This is the function each agent wants to maximise, i.e., what they think they will gain or lose by their decision. This could be happiness, health, ease of life, money, or pleasure. The utility $U_i$ will depend on the incentives provided to make a particular choice, the personal inclination of the agent, and possibly the state of other agents.

*Personal inclination.* Let $f_i$ be a number representing the tendency for agent i to choose $S_i=+1$.

*Incentives.* All individuals make their decision based on the incentives offered. Knowledge of incentives is informed by public information. This incentive F(t) may change with time. For example, the price of a product may decrease due to an advance in technology or a government may run an advertising program for a public health initiative.

*Noise.* No agent has access to perfect information to make their decision. This uncertainty can be modelled by a parameter beta, which increases with decreasing noise. According to the log-it rule[404–406] the probability that a particular decision is made has the same form as the Fermi-Dirac probability distribution where $1/\beta$ is the analogue of temperature.

$$P(S_i = +1; U_i) = \frac{1}{1 + e^{-\beta U_i}}$$

*Social interactions.* No human is an island. Social pressure and imitation play a role in making choices. Even the most "independent-minded" individual makes decisions that are influenced somewhat by the decisions of others they interact with. These "neighbours" may be friends, newspaper columnists, relatives, advertisers, or participants in an internet forum. The utility for an individual may depend on the choices of others. The interaction parameter $J_{ij}$ is the strength of the influence of agent j on agent i.

*Heterogeneity.* Everyone is different. People have different sensitivities to different incentives. This diversity reflects different personalities, values, and life circumstances. This heterogeneity can be modelled by assigning a probability distribution $P(f_i)$ for the personal inclination.

Putting all the ideas above together the utility function for agent i is the following.

$$U_i(t) = f_i + F(t) + \sum_{j \in \mathcal{V}_i} J_{ij} S_j(t-1)$$

This means that the minimal model to investigate is a Random Field Ising model. I now briefly review socio-economic insights discussed by Bouchaud.[403]

Bouchaud first considered a homogeneous population which reaches an equilibrium state. This is then described by an Ising model with an interaction (between agents) J, in an external field, F that describes the incentive for the agents to make one of the choices. The state of the model (in the mean-field approximation) is then found by solving the Curie-Weiss equation. In the sociological context, this was first derived by Weidlich and in the economic context re-derived by Brock and Durlauf.[407]

As first noted by Weidlich, a spontaneous "polarization" of the population occurs in the low noise regime $\beta > \beta_c$, i.e. [the average equilibrium value of $S_i$] $\phi^* \neq 0$ even in the absence of any individually preferred choice (i.e., F=0). When $F \neq 0$, one of the two equilibria is exponentially more probable than the other, and in principle the population should be locked into the most likely one: $\phi^* > 0$ whenever F>0 and $\phi^* < 0$ whenever F<0.

Unfortunately, the equilibrium analysis is not sufficient to draw such an optimistic conclusion. A more detailed analysis of the dynamics reveals that the time needed to reach equilibrium is exponentially large in the number of agents. As noted by Keynes, "in the long run, we are all dead." This situation is well-known to physicists but is not so well appreciated in other circles. For example, it is not discussed by Brock and Durlauf. Bouchaud discussed the meta-stability associated with the two possible polarisations, as occurs in a first-order phase transition. From a non-equilibrium dynamical analysis, based on a Langevin equation, the time τ needed for the system, starting around ϕ=0, to reach $\phi^* \approx 1$ is given by: $\tau \propto \exp[AN(1-F/J)]$, where A is a numerical factor. This means that whenever 0<F<J, the system should really be in the socially good minimum $\phi^* \approx 1$, but the time to reach it is exponentially large in the population size. The important point about this formula is the presence of the factor N(1−F/J) in the exponential. In other words, the system has no chance of ever getting there on its own for large populations. Only when F reaches J, i.e. when the adoption cost becomes zero will the population be convinced to shift to the socially optimal equilibrium. Bouchaud noted that this is very different from a highly-influential model in management science for the diffusion of innovation diffusion, based on a simple differential equation proposed by Bass in 1969.[408]

In physics, the existence of mutually inaccessible minima with different energy is a pathology of mean-field treatments that disappears when the interaction is short-ranged and fluctuations are taken into account. In this case, the transition proceeds through "nucleation", i.e., droplets of the good minimum appear in space and then grow by flipping spins at the boundaries. Based on this Bouchaud suggested an interesting policy solution when social pressure resists the adoption of a beneficial practice or product: subsidize the cost locally, or make the change compulsory there, so that adoption takes place in localized spots from which it will invade the whole population. The same social pressure that was preventing the change will make it happen as soon as it is initiated and adopted somewhere.

This analysis provides concepts to understand wicked problems. Societies get "trapped" in situations that are not for the common good and outside interventions, such as providing incentives for individuals to make better choices, have little impact.
Suppose the system of interest can be modelled by some type of Ising model where the pseudospin corresponds to two choices (good and bad) for each agent in the system. The policy maker wants to change something such as increase the incentive for agents to make the "good" choice. There are two qualitatively different possible behaviours and they are shown in the Figure below (taken from Bouchaud).
The vertical axis is the "magnetisation", i.e, the fraction of agents who make the good choice. The horizontal axis is the "external field", i.e, the level of incentive provided for agents to make the good choice.

**Fig. 1** Average opinion, or aggregate demand, or overall trust (etc.) $m$, as a function of the external incentive field $F(t)$ in the RFIM. For an imitation parameter $J < J_C$, the *curve* is smooth. But when $J > J_C$ a hysteresis effect appears with two possible stable states for the same value of $F$, plus one unstable state. As $F$ decreases the overall trust remains high even for negative values of $F$, before suddenly collapsing at $-F_C(J)$ to the lower branch (as indicated by the *arrow*). The hysteresis effect means that trust can only reappear if $F$ grows back beyond $+F_C(J)$, i.e. much beyond the value where the crash happened

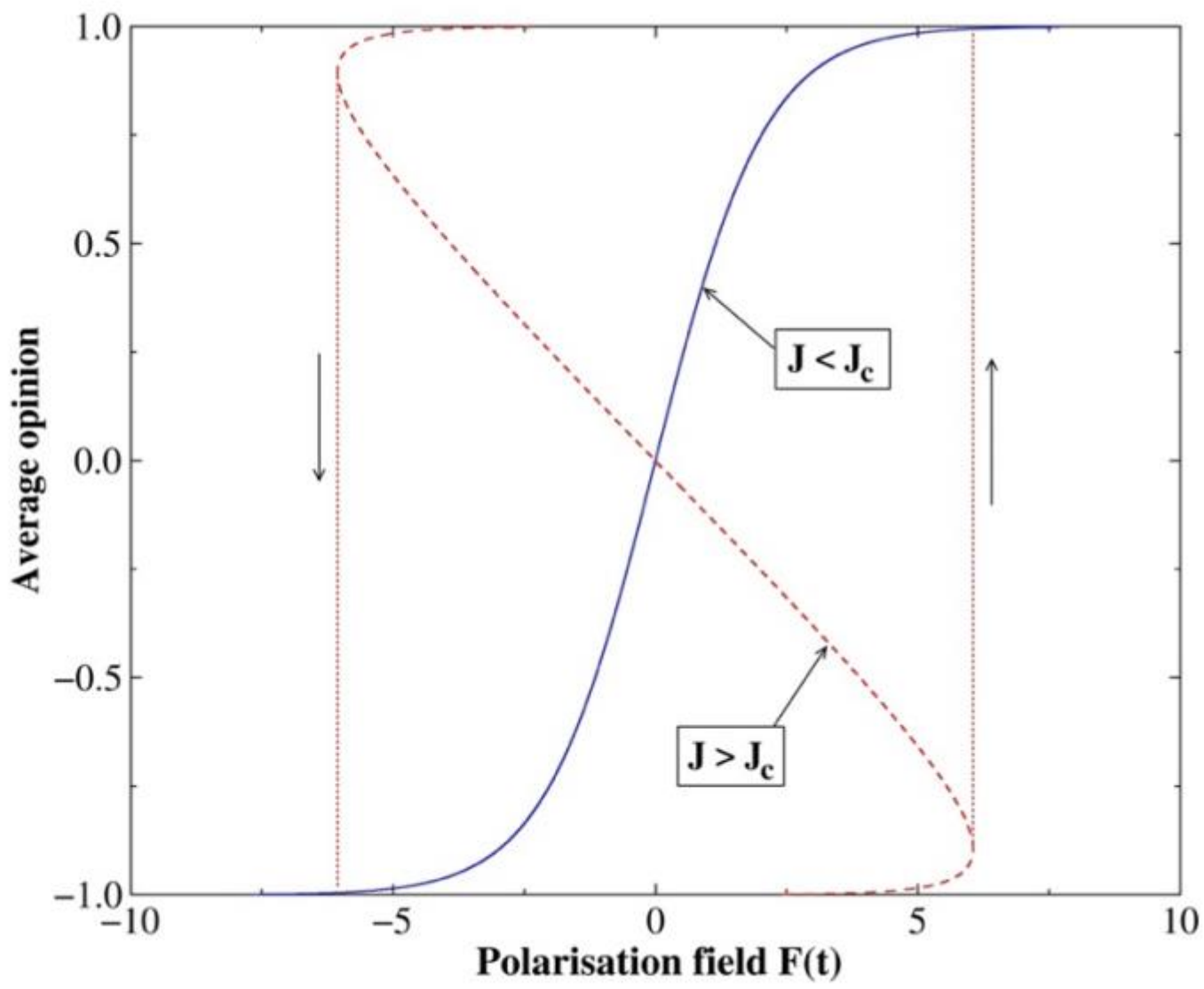

Case I. Smooth curve (blue). This occurs when the interaction between agents is weaker than some threshold strength. Suppose that a small but not insignificant minority of agents are already making the good choice and then incentive is increased slightly. If one is near the steep part of the blue curve, then this "nudge" can produce a desired outcome for the society. In economic language, there is a large multiplier.

Case II. Discontinuous curve (red). This occurs when the interaction between agents is greater than some threshold strength. People's choices are influenced more by their friends than by what the government or an NGO is telling them to do. Then large incentives must be provided to get a change in agent choice, far beyond the incentive required for a single isolated agent. The system is stuck in a state that is not good for society as a whole. It is in a metastable state.

On the other hand, if the "polarisation field" is sitting near a critical value (F ~ 5 in the figure, a tipping point), then a small nudge can lead to a dramatic change for good.

*A generalised Lotka-Volterra model*

This model provides an analogue between economic production networks and ecology. Earlier I mentioned recent work on this model, concerning how to understand the interplay of evolution and ecology. A key result is that in the large N limit (i.e., a large number of interacting species/agents) qualitatively different behaviour occurs. Ecosystems and economies can collapse. Bouchaud observed[278] that "any small change in the fitness of one species can have dramatic consequences on the whole system – in the present case, mass extinctions... most complex optimisation systems are, in a sense, fragile, as the solution to the optimisation problem is highly sensitive to the precise value of the parameters of the specific instance one wants to solve,... Small changes of these parameters can completely upend the structure of the optimal state, and trigger large-scale rearrangements, ..."

*Propagation of production delays along a supply chain*

Moran et al.[409] proposed a toy model for delay propagation and found that it exhibited a phase transition as the duration of mitigating buffers B, passed through a critical value, $B_c$.

When B is larger than $B_c$, the delays self-heal and large-scale system-wide delays are avoided. When B is smaller than $B_c$, the delays accumulate with bounds, creating system-wide disruptions. Close to the critical point, there is an "avalanche" of delays of all sizes.

*A firm network model*

For a toy model[410] the phase diagram has been mapped out as function of two parameters: the strength of forces counteracting supply/demand and profit imbalances and the perishability of goods. There are four distinct phases for the economy (a) collapse, (b) a relatively quick approach to equilibrium, (c) perpetual disequilibrium, with purely endogenous fluctuations, and (d) a deflationary equilibrium. This illustrates how a relatively simple model can describe diverse states.

## 20. Sociology

Emergence in sociology is the subject of books by Sawyer[9] , Elder-Vaas,[411] and Smith,[412,413] and reviews by Hodgson[414] and by Zahle and Kaidesoja.[415] As the discussion below will show, often the debates centre about abstract theorising and philosophical positions. Sometimes emergence is conflated with "critical realism", partly due to the legacy of Bhaskar's philosophy of social science.[416],[t] I find this conflation confusing as there are other distinctly different uses of the term "critical realism". It was originally a philosophical perspective that was advocated in the USA in the first half of the twentieth century by the philosopher Roy Wood Sellars. Today, it is also used in discussions about the relationship between science and theology.[417,418] Modestly and broadly, critical realism is the perspective that there is an real world that exists independent of our observations but we should be critical of our attempts to gain knowledge of that world. Hence, a scientist could have a reductionist orientation and still be a critical realist.

In contrast to the abstract theorising of some sociologists, emergence is central to the field of "sociophysics" which is largely populated by scientists with a background in statistical physics and is often concerned with toy models such as agent-based-models. Reviews are by Stauffer[419], Castellano et al.,[420] and Jusup et al.[421] Popular books have been published by Ball[422] and by Buchanan[423].

*Big question.* A fundamental and much debated question in sociology concerns the relationship between individual agency and social structures. Which determines which? Do individuals make choices that then lead to particular social structures? Or do social structures constrain what choices individuals make? In sociology, this is referred to as the debate between voluntarism and determinism.

Emergence is central to the debate about the tension between structure and agency. Anthony Giddens is a prominent sociologist, being the author of a widely used textbook. He has been a Director of the London School of Economics and is now a member of the House of Lords. In 2009, Giddens was listed as the fifth most cited author of books in the humanities and social sciences. Giddens proposed a social theory known as structuration, which has been criticised by Margaret Archer,[424] who in a 1995 book,[425] (p. 135) claimed that "there is a glaring absence of bold social theories which uncompromisingly make 'emergence' their central tenet". In the book Archer coined the term, elisionism, to characterise Gidden's structuration theory. Sawyer reviewed the relationship between elisionism and emergentism, in the following terms.[9]

[t] https://web.archive.org/web/20230127191619/http://www.asatheory.org/current-newsletter-online/what-is-critical-realism

In this chapter I compare the emergentist view of the social world with a contemporary alternative: elisionist theories, including both Anthony Giddens's structuration theory and socioculturalism in contemporary psychological theory. Elisionist theories share two foundational theoretical assumptions: They assume a process ontology, and they assume the inseparability of individual and social levels of analysis. A process ontology holds that only processes are real; entities, structures, or patterns are ephemeral and do not really exist. As for inseparability, the assumption is that the individual and the social cannot be methodologically or ontologically distinguished; thus the name "elisionism"…

Sawyer also provided another example.[9]

> "…socioculturalists in education argue that the individual learner cannot be meaningfully separated from the social and cultural context of learning, and they reject a traditional view of learning in which the learner is presumed to internalize knowledge presented from the external world. Rather than internalizing knowledge, the learner should be conceived of as appropriating or mastering patterns of participation in group activities. Learning involves a transformation of the social practices of the entire group and thus cannot be reduced to an analysis of what any one participant in the group does or knows."

A more balanced perspective on the relationship between structure and agency was given by Hayek, whose views Lewis described as follows.[426]

> "the relationship between structure and agency is one of non-reductionist codevelopment: both social structure and human agency possess their own sui generis, emergent causal powers, so although each depends on the other neither has ontological or analytical priority (cf. Hayek, 1967a: 76–77). In this way, Hayek avoids the reductionist extremes of voluntarism/individualism, whereby social structure are ontologically reducible to and determined by creative human agency, and holism/determinism, whereby people's actions are entirely determined by, and are therefore ontologically reducible to, over-arching social wholes. On the contrary, in order to understand the latter, therefore, the interplay of those powers must be examined."

This perspective has similarities to Noble's principle of biological relativity which claims there is no preferred level of causality.[339] In terms of scientific methodology, there are similarities to Mayr's perspective that in biology a balance is needed between reduction and analysis,[343] and Coleman's view that condensed matter physics involves an interplay between reductionism and emergentism.[78]

Social theorists who give primacy to social structures will naturally advocate solving social problems with large government schemes and policies that seek to change the structures. On the other side, those who give primacy to individual agency are sceptical of such approaches, and consider progress can only occur through individuals, and small units such as families and communities make better choices. The structure/agency divide naturally maps onto political divisions of left versus right, liberal versus conservative, and the extremes of communist and libertarian. An emergentist perspective is balanced, affirming the importance of both structure and agency.

*Discontinuities.* Lamberson and Page discussed the role of tipping points in the social sciences.[36] Gladwell popularised such ideas.[427,428]

*Modular structures*

Simon discussed the near decomposability of social systems, e.g., structures of organisations.[44] Following this a classic paper by Lawrence and Lorsch, considered the problem of differentiation and integration of units in business organisations affected productivity.[79]

Palla et al.[269] used statistical methods to uncover modular structures hidden in complex social networks, finding overlapping community structure of complex networks society interwoven sets of overlapping communities.

Zhou et al. argued that there were discrete hierarchical organisation of social group sizes.[429] Rather than a single or a continuous spectrum of group sizes, humans spontaneously form groups of preferred sizes organized in a geometrical series approximating 3-5, 9-15, 30-45, and so on. These different modules they labelled support cliques, sympathy groups, bands, cognitive groups, small tribes, and large tribes. They noted similarities to the organisation of land armies.

> "In the land armies of many countries, one typically finds sections (or squads) of ca. 10-15 soldiers, platoons (of three sections, ca. 35), companies (3-4 platoons, ca. 120-150), battalions (usually 3-4 companies plus support units, ca. 550-800), regiments (or brigades) (usually three battalions, plus support; 2500+), divisions (usually three regiments) and corps (2-3 divisions). This gives a series with a multiplying factor from one level to the next close to three."

Dunbar's number (150) is a suggested cognitive limit to the number of stable social relationships an individual can maintain, generally thought to be around 150. This is based on the correlation between primate brain size and average social group size. This is known as the social brain hypothesis.[430],[431] the need to live in large groups that are functional and cohesive selected for increased brain size among primates.

*A toy model for social segregation*

Individuals tend to like to associate with people with whom they have some commonality. This may involve profession, education, hobbies, gender, political views, language, age, wealth, ethnicity, religion, or values. But many people also enjoy or value some diversity, at least in certain areas of life. We all have varying amounts of tolerance for difference. A common social phenomenon is segregation: groups of people clump together in spatial regions or social networks with those similar to them. Examples include ethnic ghettos, cliques among teenagers, and "echo chambers" on the internet.

In 1971 Thomas Schelling published a landmark paper in the social sciences.[432] It surprised many because it showed how small individual preferences for similarity can lead to large scale segregation. The motivation for his work was to understand how racially segregated neighbourhoods emerged in cities in the USA.

Schelling's model is a nice example of emergence in a social system. A new entity [highly segregated neighbourhoods] emerges in the whole system that is qualitatively different from the properties of the components [which do not have a preference for segregation]. Besides novelty, it illustrates unpredictability as the macrobehaviour was not anticipated based on a knowledge of the properties of the components of the system.

One version of Schelling's model is the following. Take a square grid and each square can be red, blue or white (vacant). Fix the relative densities of the three quantities and begin with a random initial distribution. A person is "unhappy" if less than a certain percentage of their 8 neighbours on the lattice are like them. Then they then move to a nearby vacancy. After many iterations (moves) an equilibrium is reached where everyone is "happy". The Figure below shows how the final configuration varies significantly depending on the level of preference. What is striking and surprising is that the critical threshold of preference for segregation to occur is relatively low.

The model also illustrates how quantitative changes can lead to qualitative changes. In other words, there is a tipping point for segregation.

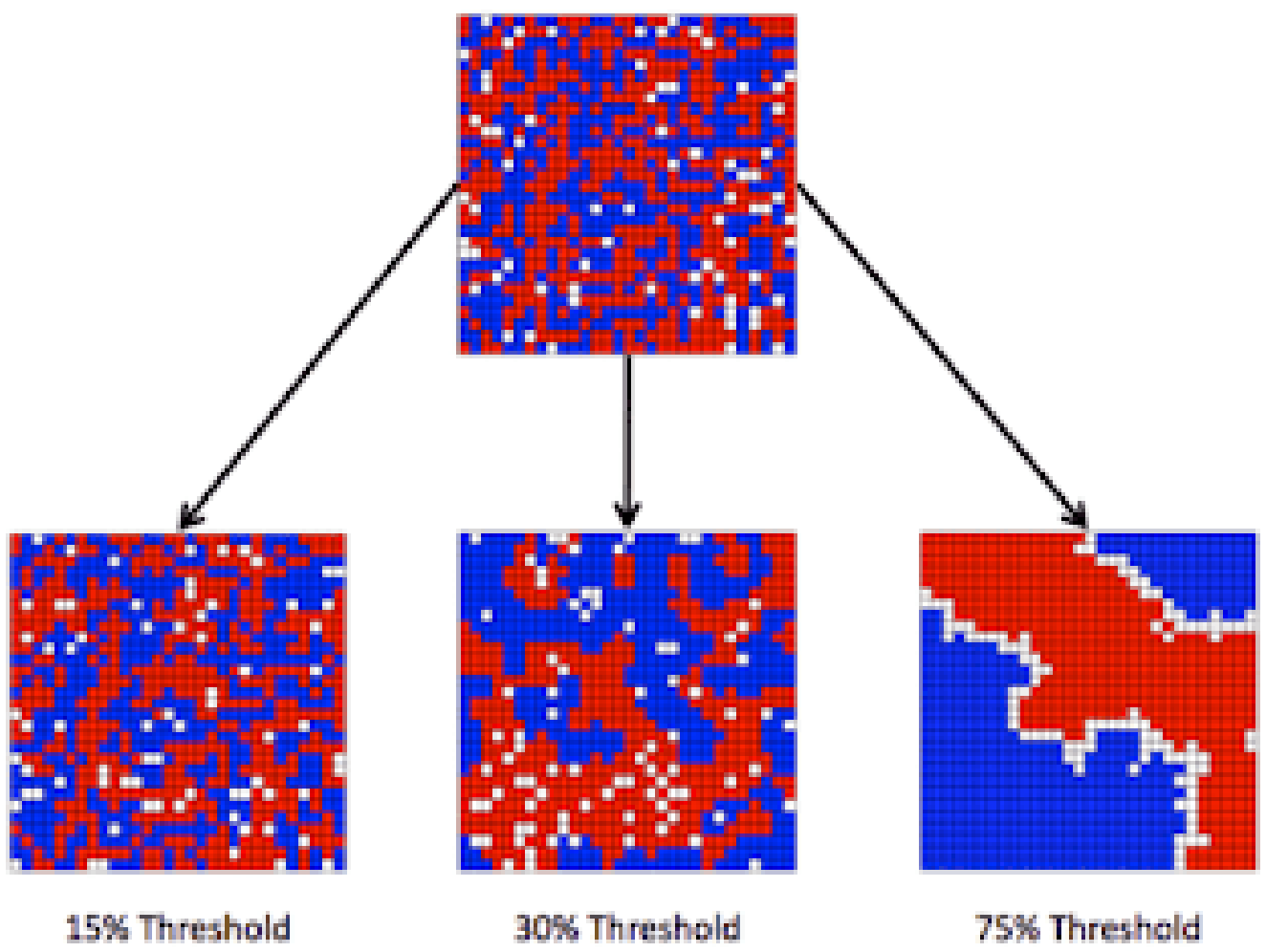

Figure 8. Dependence of segregation on the level of local preference for similarity. In Schelling's model in the initial configuration there is a random distribution of red and blue residents with white denoting a vacancy (top). All residents have a level of preference: they will move if a certain percentage of their nearest neighbours falls below a certain threshold. The bottom three panels show the final equilibrium neighbourhood configurations for different levels of preference. At low levels of preference, a mixed neighbourhood occurs with small regions of similarity (left). At high levels of preference, segregation occurs with a "no-mans-land" between the neighbourhoods (right). The surprising result of the model is that only a preference of 30 percent is required to produce segregation (centre). The figure is taken from Jackson et al.[96]

A major conclusion is that motives at the individual level are not the same as the outcomes at the macro level. People may be very tolerant of diversity (e.g. only have a preference that 30 per cent of their neighbours be like them) but collectively this still results in them living in very segregated neighbourhoods. Macrobehaviour and micromotives are distinct and may be different. This was the basis for the title of an influential book, *Micromotives and Macrobehavior*, that Schelling published in 1978.[8] Actions at the micro- level may not produce the macro- outcome that is desired or anticipated. In the other direction, if a particular phenomenon is observed at the macro- level the true explanation in terms of behaviour at the micro- level may not be what was anticipated. For example, if people are racist and only want neighbours just like themselves then segregated neighbourhoods will occur. However, the model shows that observation of segregation does not necessarily imply racism. In his book, Schelling warns about jumping to conclusions about individual intentions based on properties of the social aggregate.

Physicists have considered variants of Schelling's model that can be connected to lattice models from statistical mechanics,[433] including the Ising model[434] and classical spin-1 models such as the Blume-Capel model.[435]

Bouchaud recently argued[81] for the need for the phase diagrams of agent-based models to be mapped out. The need is illustrated by the fact that the diagram for the Schelling segregation model was only mapped out in 2009.[435]

Unlike in physics models, such as the Ising model, there is no global function (such as free energy) to minimise or maximise to determine the stable state of the system. In contrast, individual agents maximise their utility. Some work has been done on the Schelling model for intermediate situations.[436]

Schelling's model can be considered to be the first agent-based model. There is a simulation of the model in NetLogo.[u] It is fascinating that Schelling did not use a computer but rather did his "simulation" manually on a checkerboard!

*Granovetter's models*

Mark Granovetter has proposed two influential toy models. One paper shows how changes in social heterogeneity can induce a transition between two different collective states.[437] This has similarities to a transition in the Random Field Ising model. A second paper shows how "weak links" are necessary for social cohesion.[355] The model was inspired by hydrogen bonding in water.

## 21. Computer science

---

[u] Wilensky, U. (1997). NetLogo Segregation model. http://ccl.northwestern.edu/netlogo/models/Segregation. Center for Connected Learning and Computer-Based Modeling, Northwestern University, Evanston, IL.

Holland[1] and Fromm[12] discussed emergence in computer science. Hopfield's toy model for associative memory was influential in the development of neural networks in computer science. This was recognised by Hopfield sharing the Nobel Prize in Physics with Geoffrey Hinton, who together with Hopfield's former student Terry Sejnowski, proposed a Boltzmann machine (equivalent to a Sherrington-Kirkpatrick spin-glass model in a magnetic field) for machine learning.

## 19.1 Neural networks

John Hopfield received half of the 2024 Nobel Prize in Physics "for foundational discoveries and inventions that enable machine learning with artificial neural networks". The award was largely based on a single paper entitled, "Neural networks and physical systems with emergent collective computational abilities," that Hopfield published in 1982.[438] This work and its significance reflects many of the issues I highlight in this review.

Associative memory is the ability of a brain to retrieve stored information when provided part, or a corruption version, of that information. Hopfield viewed the function of associative memory as an emergent property of the brain. The paper proposed and studied a *toy model.* The model is essentially an Ising model. Neurons are represented by spins that can be in state +1 or -1, corresponding to whether the neuron is firing or not. Synapses (interactions) are represented by the inter-spin interaction strengths $J_{ij}$. They define a neural network. The $J_{ij}$ are not fixed but are trained by input data. This is related to Hebb's idea that neural systems learn associations through selective modification of synaptic connections.

Hopfield drew inspiration from Anderson's study of spin glasses and associated concepts, particularly that of rugged energy landscapes. This is nicely illustrated in the figure below produced by the Nobel Foundation.[v]

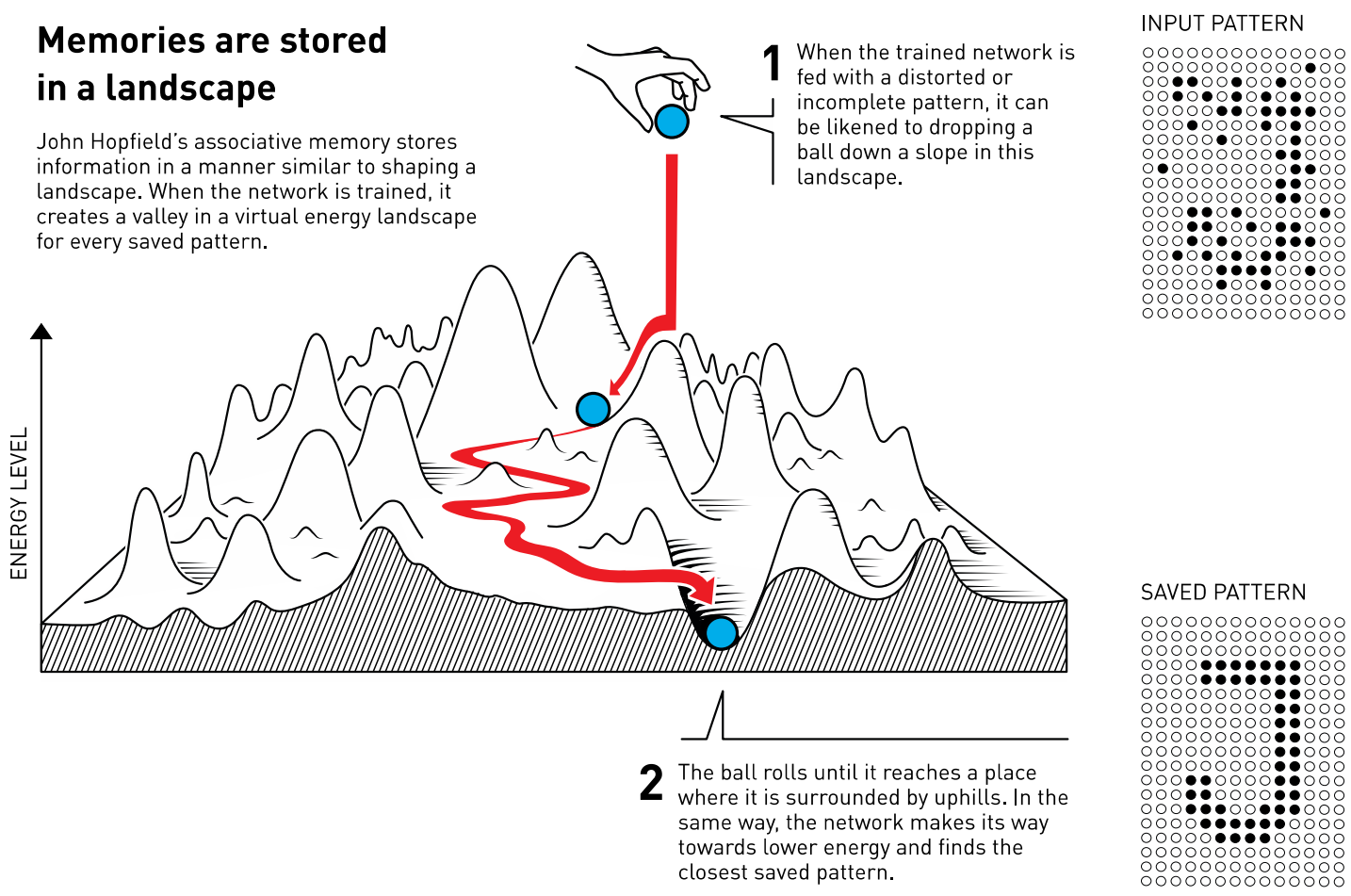

*Universality* (many details don't matter) plays out in two ways. First, it provides a basis for the relevance of a toy model. It overlooks a multitude of details that neuroscientists are concerned with. But it may still capture essential aspects of the phenomena. For example, in Hopfield's model, the interactions $J_{ij}$ are symmetric, i.e., the effect of neuron *i* on neuron *j* is the same as the effect of neuron *j* on *i*. This is inconsistent with experimental observations and Hepp's rule does not assume this symmetry. Making this assumption makes analysis of model much easier. Later studies considered a model with asymmetric interactions and found that the essential properties of the model were unchanged.

[v] https://www.nobelprize.org/uploads/2024/10/fig3_fy_en_24.pdf

Second, the universality associated with Hopfield's model gives insight into how associative memory may work, particularly how the brain might perform error correction. This is nicely illustrated in the figure above. An input (a rough memory) may contain errors, but the brain can process this to obtain an output (the associated memory) which is correct.

Hopfield's work involves several shifts in perspective, particularly from a reductionist view concerned with microscopic details, down to the molecular level, of how neurons work. Many details do not matter. The whole system matters: information is processed using the entire network structure. Unlike in a conventional computer information is stored in the $J_{ij}$ , not in the spins. The dynamics are asynchronous. Error correction is not explicitly programmed. Related shifts in perspective are involved in the design of neural networks in computer science: rather than a computational task being explicitly programmed the computer is trained to do the task.

*Interdisciplinarity.* A benefit of Hopfield's model was that it allowed theoretical methods developed for the study of spin glasses to be applied to understand the properties of the model. The figure below shows a *phase diagram* calculated accordingly.[439] It shows under what conditions reliable memory is emergent and when it does not emerge.

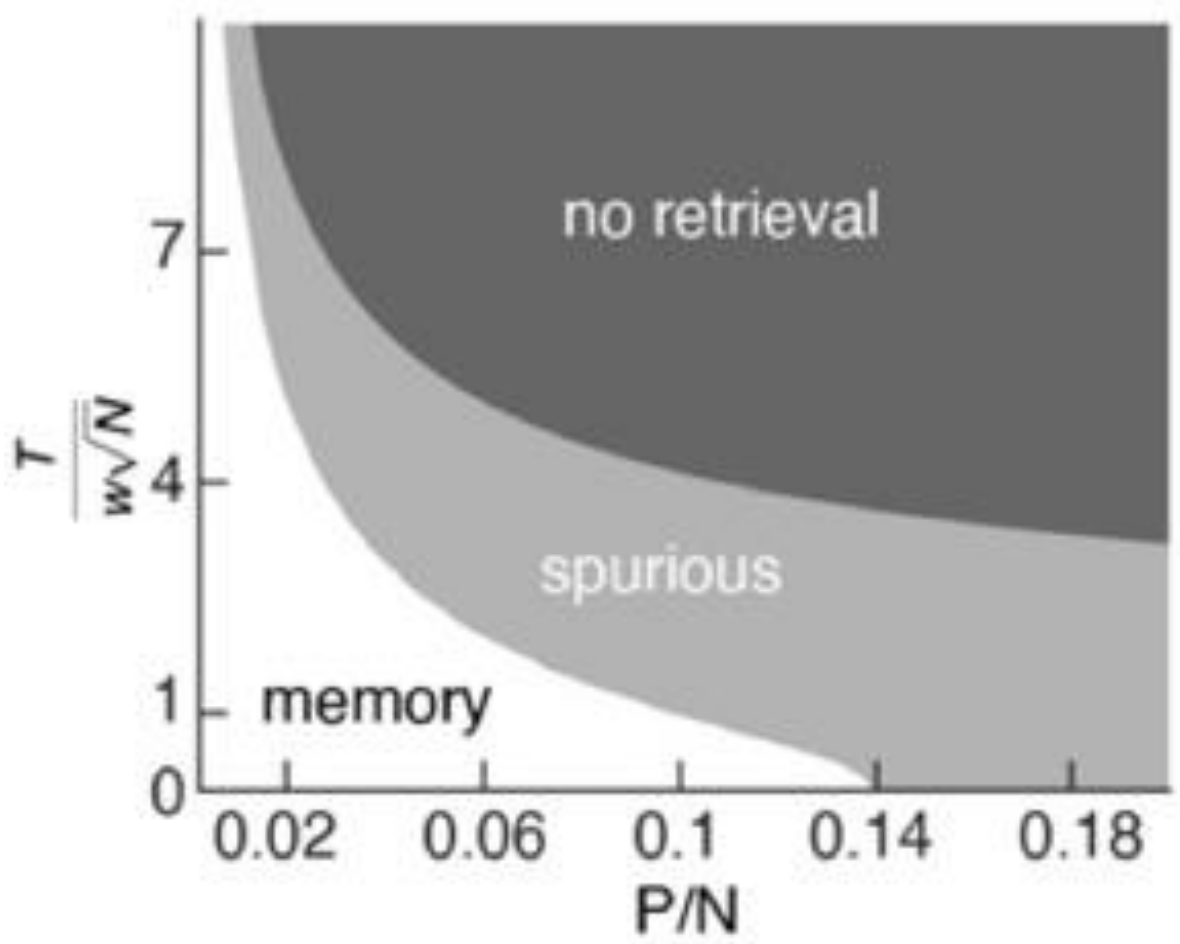

A specific memory is a state of the system, i.e., a specific spin configuration. N is the number neurons and P is the size of the training set, i.e., the number of different memories that the brain wants to store. T is a measure of noise and corresponds to temperature in the spin-glass model. w is the average strength of the synaptic connections, i.e., the $J_{ij.}$ 's. The phase diagram shows that the memory capacity, $P_c = 0.14$ N. This is close to Hopfield's original estimate. When P/N is larger than 0.14 spurious memories (confabulation) occur, i.e., the output is unrelated to the output.

Hopfield's model illustrates the adage that "All models are wrong, some are useful". The model also illustrates the role that physicists can play in inter-disciplinary research. In 1989 Sompolinsky, reviewed Hopfield's model and the associated work using ideas from spin glasses and stated that its "impact on neuroscience is marginal"[440]. However, things gradually changed. Hopfield's work may have had a bigger impact on computer science than neuroscience. Nevertheless, the work provided a slow route into neuroscience for many physicists. For example, Sompolinsky is now leader of neuroscience research groups at Harvard and the Hebrew University of Jerusalem, and was recently awarded The Brain Prize, the largest prize for brain research in the world. His two co-awardees also had Ph.D.'s in physics.

### 19.2 Large Language Models

The public release of ChatGPT was a landmark that surprised many people, both in the public and researchers working in Artificial Intelligence. Suddenly, it seemed Large Language Models had capabilities that some people thought were a decade away or even not possible. It is like the field underwent a "phase transition." That perspective turns out to be more than just a physics metaphor. It was made concrete and rigorous in a paper, "Emergent Abilities of Large Language Models," that was published in June 2022, by Wei et al.[30] Four years later it had been cited more than 5500 times. They used the definition, "Emergence is when quantitative changes in a system result in qualitative changes in behavior," citing Anderson's "More is Different" article. With regard to large language models: "An ability is emergent if it is not present in smaller models but is present in larger models. Size or scale is defined "primarily along three factors: amount of computation, number of model parameters, and training dataset size."

The essence of their analysis is summarised as follows.

> "We first discuss emergent abilities in the prompting paradigm, as popularized by GPT-3 (Brown et al., 2020). In prompting, a pre-trained language model is given a prompt (e.g. a natural language instruction) of a task and completes the response without any further training or gradient updates to its parameters… Brown et al. (2020) proposed few-shot prompting, which includes a few input-output examples in the model's context (input) as a preamble before asking the model to perform the task for an unseen inference-time example.
>
> The ability to perform a task via few-shot prompting is emergent when a model has random performance until a certain scale, after which performance increases to well-above random."

An example is shown in the Figure below.

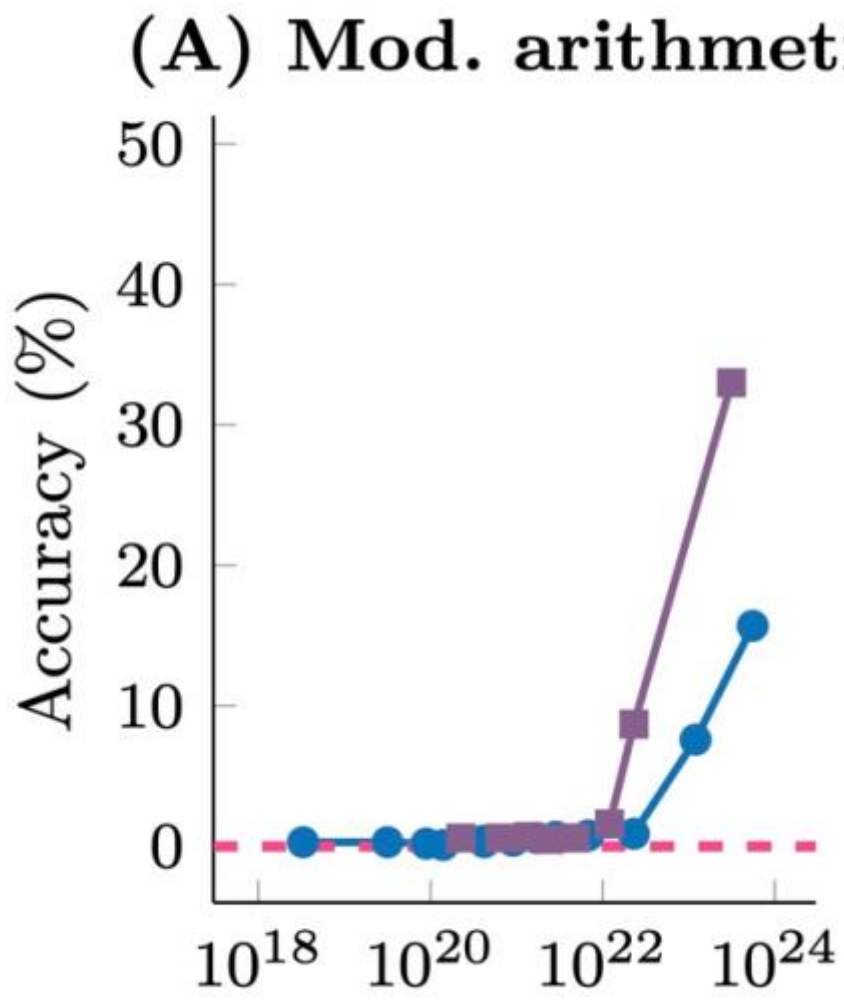

Figure. The horizontal axis is the number of training FLOPs [Floating Point Operations Per second] for the model, a measure of model scale. The vertical axis measures the accuracy of the model to perform a task, Modular Arithmetic, for which the model was not designed, but just given two-shot prompting. The red dashed line is the performance for a random model. The purple data is for GPT-3 and the blue for LaMDA. Note how once the model scale reaches about $10^{22}$ there is a rapid onset of ability.

The Figure above is taken from their Figure 2, which summarises results from a range of research groups studying five different language model families. It shows eight different emergent abilities.

Wei et al., pointed out that "there are currently few compelling explanations for why such abilities emerge the way that they do".

On the positive side, this paper presents hope that computational science and technology are at the point where AI may produce more exciting capabilities. On the negative side, there is also the possibility of significant societal risks such as having unanticipated power to create and disseminate false information, bias, and toxicity.

In May 2023, Schaeffer et al., asked, “Are Emergent Abilities of Large Language Models a Mirage?”[441]

> “we present an alternative explanation for [the claimed] emergent abilities: that for a particular task and model family, when analyzing fixed model outputs, emergent abilities appear due to the researcher's choice of metric rather than due to fundamental changes in model behavior with scale. Specifically, nonlinear or discontinuous metrics produce apparent emergent abilities, whereas linear or continuous metrics produce smooth, continuous predictable changes in model performance... we provide evidence that alleged emergent abilities evaporate with different metrics or with better statistics, and may not be a fundamental property of scaling AI models.”

One of the issues they suggest is responsible for the smooth behaviour is “the phenomenon known as neural scaling laws: empirical observations that deep networks exhibit power law scaling in the test loss as a function of training dataset size, number of parameters or compute" One of the papers they cite on power law scaling is by Hestness et al. from 2017.[442] They note that these empirical power laws are yet to be explained.

The concerns of Schaeffer et al. did not quench interest in emergence in LLMs. In March 2025, Berti et al., made a helpful review of the state of the field.[443] It begins with a discussion of what emergence is, quoting from Anderson's "More is Different" article and Hopfield's statement[438], "Computational properties of use to biological organisms or the construction of computers can emerge as collective properties of systems having a large number of simple equivalent components (or neurons)."

Berti et al. observed, "Fast forward to the LLM era, notice how Hopfield's observations encompass all the computational tasks that LLMs can perform." They discuss emergent abilities as in-context learning, defined as the "capability to generalise from a few examples to new tasks and concepts on which they have not been directly trained."

From a complex systems perspective, Krakauer, Krakauer, and Mitchell,[444] argued that LLMs demonstrate emergent capabilities but not "emergent intelligence." The latter is defined as "a consummate emergent property manifesting increasingly efficient -- cheaper and faster -- uses of emergent capabilities to solve problems." The ability to make analogies and "do more with less" is integral to intelligence.

*Scales.* Simple scales that describe how large an LLM is include the amount of computation, the number of model parameters, and the size of the training dataset. More complicated measures of scale include the number of layers in a deep neural network and the complexity of the training tasks.

Berti et al. noted that the emergence of new computational abilities does not just follow from increases in the simple scales but can be tied to the training process. I note that this subtlety is consistent with experience in biology. Simple scales would be the length of an amino acid chain in a protein or base pairs in a DNA molecule, the number of proteins in a cell or the number of cells in an organism. More subtle scales include the number of protein interactions in a proteome or gene networks in a cell. Deducing what the relevant scales are is non-trivial. Furthermore, as discussed earlier, Ball, Noble and Bishop, independently all emphasised that context matters, e.g., a protein may only have a specific function when the protein is located in a specific cell.

*Novelty.* When they become sufficiently "large", LLMs have computational abilities that they were not explicitly designed for and that "small" versions do not have. An alternative to defining novelty in terms of a comparison of the whole to the parts is to compare properties of the whole to those of a random configuration of the system. The performance of some LLMs is near-random (e.g., random guessing) until a critical threshold is reached (e.g., in size) when the emergent ability appears.

*Diversity.* The emergent abilities range "from advanced reasoning and in-context learning to coding and problem-solving." Wei et al. listed 137 emergent abilities in an Appendix! Berti et al. gave another example.

> "Chen et al. [15] introduced a novel framework called AgentVerse, designed to enable and study collaboration among multiple AI agents. Through these interactions, the framework reveals emergent behaviors such as spontaneous cooperation, competition, negotiation, and the development of innovative strategies that were not explicitly programmed."

*Discontinuities.* Are there quantitative objective measures that can be used to identify the emergence of a new computational ability? Researchers are struggling to find agreed-upon metrics that show clear discontinuities. That was an essential point of Schaeffer et al. In condensed matter physics, the emergence of a new state of matter is (usually) associated with symmetry breaking and an order parameter. Figuring out what the relevant broken symmetry is and the associated order parameter often requires brilliant insight and may even lead to a Nobel Prize (e.g., Neel, Josephson, Ginzburg, Leggett, Parisi,...) A similar argument can be made with respect to the development of the Standard Model of elementary particles and gauge fields. Identifying the relevant broken symmetries and effective theories led to Nobel Prizes. Another problem is that the discontinuities only exist in the thermodynamic limit (i.e., in the limit of an infinite system), and there are many subtleties associated with how the data from finite-size computer simulations should be plotted to show that the system really does exhibit a phase transition.

*Unpredictability.* The observation of new computational abilities in LLMs was unanticipated and surprised many people, including the designers of the specific LLMs involved. This is similar to what happens in condensed matter physics, where new states of matter have mostly been discovered by serendipity. Some authors seem surprised that it is difficult to predict emergent abilities. Berti et al. stated that while "early scaling laws provided some insight, they often fail to anticipate discontinuous leaps in performance" and this has been discussed in detail by Schaeffer et al.[445] Given the largely "black box" nature of LLMs, I don't find the unpredictability necessarily surprising. Predicting emergence is extremely hard for condensed matter systems, and they are much better characterised and understood.

*Modularity at the mesoscale.* Berti et al. did not mention the importance of this issue. However, they do mention that "functional modules emerge naturally during training" [Ref. 7,43,81,84] and that "specialised circuits activate at certain scaling thresholds [24]". Modularity may be related to a connection between a deep learning algorithm, known as the "deep belief net" of Geoffrey Hinton, and renormalisation group methods (which can be key to identifying modularity and effective interactions).[446]

*Is emergence good or bad?* Along with useful capabilities, undesirable and dangerous capabilities can emerge. Those observed include deception, manipulation, exploitation, and sycophancy. These concerns parallel discussions in economics. Libertarians, the Austrian school, and Hayek tend to see the emergence as only producing socially desirable outcomes, such as the efficiency of free markets [the invisible hand of Adam Smith]. However, emergence also produces bubbles, crashes and recessions.

*Toy models.* These were not discussed in the review of Berti et al. A useful example may be recent work of Nam et al. who studied, *"*An exactly solvable model for emergence and scaling laws in the multitask sparse parity problem."[447] In a similar vein, the review*, "*Statistical Mechanics of Deep Learning," by Bahri et al.[448] They considered a toy model for the error landscape for a neural network, and show that the error function for a deep neural net of depth D corresponds to the energy function for a D-spin spherical spin glass. [Section 3.2 in their paper].

## 22. Philosophical issues

Emergence has attracted considerable attention from philosophers as it relates to questions such as theory reduction in science and the nature of consciousness. Extensive discussions can be found in the article on "Emergent Properties" in *The Stanford Encyclopedia of Philosophy*[14], articles in *The Routledge Handbook of Emergenc*e,[15] books by Batterman[10] and Bishop,[11] Bishop, Silberstein, and Pexton,[64] and the thesis of Mainwood.[449] Hu was given a brief review[41] of perspectives on emergence by philosophers and highlighted how they differ from physicists in some terminology and criteria for emergence. For example, philosophers are often concerned with "strong emergence", "supervenience" and "downward causation." Hu points out that many philosophical discussions focus on consciousness, whereas it would be better to focus on less ill-defined emergent phenomena, such as occur in condensed matter. A similar point was made by McLeish, who considered biological physics as providing model systems to discuss downward causation and strong emergence.[450]

I now briefly discuss some of the important philosophical questions from a physics perspective.

*What is an explanation?*

Reductionists tend to claim that the ultimate explanation for a phenomenon is the one that operates at the lowest possible level. As mentioned earlier, Weinberg claims that the chain of explanation always points down. However, the question, "Why does a car travel between two cities?'' has several different answers, ranging from "because the owner wanted to visit his mother'' to "because the driver chose that route" to "because combustion of fuel produces expansion of gas that drives a piston in the engine." Determining which answer is "correct'' depends on the context of the original question. Different individuals, social groups, and contexts may have different standards and values that determine what is a "satisfactory'' explanation for a phenomenon.

*What is fundamental?*

Reductionists claim that the constituent particles of a system and the interactions between them are fundamental (i.e., the microscopic is more fundamental than the macroscopic). Genetics is more fundamental than cell biology, which is more fundamental than physiology. Physics is claimed to be more fundamental than chemistry. Elementary particle physicists, such as Steven Weinberg,[20,111] have sometimes claimed that their research is more fundamental than condensed matter physics. Laughlin argued that certain emergent properties are *exact* (such as quantisation of magnetic flux in a superconductor, hydrodynamics, and sound waves) and so they are more fundamental than microscopic theories.[4] (pp. 36-40).

In 2017-8, the Foundational Questions Institute (FQXi) held an essay contest to address the question, "What is Fundamental?" Of the 200 entries, 15 prize-winning essays were published in a single volume.[451] The editors gave a nice overview in their Introduction. In her prize-winning essay, Emily Adam argued[452] that what is fundamental is subjective, being a matter of values and taste. She raised questions about the vision and hopes of scientific reductionists, stating "the original hope of the reductionists was that things would get simpler as we got further down, and eventually we would be left with an ontology so simple that it would seem reasonable to regard this ontology as truly fundamental and to demand no further explanation." However, she noted how efforts Beyond the Standard Model [BSM], involve introducing more entities, particles, and parameters. She concluded:

> "... the messiness deep down is a sign that the universe works not 'bottom-up' but rather 'top-down,' ... in many cases, things get simpler as we go further up.
>
> But one might argue that this [reductionist fundamentalism] is getting things the wrong way round: the laws of nature don't start with little pieces and build the universe from the bottom up, rather they apply simple macroscopic constraints to the universe as a whole and work out what needs to happen on a more fine-grained level in order to satisfy these constraints."

*What is real?*

On the one hand, reductionists tend to claim that the fundamental constituent particles are what is real. Water may appear to be a continuous static fluid or uniform density. On the other hand, it can be viewed a rapidly fluctuating collection of molecules which are separated by empty space. Is the fluid real or are the molecules real?

*Are quasi-particles real?*

A fundamental question concerning quantum-many body systems is whether quasi-particles are real. Axel Gelfert argued[453] that quasi-particles were an "illusion" and used them as counter-example to a the entity realism, of the influential philosopher of science, Ian Hacking, "If you can spray them, they exist." Brigitte Falkenberg presented a counter-argument[454] (p. 245) to Gelfert stating:

> "quasi-particles are as real as a share value at the stock exchange. The share value is also due to a collective effect ..., namely to the the collective behavior of all investors. It is also possible to `spray' the share value in Hacking's sense, that is, to manipulate its quotation by purchase or sale for purposes of speculation. Its free fall can make an economy crash, its dramatic rise may make some markets flourish. And the crash as well as the flourishing may be local, i.e., they may only affect some local markets. But would we conclude that the share value does not exist, on the sole grounds that it is a collective effect? Obviously, share values as well as quasi-particles have another ontological status than, say, Pegasus. Pegasus does not exist in the real world but only in the tales of antique mythology. But quasi-particles exist in real crystals, as share values exist in real economies and markets. Indeed, both concepts have a well-defined operational meaning, even though their cause cannot be singled out by experiments or econometric studies."

Wallace has also argued that quasi-particles are real.[455] Guay and Sartenauer have given a philosophical perspective on whether anyons associated with the Fractional Quantum Hall effect are real.[456]

*Ontological versus epistemological emergence*

In philosophy, ontology concerns the nature of objects and what makes them "real". Epistemology concerns the process whereby we establish what is "true". Philosophers distinguish between *epistemological* emergence and *ontological* emergence. They are associated with prediction that is "possible in principle, but difficult in practice" and "impossible in principle" respectively. Silberstein and McGeever[57] discussed how the distinction between "weak" and "strong" emergence can also be viewed as a distinction between epistemological emergence and ontological emergence:

> "A property of an object or system is epistemologically emergent if the property is reducible to or determined by the intrinsic properties of the ultimate constitutents of the object or system, while at the same time it is very difficult for us to explain, predict or derive the property on the basis of the ultimate constituents. Ontologically emergent features are neither reducible to nor determined by more basic features. Ontologically emergent features are features of systems or wholes that possess causal capacities not reducible to any of the intrinsic causal capacities of the parts nor to any of the (reducible) relations between the parts."

Blundell[271] gave examples from condensed matter that he considered contrary to the claim of Silberstein and McGeever[57] that "epistemological emergence does not have any obvious ontological implications; but ontological emergence does."

*Weak versus strong emergence*

Chalmers gave the following definitons.[58]

> "*Weak Emergence of Phenomena*
> We can say that a high-level phenomenon is weakly emergent with respect to a low-level domain when the high-level phenomenon arises from the low-level domain, but truths concerning that phenomenon are unexpected given the principles governing the low-level domain ... It often happens that a high-level phenomenon is unexpected given principles of a low-level domain, but is nevertheless deducible in principle from truths concerning that domain.
>
> *Strong Emergence of Phenomena*
> We can say that a high-level phenomenon is strongly emergent with respect to a low-level domain when the high-level phenomenon arises from the low-level domain, but truths concerning that phenomenon are not deducible even in principle from truths in the low-level domain."

To illustrate some of the subtleties involved, consider the following clear statements from McLeish, Pexton, and Lancaster.[457]

> "We do not claim that our definition of ontological emergence will meet all tests of strength in comparison with accounts in the literature. For example, whether topologically ordered states should be described as examples of top down causation or not is not central to our claims. This is in part because all metaphysical assessments of the nature of the causal structure of a system depend on what philosophical account of causation one starts with. We are largely agnostic as to whether one wants to think of these cases in causal terms or not. In a similar vein, we argue that these cases of emergence are ontological (in the sense they are not artefacts of epistemology alone) but, since the philosophical literature has no universally agreed upon definitions of terms such as 'strong' or 'weak' emergence, these terms carry too much baggage to be useful. So, whether our account of emergence is strong or weak simply reflects how the reader typically uses the terms. What we are clear about is our claim that topological ordering is an ontological category in the world that is ineliminable in principle from any explanans of certain phenomena of the FQHS [Fractional Quantum Hall State] and some polymer systems."

My impression is that most physicists would subscribe to a view of "weak" emergence. In contrast, Ellis[313] and Drossel[314] have both argued that emergent phenomena in condensed matter support strong emergence. They argue that microscopic explanations of emergent phenomena are often dependent on the use of macroscopic concepts that are foreign to microscopic theory. Thus they fulfil Chalmer's definition of strong emergence that "truths concerning that [high-level] phenomenon are not deducible even in principle from truths in the low-level domain."[58]

*Logical relationship between different characteristics of emergence*

This article has discussed thirteen different characteristics of emergence and suggested that novelty should be the primary characteristic. An open question concerns the logical relationships between the different characteristics. Specifically, is the existence of one particular characteristic necessary, sufficient, or neither, for the existence of another characteristic? Philosophers have addressed some of these issues.

For example, Butterfield[42] argued he could "rebut two widespread philosophical doctrines about emergence. The first, and main, doctrine is that emergence is incompatible with reduction. The second is that emergence is supervenience; or more exactly, supervenience without reduction…I take emergence as behaviour that is novel and robust relative to some comparison class. I take reduction as, essentially, deduction…"

Mainwood summarized, "... systemic properties are novel, if and only if it is practically impossible to derive them from the microphysical properties mentioned in microphysical supervenience." (p. 30)[449] He argued that there were three entirely separate distinctions between wholes and parts (i.e., criteria for emergence) (Sec. 1.5): "1) a failure of intertheoretic reduction; 2) an impossibility of deducing the systemic properties from the properties of the parts; or 3) a failure of mereological supervenience."

*Structuralism*

Earlier I discussed structuralism in biology and sociology. In his book, *From Current Algebra to Quantum Chromodynamics: A Case for Structural Realism*, Tian Yu Cao stated,[458] (p.216)

> "Structuralism as an influential intellectual movement of the twentieth century has been advocated by Bertrand Russell, Rudolf Carnap, Nicholas Bourbaki, Noam Chomsky, Talcott Parsons, Claude Levi-Strauss, Jean Piaget, Louis Althusser, and Bas van Fraassen, among many others, and developed in various disciplines such as linguistics, mathematics, psychology, anthropology, sociology, and philosophy."

In different words, structuralism and post-structuralism are a big deal in the humanities and social sciences. Structuralism has been central to the rise and fall of a multitude of academic fashions, careers, and reputations. Cao defines structuralism as follows.

"As a method of enquiry, it takes a structure as a whole rather than its elements as the major or even *the only legitimate subject for investigations*. Here, a structure is defined either as a system of stable relations among a set of elements, or as a self-regulated whole under transformations, depending on the specific subject under consideration. The structuralist maintains that the character or even the reality of a whole is mainly determined by its structuring laws, and *cannot be reduced to its parts*; rather, the existence and essence of a part in the whole *can only be defined* through its place in the whole and its relations with other parts."

Structuralism and its historical development has been reviewed by Barbosa de Almeida.[459] In a sense, structuralism favours emergence over reductionism. However, it is more. Note some of the strong exclusivist language that I have put in italics in the quotation above. Structuralism is an overreaction to extreme reductionism.

There are also controversies in biology about structuralism, where it is put in conflict with perspectives labelled as adaptationism and functionalism. Ernst Mayr described these conflicts [54](p.118) as an emphasis on phenotype versus genotype, proximate vs. ultimate causes, what? versus why? questions, and on mechanics versus chance. Dwyer[460] argued that debates about structuralism versus functionalism were not just philosophical but concerned different methods of relating map and territory, and of object and environment.

Condensed matter physics has something concrete to contribute to debates about the merits and weaknesses of structuralism. Consider the case of Ising models defined on a range of lattices. We do not have an exclusive interest in the whole system or in the parts of the system. Rather, we want to know the relationship between macroscopic properties [different ordered states], mesoscopic properties [domains, long-range correlations, networks], and microscopic properties [the individual spins and their local interactions]. A greater importance or reality should not be assigned to the macro-, meso-, or micro-scale descriptions. All are needed and full understanding requires understanding the relationship between them. Hence, structuralism is an intellectual position that is hard to justify.

## 23. Notes on the history of the concept of emergence

There is a long history of philosophical discussion about emergence. Aristotle (384-322 B.C.) made comments that some interpret along the lines of "the whole is greater than the parts." The *sorites* paradox (also known as the paradox of the heap) concerns a heap of sand from which grains are removed one by one. Removing a single grain does not transform a heap into a non-heap. However, if the process is repeated enough times one is left with a single grain. It is not a heap. So, when does the transition from heap to non-heap occur?

The word *emergence* as a scientific term was first used in 1875 by George Henry Lewes (1817-1878), a philosopher and amateur physiologist, who wrote a five-volume treatise, *The Problems of Life and Mind*. Following Lewes, British philosophers such as C. Lloyd Morgan (1852-1936), Samuel Alexander (1859-1938), and C.D. Broad (1887-1971) initiated extensive discussions about emergence, particularly with regard to consciousness and biological evolution.[461]

The idea that quantitative changes can lead to qualitative changes has a long history and a controversial heritage. Using the example of transitions between distinct states of matter, the philosopher Georg Friedrich Hegel (1770-1831), proposed the first of three "laws of dialectics": "The law of transformation of quantity into quality and *vice versa*." This "law" was promoted by Friedrich Engels (1820-1895) and was claimed to be part of the scientific basis of Marxist-Leninist ideology, including claims about the qualitative difference between workers and people with capital, the instability of capitalism, and the inevitability of political revolution. For a summary of the history see Carneiro[462] and A.G. Spirkin, *The Great Soviet Encyclopedia*, 3rd Edition. S.v. "Transformation of Quantitative into Qualitative Changes."[387] That encyclopedia entry illustrates the role this "law" played in Soviet Ideology.

Discussions of emergence in condensed matter physics usually cite the seminal article, *More is Different* published in 1972 by Phil Anderson. The concept of spontaneous symmetry breaking was central to his discussion of the hierarchical nature of reality and the limitations of reductionism. However, the word "emergence" does not appear in the article.
Following Anderson's Nobel Prize in 1977, he received many invitations to speak to groups outside the physics community, including biologists, some of whom were fans of "More is Different". This then exposed Anderson to the thinking and terminology of the biologists. He first used the term "emergence" in print in 1981. This history is nicely discussed by Andrew Zangwill in Chapter 12 of his biography of Anderson, *Mind over Matter*.[463]

In the social sciences the intellectual origin of the concept of emergence are usually attributed to the work of Michael Polanyi in the 1960s. Polanyi began his academic career as a physical chemist, using quantum theory to understand how chemical reactions proceed. After fleeing Nazi Germany in 1933, his interests gradually moved to the social sciences, mostly economics, and then the philosophy of science. Emergence is the title of one of three chapters in his small book *The Tacit Dimension* published in 1966. The book is based on the Terry Lectures that Polanyi gave in 1962 at Yale University. Emergence is discussed in the context of the relationship between the nature of objects (ontology) and what we know about them (epistemology), particularly in biology. A measure of the enduring influence of Polanyi's book is that in 2009 it was republished by University of Chicago Press, with a new Forward by Amartya Sen, a Nobel Laureate in Economics.[17]

## 24. Conclusions

### *Scientific strategy and resource allocation*

Every scientist has limited resources: time, energy, money, lab space, equipment, facilities access, social networks, and collaborators. Where to use these resources requires strategic decisions about the relative priority of different questions, goals, and methods of investigation. Research groups, departments, institutions, professional societies, and funding agencies must and do also make decisions about such priorities. The decision outcomes are emergent properties of a system with multiple scales from the individual scientist to global politics. I wonder whether these weighty decisions are too often made implicitly, rather than explicitly following debate and deliberation.
There are many elements to the strategy that is chosen by an individual scientist or a community to study and understand a particular emergent phenomenon. Some elements may be implicit since they reflect choices and discoveries that are well-established and accepted. Pioneering a new method may require significant insight or technical skill. Elements include choice of scale to focus on, differentiation and integration of parts, effective theories, toy models, intellectual synthesis, discovering new systems, developing new experimental probes, and interdisciplinarity. Given these multiple elements, individuals and communities must prioritise the relative amounts of resources (time, personnel, money, infrastructure) that will be allocated to the different elements.

Joseph Martin argued[464] that in the period 1939-1993 philosophical perspectives about emergence versus reductionism had a significant impact on the physics community in the USA. This issue came to a head with conflicting congressional testimony from Phil Anderson and Steven Weinberg concerning the Superconducting Super Collider.[20,463]

A key issue is big science versus tabletop science. Or in different words, balancing large teams with a single ambitious goal versus many small independent teams driven by curiosity. Increasingly government funding agencies seem to favour the former over the latter. The 2023 Nobel Prizes illustrate the dilemma of size. John Hopfield shared the Physics prize based on one single author paper that proposed a toy model which he studied using a state-of-the-art supercomputer. In contrast, John Jumper and Demis Hassabis shared the Chemistry prize based on several papers that each involved approximately fifty authors, mostly employed by Google.

There is an increasing trend to fund big computational initiatives that tackle emergent problems largely by brute force such as the Materials Genome Initiative,[465] or The Human Brain Project. The latter was an approximately $1.6 billion effort to build "a complete computer simulation of the human brain", funded by the European Commission. The US has also funded a massive project, The Brain Initiative, focussed on developing new measurement techniques. In *The New York Times*, Gary Marcus, a psychologist, criticised these initiatives in 2014.[w]

> "The controversy serves as a reminder that we scientists are not only far from a comprehensive explanation of how the brain works; we're also not even in agreement about the best way to study it, or what questions we should be asking.
> .... a critical question that is too often ignored in the field: What would a good theory of the brain actually look like?
> ..... biological complexity is only part of the challenge in figuring out what kind of theory of the brain we're seeking. What we are really looking for is a bridge, some way of connecting two separate scientific languages — those of neuroscience and psychology....
> We know that there must be some lawful relation between assemblies of neurons and the elements of thought, but we are currently at a loss to describe those laws.......
> The problem with both of the big brain projects is that too few of the hundreds of millions of dollars being spent are devoted to spanning this conceptual chasm."

The rise of Artificial Intelligence has led to hype, grand claims, and ambitious projects concerning its application in science. These have been critiqued by Peter Coveney and collaborators.[466,467] Hoffmann and Malrieu pointed out the merits and the pitfalls of AI in theoretical chemistry.[468] Leeman et al. recently pointed out how some recent claims of AI guided materials prediction are mistaken and suggest constraints from solid state chemistry that need to be taken into account.[469] Although, the achievements of AlphaFold are impressive, questions have been raised about how much it has actually learned about protein

[w] https://www.nytimes.com/2014/07/12/opinion/the-trouble-with-brain-science.html

folding energy landscapes and whether its successes largely come from structure memorisation.[470]

Funding agencies should make a balanced portfolio of investments. Scientific communities should appreciate and promote a diversity of approaches.

*A holy grail: manipulation and control of emergent properties*

> "The philosophers have only interpreted the world, in various ways. The point, however, is to change it."
> Karl Marx[x]

> "The curious task of economics is to demonstrate to men how little they really know about what they imagine they can design."
> Friedrich A. Hayek, *The Fatal Conceit: The Errors of Socialism*

Understanding complex systems with emergent properties is an ambitious scientific challenge. A different challenge is to use any understanding gained to modify, manipulate, or control the properties of systems with emergent properties. A related challenge is to design systems with desired properties. This enticing prospect appeals to technologists, investors, activists, and governments. Some may be motivated by a desire to promote the common good. On the other hand, some may be motivated by greed or malice. Promises of controlling emergent properties feature prominently in grant applications, press releases, and reports from funding agencies [471]. Examples include chemical modification of known superconductors to produce room-temperature superconductivity, drug design, "nudges" in social policy, leadership of business corporations, and governments attempting to manage the economy.

It would be nice to understand superconductivity well enough to design a room-temperature superconductor. But this pales in significance compared to the "holy grail" of being able to manage economic markets to prevent bubbles, crashes, and recessions.
There are fundamental questions as to what extent control of emergent properties is possible. For example, Bouchaud argued that in systems with self-organised criticality optimisation and fragility are in tension with one another.[278] He illustrated this with the balancing stick problem.[472] An increase in ability to stabilize the system makes it more difficult to predict its future evolution!

On the quest of managing the economy, Bouchaud argued[278] that the quest for efficiency and the necessity of resilience may be mutually incompatible. This is because markets may tend towards self-organised criticality which is characterised by fragility and unpredictability (Black swans). He concluded that

> "the main policy consequence of fragility in socio-economic systems is that any welfare function that system operators, policy makers of regulators seek to optimize should contain a measure of the robustness of the solution to small perturbations, or to the uncertainty about parameters value.

[x] This statement is engraved on his tombstone.

> Adding such a resilience penalty will for sure increase costs and degrade strict economic performance, but will keep the solution at a safe distance away from the cliff edge. As argued by Taleb… good policies should ideally lead to "anti-fragile" systems, i.e., systems that spontaneously improve when buffeted by large shocks."

Materials by design is a grand challenge in materials science. The dream is that if an engineer wants a material with specific properties, such as room temperature superconductivity and to be processible into durable wires that can sustain large electrical currents, these requirements would be entered into a computer that would then calculate the chemical composition, synthesis method, structure, and physical properties of the best candidate materials. Given the unpredictability characteristic of emergent properties this is a formidable and controversial challenge.

It was claimed that the Human Genome project would revolutionise medicine and drug discovery. In 2002, Richard Sykes, Chairman of the pharmaceutical giant, GlaxoSmith Kline (GSK) claimed[y] that "The past in drug discovery was about serendipity; the future is about predictability" and that GSK will become the "Microsoft of pharmaceutical industry". [z]Since 2002, the share price of GSK has changed little, whereas that of Microsoft has increased by a factor of 20! Arguably, the hype ignored the fact that the Human Genome project only provided information about the components of the system (genes) and not the interactions between them (gene networks). Furthermore, these interactions can be dependent on context, as emphasised by Ball[56] and Noble.[48,339]

We should honestly reflect on decades of "scientifically informed" and "evidence-based" initiatives in materials science, medicine, poverty alleviation, government economic policy, business management, and political activism. Unfortunately, the fruit of these initiatives is disappointing compared to what has been promised.

My goal is not to promote despair but rather to prevent it. With more realistic expectations, based on reality rather than fantasy and hype, we are more likely to make significant progress in finding ways to make some progress (albeit modest but worthwhile) in learning how to manipulate these complex systems for better outcomes. Understanding complex systems is hard. Controlling them is much harder.

*The disposition of the scientist*

All scientists are human. In our professional life, we have hopes, aspirations, values, fears, attitudes, expectations, and prejudices. These are shaped by multiple influences from the personal to the cultural to the institutional. We should reflect on the past century of our study of emergent systems from physics to biology to sociology. Our successes and failures may lead us to certain dispositions.

*Humility.* There is so much we do not understand. Furthermore, we usually fail at predicting emergent properties. This is not surprising. Unpredictability is one of the characteristics of emergent properties. Nevertheless, there is sometimes hubris associated with grand initiatives such as “The Theory of Everything”, the Human Genome Project, “materials by design”, “Human Brain Project” (Europe), “The Brain Initiative” (USA), Human Connectome Project[473] and macroeconomic modelling.

---

[y] Nils Pratley, "Drugs cocktail is a downer" The Guardian, July 24, 2002. https://www.theguardian.com/business/2002/jul/24/businessofresearch.research

Specialists need to be open to toy models that ignore almost all their hard-earned knowledge and expertise. Universality tells us many details may not matter. We need to open to a change in perspective concerning how we look at a system.[474] This may mean focusing on a different scale or on the interactions rather than the units.

*Quiet optimism.* Based on previous successes at identification of relevant scales, modularity at mesoscale, toy models, and universality.

*Expect surprises.* There are many exciting discoveries waiting. Based on past experience, they will be found by curiosity and serendipity, not by computational brute force or uncritical collections of massive data sets.

*Wonder.* Emergent phenomena are incredibly rich and beautiful to behold, from physics to biology to sociology. Furthermore, the past century has seen impressive levels of understanding. But this is a “big picture” and “coarse-grained” understanding, not the description that the reductionists lust for and claim possible.

*Realistic expectations.* Given the considerations above I think we should have modest expectations of the levels of understanding possible, and what can be achieved by research programs, whether that of individual scientists or billion-dollar initiatives. We need to stop the hype. Modest expectations are particularly appropriate with respect to our ability to control emergent properties.

*Open questions*

There are many fascinating and important questions about emergent phenomena in specific systems and sub-disciplines. Many of the philosophical questions about emergence remain open. Although, I consider them fascinating and important, I do not consider them below. Rather, my concern is about questions of greater relevance to scientific practise.
Different types of questions concern the extent to which there may be a general “theory” of emergence. Or more modestly, are there some general concepts, principles and methods that are universal in some sense, i.e., they are relevant across disciplines. Examples include symmetry breaking, renormalisation, universality, coarse-graining, and effective theories. In the past organising principles such as self-organised criticality, protectorates, order at the edge of chaos, and catastrophe theory, have been touted. However, they remain contentious.
I have argued that novelty is the defining characteristic of emergence but described more than ten other possible characteristics.

An open question is what is the logical relationship between each of these characteristics? Is one characteristic necessary, sufficient, or independent of the another? I have provided a few counter examples to some of these logical connections.

I have emphasised that often modularity occurs at a meso-scale. A system can be described in terms of collections of weakly interacting modules. A question is whether for a specific system one such modules must exist and is there a method that provides objective quantitative criteria that allows determination of what those modules are. For example, in quantum many-body systems are much study with profound insight the quasi-particles are discovered. But do quasi-particles exist for any system with physically reasonable interactions? Is there a general method to identify them just starting from the Hamiltonian? I briefly mentioned work by Hoel et al.,[341] Rosas et al.,[46] and Palla et al.[269] on systematic approaches to identifying modular structures. It would be helpful to have a comparison of these methods and an attempt to integrate them into a more general framework.

How does one connect strata? Are there optimal ways for doing coarse graining? Is there an objective way to identify distinct levels that occur in the stratification of reality? Crutchfield claimed[275] that this could be made operational, stating that "different scales are delineated by a succession of divergences in statistical complexity at lower levels."

**Acknowledgements**

I acknowledge the Traditional Custodians of the land on which all this work was done, the Turrbal and Yuggera peoples. I pay my respects to their Elders past, present and emerging.

Several key references were introduced to me by Peter Evans, Luke McKenzie, Gerard Milburn, and Stephen Ney.

To quote B.L. Hu,[41] "This kind of non mission-driven, non utilitarian work addressing purely intellectual issues is not expected to be supported by any U.S. grant agency." Nevertheless, I am grateful for generous and substantial financial support over many years for projects that produced relevant ideas and experiences that helped shape this review. I acknowledge the University of Queensland, Australian Research Council, and John Templeton Foundation.

My interest in and understanding of emergence was enhanced by involvement in the International Institute for Complex Adaptive Matter, funded by the National Science Foundation.

I thank the "holy" scribblers writers collective for their interest and support.